\documentclass[aps,prl,reprint,groupedaddress,showpacs,amsmath,amssymb,ocolumn,floatfix,longbibliography]{revtex4-1}
\usepackage{graphicx}
\usepackage{bm}
\usepackage{color}
\usepackage[normalem]{ulem}
\usepackage{amsmath}
\usepackage{tabularx}
\usepackage{lineno}

\usepackage[colorlinks=true]{hyperref}  
\hypersetup{
    unicode=false,          
    pdftoolbar=true,        
    pdfmenubar=true,        
    pdffitwindow=false,     
    pdfstartview={FitH},    
    pdftitle={XXX},    
    pdfauthor={XXX et al.},     
    pdfsubject={},   
    pdfcreator={},   
    pdfproducer={}, 
    pdfkeywords={} {} {}, 
    pdfnewwindow=true,      
    colorlinks=true,       
    linkcolor=magenta, 
    citecolor=blue,        
    filecolor=magenta,      
    urlcolor=blue           
} 
\usepackage{cleveref}

\newcommand{\Rpara}{$R_{\parallel}$}
\newcommand{\Rperp}{$R_{\perp}$}

\usepackage{pifont}

\usepackage{bbm}

\renewcommand{\vec}[1]{\boldsymbol{#1}}

\begin{document}

\title{Stripe Order in the Metallic and Superconducting Phases of Rhombohedral Hexalayer Graphene}

\author{Peiyu Qin$^{1}$}
\thanks{These authors contributed equally to this work.}
\author{Hai-Tian Wu$^{2}$}
\thanks{These authors contributed equally to this work.}
\author{Ron Q. Nguyen$^{2}$}
\thanks{These authors contributed equally to this work.}
\author{Erin Morissette$^{2}$}
\thanks{These authors contributed equally to this work.}
\author{Naiyuan J. Zhang$^{2}$}
\author{K. Watanabe$^{3}$}
\author{T. Taniguchi$^{4}$}
\author{J.I.A. Li$^{1,2}$}
\email{jia.li@austin.utexas.edu}

\affiliation{$^{1}$Department of Physics, University of Texas at Austin, Austin, TX 78712, USA}
\affiliation{$^{2}$Department of Physics, Brown University, Providence, RI 02912, USA}
\affiliation{$^{3}$Research Center for Functional Materials, National Institute for Materials Science, 1-1 Namiki, Tsukuba 305-0044, Japan}
\affiliation{$^{4}$International Center for Materials Nanoarchitectonics,
National Institute for Materials Science,  1-1 Namiki, Tsukuba 305-0044, Japan}

\date{\today}

\maketitle

\textbf{
In strongly correlated electronic systems, Coulomb interactions frequently give rise to emergent electronic orders that spontaneously break rotational symmetry. Understanding how such symmetry breaking intertwines with other collective phenomena—such as unconventional superconductivity—and how it shapes experimental observables, particularly transport responses, remains a central challenge in modern condensed-matter physics. Here we report experimental signatures of charge stripe order, with a transport anisotropy rivaling that of quantum Hall stripe phases, coexisting with superconductivity and magnetism in rhombohedral hexalayer graphene. Strikingly, the low-temperature superconducting state not only inherits strong anisotropy but also exhibits a wide range of hysteretic transitions arising from the tunability of the underlying stripe order. Together, these findings reveal a previously unrecognized coexistence between superconductivity and charge stripe, shedding new light on the role of rotational symmetry breaking in shaping unconventional superconductivity in rhombohedral graphene.
}


\begin{figure*}
\includegraphics[width=1\linewidth]{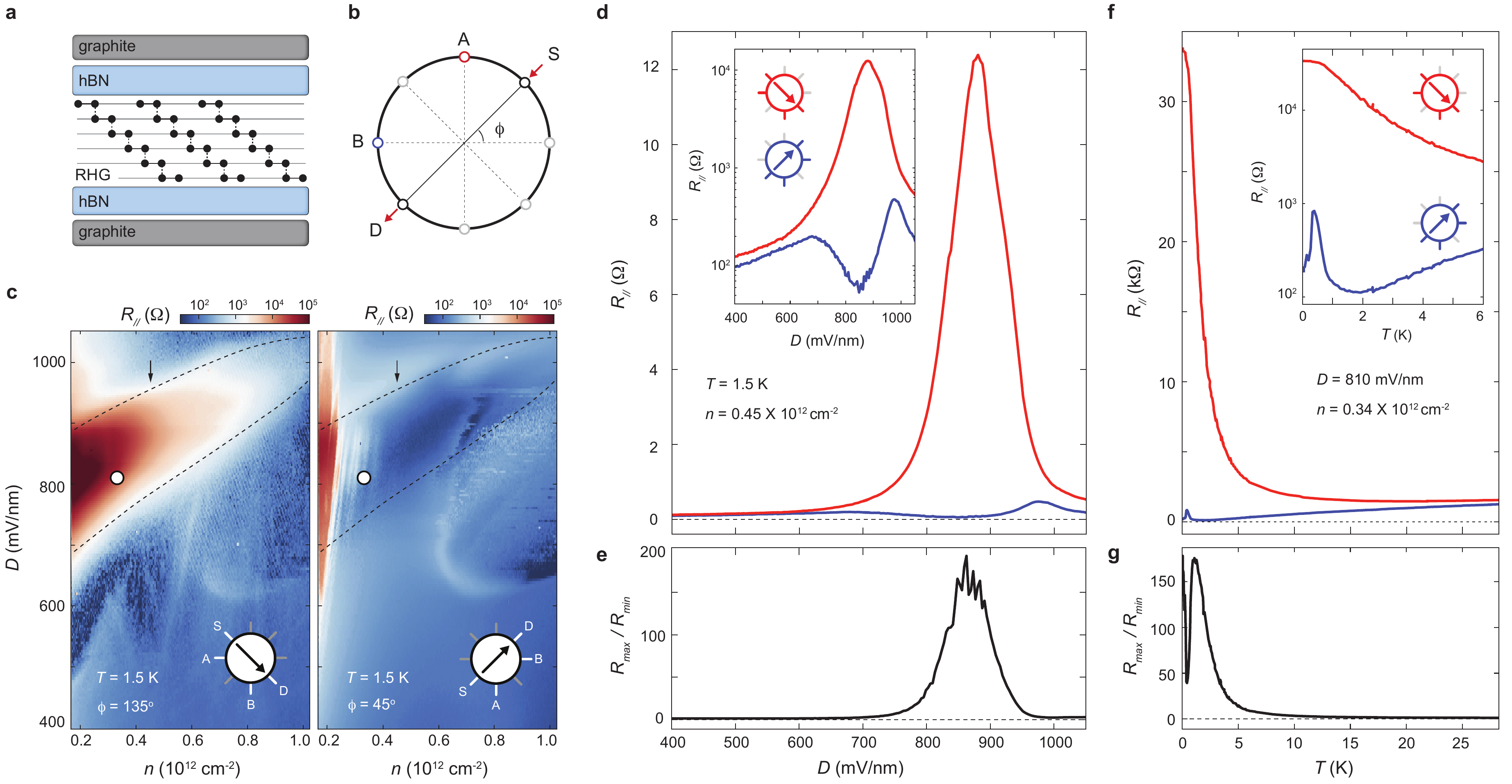}
\caption{\label{fig1} \textbf{Extreme transport anisotropy in rhombohedral hexalayer graphene.} 
(a) Schematic diagram of the R6G heterostructure. 
(b) Schematic diagram of the sunflower sample geometry with a disk-shaped channel and eight evenly spaced leads. 
(c) Color-scale map of the longitudinal resistance \( R_{\parallel} \), measured at \( T = 1.5\,\mathrm{K} \) and \( B = 0 \), along \( \phi = 135^{\circ} \) (left) and \( \phi = 45^{\circ} \) (right), shown as a function of carrier density \( n \) and displacement field \( D \). Inset: measurement configuration. 
(d) \Rpara\ measured along \( \phi = 135^{\circ} \) (red trace) and \( \phi = 45^{\circ} \) (blue trace) as a function of \( D \) at \( n = 0.45 \times 10^{12}\,\mathrm{cm}^{-2} \). Inset: logarithmic scale. 
(e) Ratio between the red and blue traces in panel (d). 
(f) \( R \)—\( T \) curves measured along \( \phi = 135^{\circ} \) (red trace) and \( \phi = 45^{\circ} \) (blue trace) at \( n = 0.34 \times 10^{12}\,\mathrm{cm}^{-2} \) and \( D = 810\,\mathrm{mV/nm} \) (location marked by the open circle in panel (c)). Inset: logarithmic scale.
(g) Ratio between the red and blue traces in panel (f).
}
\end{figure*}

\begin{figure*}
\includegraphics[width=0.94\linewidth]{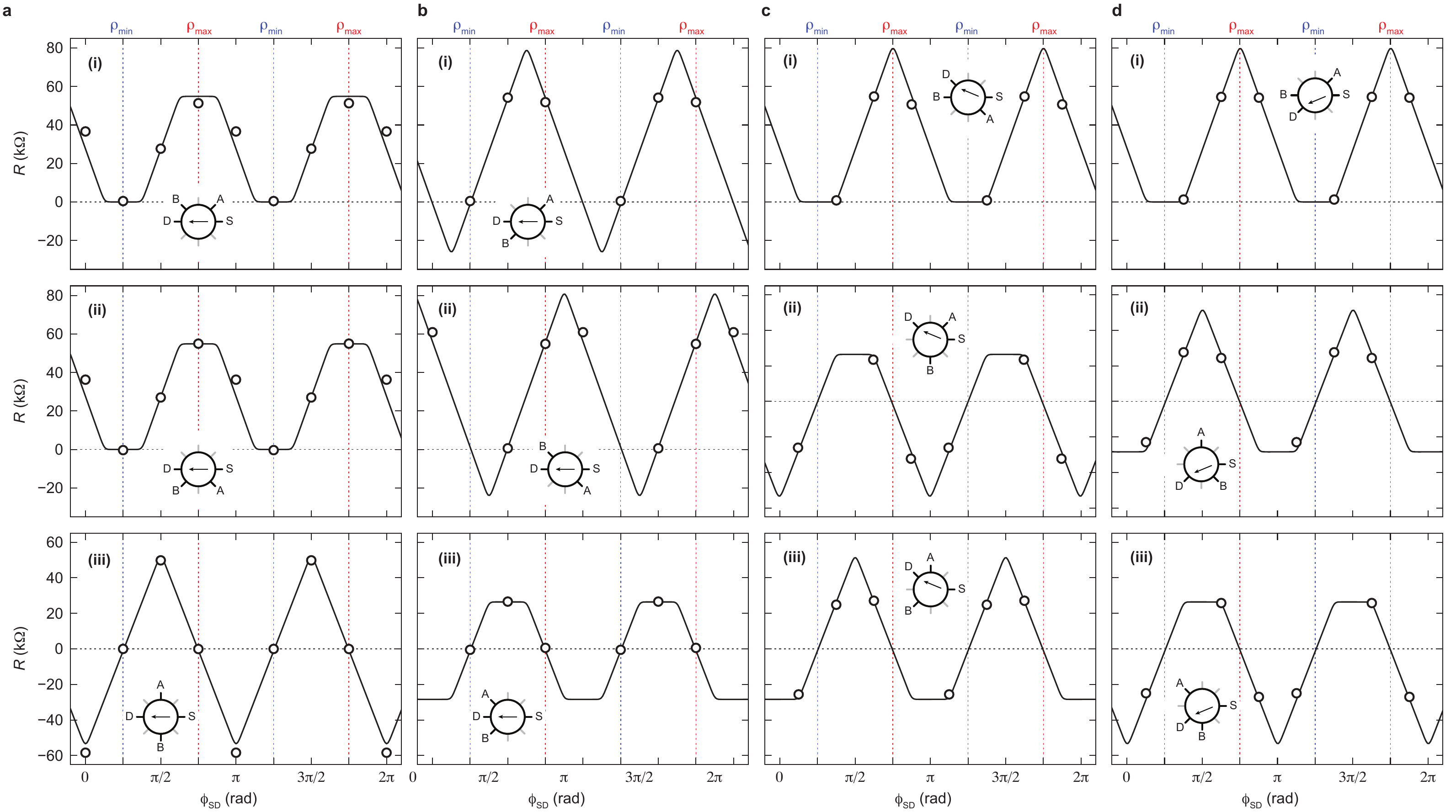}
\caption{\label{fig2}{\bf An extensive angle-resolved transport measurement.} 
Angle-resolved transport measured from an extensive set of measurement configurations.  The inset in each panel shows a schematic of the corresponding configuration. The black circles represent the measured response for each configuration at given values of $\phi$ (see Methods for a detailed discussion of the measurement procedures), while the black solid line denotes the expected angular dependence of each configuration, computed from a single conductivity matrix according to the theoretical framework of Ref.~\cite{Vafek2023anisotropy}. All measurements are performed at \( T = 1.5\,\mathrm{K} \), \( D = 850\,\mathrm{mV/nm}\), and \( n = 0.31 \times 10^{12}\,\mathrm{cm}^{-2} \).
}
\end{figure*}

\begin{figure*}
\includegraphics[width=0.98\linewidth]{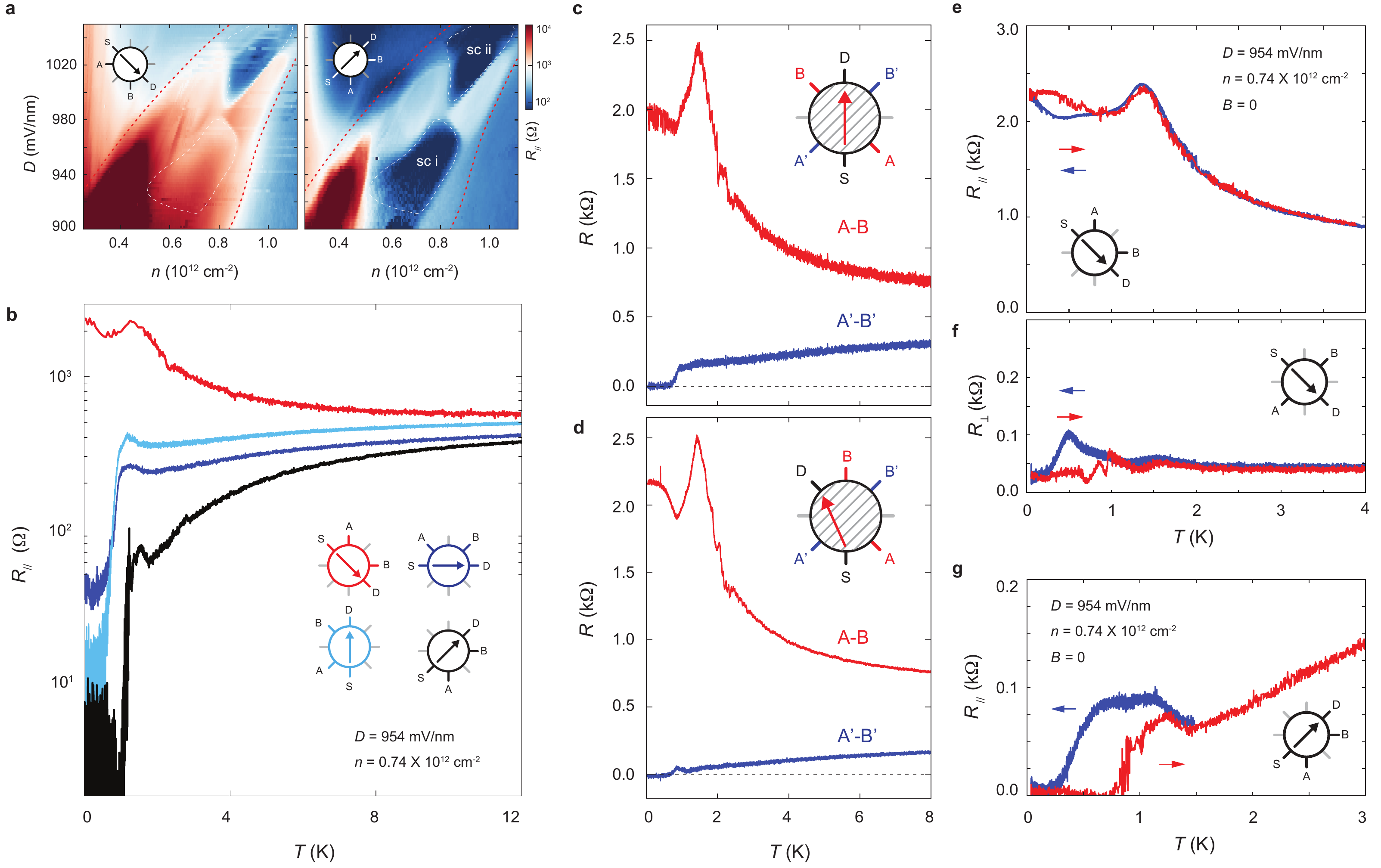}
\caption{\label{fig3} \textbf{Stripe-like superconducting transport and thermal hysteresis.} 
(a) Color-scale map of $R_{\parallel}$, measured at $T = 30\,\mathrm{mK}$ along $\phi = 135^{\circ}$ (left) and $\phi = 45^{\circ}$ (right), shown as a function of carrier density $n$ and displacement field $D$. Inset: measurement configuration. 
(b) Temperature dependence of \Rpara\ measured with current flowing along different directions. Inset: measurement configurations. 
(c--d) $R$--$T$ traces measured in alternative configurations distinct from \Rpara. The observation of superconducting or insulating responses is solely determined by the alignment of voltage probes relative to the orientation of the easy axis. 
All measurements in panels (b--d) are performed at $n = 0.74 \times 10^{12}\,\mathrm{cm}^{-2}$ and $D = 954\,\mathrm{mV/nm}$, within the \textit{SC\,i} region. 
(e--f) Temperature sweeps of (e) \Rpara\ and (f) \Rperp\ measured continuously on warming (red) and cooling (blue), with current flowing along \( \phi = 135^{\circ} \), which is perpendicular to the easy axis. 
(g) \Rpara\ as a function of temperature measured continuously on warming (red) and cooling (blue), with current flowing along the easy axis at \( \phi = 45^{\circ} \). All measurements in panels (b--g) are performed at $n = 0.74 \times 10^{12}\,\mathrm{cm}^{-2}$ and $D = 954\,\mathrm{mV/nm}$, within the \textit{SC\,i} region.
}
\end{figure*}

\begin{figure*}
\includegraphics[width=0.82\linewidth]{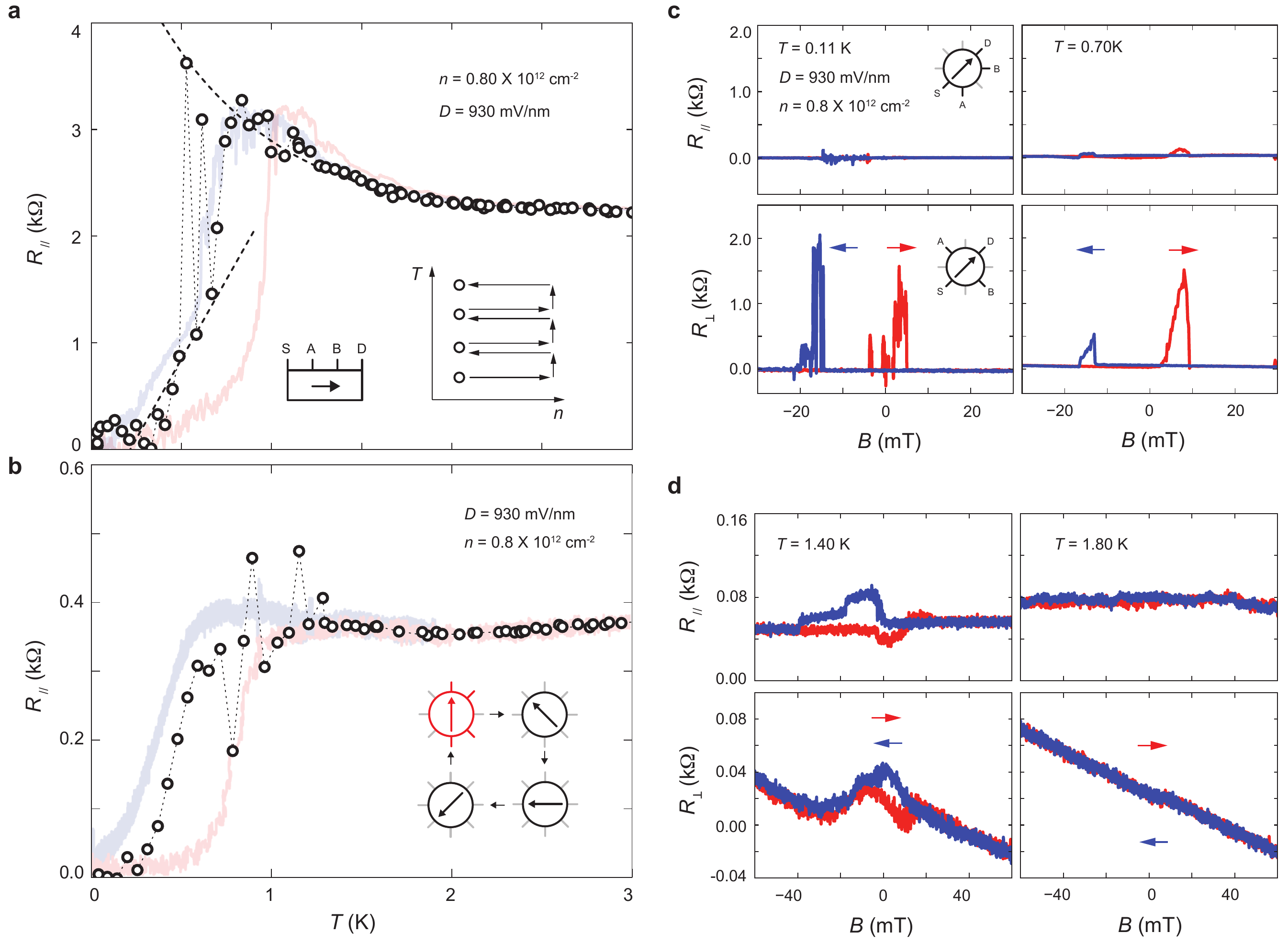}
\caption{\label{fig4} \textbf{Tunable hysteresis and nonlinear transport emerging at low temperature. } 
(a) $R$--$T$ curves measured in the superconducting phase of the Hall-bar sample (HB1). 
The shaded red and blue traces correspond to continuous warming and cooling sweeps, respectively, whereas the black open circles are measured at the same $(n, D)$ point but with $n$ swept outside the anisotropy regime between consecutive temperature points. 
(b) $R$--$T$ curves measured in the superconducting phase of a sunflower sample (SF1). 
The shaded red and blue traces correspond to continuous warming and cooling sweeps, respectively, whereas the black open circles are measured at the same $(n, D)$ point but with the current applied along different directions between consecutive temperature points. 
(c--d) $R_{\parallel}$ (top) and $R_{\perp}$ (bottom), measured from SF1 (c) below, and (d) above the superconducting transition, as the out-of-plane magnetic field $B$ is swept back and forth, with current flowing parallel to the easy axis. 
}
\end{figure*}

Smectic order represents a distinctive class of emergent electronic phases~\cite{Gruner1988CDW}. A paradigmatic example is the quantum Hall stripe phase, which typically appears at half filling of the third and higher Landau levels. Its formation is marked by extreme transport anisotropy for currents applied along orthogonal crystallographic directions ~\cite{Fogler1996stripe,Koulakov1996stripe,Lilly1999stripe,Lilly1999stripe2,Du1999stripe,Pan1999stripe}. This phase provides a canonical illustration of how strong electronic interactions can drive an instability that breaks both rotational and translational symmetries. 

An even more exotic manifestation occurs in the cuprate superconductors, where rotational and translational symmetries are spontaneously broken within the superconducting phase 
itself~\cite{Berg2009stripe,Fradkin2010nematic,Berg2009PDW,Kivelson2003stripe,Vojta2009stripe}. This gives rise to a pair-density-wave order, whose identification and characterization have been central topics of extensive research. Although the interplay between smectic order and unconventional superconductivity offers a fertile platform for new 
discoveries, clear and robust coexistence between these phases remains rare and poorly understood.

In this work, we investigate a potential coexistence between smectic order and superconductivity in rhombohedral hexalayer graphene (R6G). Multilayer graphene with rhombohedral stacking order has recently emerged as a powerful platform for discovering novel quantum phases ~\cite{Zhou2021RTG,Zhou2021RTG_SC,Zhou2022BLG,Han2025chiral,Choi2025SC}. Notably, in rhombohedral tetra- and pentalayer graphene (R4G and R5G), superconductivity is found to emerge from a quarter-metal phase in which the presence of time-reversal symmetry breaking is evidenced by the anomalous Hall effect ~\cite{Han2025chiral}. The resulting superconducting state exhibits hysteretic transitions between distinct ground states when tuned by an out-of-plane magnetic field, behavior that has been interpreted as evidence for chiral superconductivity
~\cite{Parra2025nematicQM,Chen2025chiralSC,Chou2024chiralSC,Yang2024chiralSC,Christos2025chiralSC,Yoon2025QMSC,Geier2025chiralSC,Patri2025chiralSC,Li2025trashcan}.

Here, the R6G sample is assembled into a van der Waals heterostructure consisting of hexagonal boron nitride and graphite encapsulation, as illustrated in Fig.~\ref{fig1}a
~\cite{Zibrov2017,Li.17b,Lei.13}. The dual-gated structure enables independent control of the carrier density $n$ and displacement field $D$, which is essential for accessing the low-temperature phase space. In this work, we report measurements from three R6G samples patterned into different device geometries, labeled SF1, SF2, and HB1 for simplicity (see Table~\ref{table:1}).

Angle-resolved transport measurement requires a sunflower sample geometry, in which the heterostructure is patterned into a disk, $2.5~\mu\mathrm{m}$ in diameter, with eight evenly spaced electrical contacts ~\cite{Zhang2024nonreciprocity,Chichinadze2024nonlinearHall,Morissette2025rhombohedral,Zhang2025angular}. Fig.~\ref{fig1}b illustrates the measurement configuration, in which the voltage response is probed between contacts aligned parallel to the direction of current flow. The resulting resistance, denoted $R_{\parallel}$, is comparable to the longitudinal resistance. The angular dependence of the transport response can be obtained by changing the angular location of the source--drain contacts and the voltage probes, while keeping their relative location fixed ~\cite{Wu2017nematic,Wu2020nematic,Vafek2023anisotropy,Chichinadze2024nonlinearHall,Zhang2024nonreciprocity,Morissette2025rhombohedral,Zhang2025angular}.

\vspace{0.1 in}
\noindent\textbf{Extreme anisotropy and stripe order in the metallic phase}

We begin by characterizing the metallic phase above the superconducting transition temperature. Fig.~\ref{fig1}c shows a color-scale map of $R_{\parallel}$ as a function of $n$ and $D$. The left and right panels compare $R_{\parallel}$ measured for two orthogonal current-flow directions. When current flows along $\phi = 135^{\circ}$, the sample is highly resistive, appearing as red and white in the chosen color scale, whereas a markedly more conductive response is observed when current flows along $\phi = 45^{\circ}$, 
appearing as dark blue. This comparison reveals an extreme transport anisotropy in the regime enclosed by the black dashed lines

To better visualize the extent of the anisotropy, Fig.~\ref{fig1}d shows line traces of $R_{\parallel}$ as a function of $D$ taken at a fixed carrier density $n$. 
Measured along $\phi = 135^{\circ}$ (red) and $\phi = 45^{\circ}$ (blue), the values of $R_{\parallel}$ diverge in the range $800 < D < 950~\mathrm{mV/nm}$, while converging outside this regime. As illustrated in the inset, $R_{\parallel}$ measured along $\phi = 45^{\circ}$ exhibits a pronounced minimum around 
$D = 850~\mathrm{mV/nm}$, whereas it develops an insulating peak along $\phi = 135^{\circ}$. 

This behavior indicates the emergence of two principal axes defining the resistivity tensor: one with minimum resistivity along $\phi = 45^{\circ}$, and the orthogonal one with maximum resistivity along $\phi = 135^{\circ}$. For simplicity, throughout this work we refer to the principal axis with minimum resistivity as the easy axis, and the orthogonal axis with maximum resistivity as the hard axis. 

Fig.~\ref{fig1}f--g show the temperature dependence of the transport anisotropy, measured at the $(n,D)$ point marked by 
the open circle in Fig.~\ref{fig1}c, which lies outside the superconducting regime at low temperature. At elevated temperatures ($T > 15~\mathrm{K}$), the transport response is approximately isotropic, evidenced by the converging resistance between the two orthogonal axes. As the temperature is lowered, the onset of transport anisotropy is signaled by diverging resistance along the two directions below $T \approx 5~\mathrm{K}$.

Figures~\ref{fig1}e and \ref{fig1}g show the ratio of measured resistances between two orthogonal directions. Within the regime of pronounced anisotropy, this ratio exceeds 150, indicating an extreme transport anisotropy. It is well recognized, however, that in a square-shaped sample the ratio of measured resistances can substantially overestimate the intrinsic anisotropy due to geometric effects~\cite{Simon1999comment}. To obtain a more accurate characterization of the underlying anisotropy, we therefore account for these geometric effects by performing an extensive set of angle-resolved transport measurements.

As illustrated in Fig.~\ref{fig2}, this data set includes the angular dependence of 12 measurement configurations, with the results from each configuration shown as open black circles. Using a previously established framework~\cite{Vafek2023anisotropy}, all measurements can be fit using a single resistivity tensor, shown as solid black lines in Fig.~\ref{fig2}. The excellent agreement between the data and the tensor fit provides a comprehensive identification of the anisotropy that corroborates the orientations of the principal axes (see Fig.~\ref{resistivity} in Methods for a detailed analysis of the anisotropy in the metallic phase). More importantly, the excellent agreement of the angular fits across many configurations provides strong indication for the uniformity of the sample.



Spontaneous rotational symmetry breaking in electronic systems is commonly associated with either nematic or smectic order~\cite{Fradkin2010nematic}. Here, several observations indicate that the extreme transport anisotropy  is 
smectic in nature. Most notably, the onset of anisotropy is accompanied by a strongly insulating response perpendicular to the easy axis: not only does the low-temperature resistivity exceed $25~\mathrm{k}\Omega$, but its temperature dependence is 
well described by activated behavior (see Fig.~\ref{Activated}), with activation energies in some cases exceeding $10$~K. The simultaneous observation of insulating transport along one direction and highly conductive transport along the orthogonal direction is highly unusual, and is naturally explained by a smectic electronic order in which conductive stripes are separated by insulating regions.  


\vspace{0.1 in}
\noindent\textbf{Stripe-like response in superconducting transport}

Cooling the sample to the base temperature of the dilution refrigerator reveals superconductivity emerging within the anisotropic regime. Figure~\ref{fig3}a displays color-scale maps of $R_{\parallel}$ measured at $T = 30~\mathrm{mK}$. When current flows along the easy axis ($\phi = 45^{\circ}$), two superconducting regions are identified by their vanishing longitudinal resistance, appearing as dark blue in the chosen color scale (right panel of Fig.~\ref{fig3}a). The locations of these superconducting pockets are in excellent agreement with prior observations in R4G and R5G ~\cite{Han2025chiral}. Intriguingly, when current flows along the orthogonal hard axis, these same regions remain resistive (left panel of Fig.~\ref{fig3}a). 

Following the naming convention established in earlier studies, we label these two superconducting regions as \textit{SC\,i} and \textit{SC\,ii}, as illustrated in Fig.~\ref{fig3}a. In the remainder of this work, we concentrate on the properties of \textit{SC\,i}. Although \textit{SC\,ii} also exhibits distinctive and potentially interesting behavior, a detailed investigation is beyond the scope of the present study.

Fig.~\ref{fig3}b shows $R_{\parallel}$ as a function of temperature, measured from within the \textit{SC\,i} regime. When current flows along the principal axis of minimum resistivity, the vanishing of $R_{\parallel}$ reveals a sharply defined superconducting transition around $T_c = 1.1~\mathrm{K}$ (black trace). In contrast, when current flows along the orthogonal direction ($\phi = 135^{\circ}$), the sample remains highly resistive down to $T = 30~\mathrm{mK}$ (red trace), with no indication of superconductivity.

The extreme anisotropy observed in the superconducting phase constitutes a striking and unexpected phenomenon, which can be easily distinguished from previous observation of anisotropic superconductors  ~\cite{Chu2012divergent,Fernandes2014nematicity,Bohmer2022nematicity}.
This unique anisotropic response is reproducible across a wide variety of measurement 
configurations and multiple samples.

Figures~\ref{fig3}c and \ref{fig3}d illustrate alternative measurement configurations in which insulating and superconducting behaviors appear simultaneously at low-temperature. 
When the source and drain contacts are misaligned from the easy axis, voltage probes along the hard axis (A, B) register insulating behavior, whereas those aligned parallel to the easy axis (A$^\prime$, B$^\prime$) simultaneously exhibit vanishing resistance at low temperature (see Supplementary Information for additional measurements)~\cite{SI}.
These observations confirm that the diverging anisotropy reflects an intrinsic property of the emerging order, rather than an artifact arising from a particular measurement configuration.

The diverging anisotropy is also observed in a second R6G sample patterned in the sunflower geometry. As shown in Fig.~\ref{RTdiverge_SF2}, the measured $R$--$T$ traces exhibit either superconducting or insulating behavior depending on the alignment of the voltage probes relative to the easy axis. In this sample, labeled SF2, the superconducting transition occurs at a temperature comparable to that of SF1, whereas the onset of transport anisotropy appears at a slightly lower temperature.

\vspace{0.1 in}
\noindent\textbf{Tunability and hysteresis in the superconducting transition}

To shed further light on the nature of the superconducting phase, we show below that the superconducting transition is accompanied by pronounced hysteretic behaviors, which constitute a defining hallmark of the coexistence of superconductivity with extreme electronic anisotropy.

First, we observe pronounced thermal hysteresis around the superconducting transition temperature $T_c$. As illustrated in Figs.~\ref{fig3}e--g, the resistance measured within the \textit{SC\,i} regime displays clear hysteretic behavior between controlled warming and cooling cycles, observed along both the easy and hard axes. Similar thermal hysteresis is also observed in samples SF2 and HB1 (see Fig.~\ref{RThysteresis_SF2}). In some cases, the cooling trace exhibits sharp, discontinuous jumps, and the hysteresis persists down to the base temperature of the dilution refrigerator (Fig.~\ref{RThysteresis_SF2}c--d).

Since the superconducting transition itself is second order in nature, the observed hysteresis likely arises from the melting of stripe order, which is expected to be first order. The fact that this melting transition manifests as hysteresis in the superconducting transition suggests that the observed stripe-like behavior is not merely inherited from the normal state, nor simply a passive background on which superconductivity develops. Instead, our findings point to an intrinsic link between the stripe order and the stability of the superconducting phase. 

Second, the superconducting transition is also tunable through other experimental parameters. For instance, Fig.~\ref{fig4}a illustrates this tunability by comparing several $R$--$T$ traces measured at a fixed $(n, D)$ point in HB1. The black open circles correspond to measurements taken as the temperature is slowly increased; however, between successive temperature points, the carrier density $n$ is intentionally tuned outside the anisotropic regime. This preparation protocol produces an $R$--$T$ curve that differs markedly from those obtained through continuous warming and cooling sweeps (shaded red and blue traces). In particular, the continuous warming trace exhibits the highest superconducting transition temperature, whereas repeated $n$ sweeps introduce a pronounced suppression of the apparent $T_c$.

In addition, near the superconducting transition, sweeping $n$ between measurements gives rise to random switching between two diverging branches of transport response. One branch shows a rapidly increasing resistance upon cooling, consistent with insulating behavior, while the other branch displays a vanishing resistance at low temperature. This switching behavior suggests that the superconducting transition is directly linked to the configuration of the underlying stripe order, which is modified when the carrier density is swept.

Moreover, the superconducting transition can be modified by varying the current-flow direction. 
Figure~\ref{fig4}b shows $R$--$T$ traces measured at a fixed $(n, D)$ point in SF1. 
The black open circles correspond to measurements taken during warming. Between successive temperature points, the applied current is cycled through several orientations (inset of Fig.~\ref{fig4}b), while $n$ and $D$ are held fixed throughout. This measurement protocol produces behavior analogous to that in Fig.~\ref{fig4}a: the highest superconducting transition temperature is obtained during a continuous warming sweep, whereas cycling the current direction introduces a pronounced suppression of the apparent $T_c$.

Beyond the hysteretic behavior observed in the superconducting transition, the application of an out-of-plane magnetic field also induces a hysteretic transition in the low-temperature superconducting transport, as shown in Fig.~\ref{fig4}c. As the perpendicular magnetic field is swept back and forth, the transition is marked by resistive features separating superconducting phases with vanishing \Rpara\ and \Rperp. Above the superconducting transition temperature, the magnetic-field-driven hysteresis adopts a different transport signature, more closely resembling the hysteresis loop of the anomalous Hall effect (Fig.~\ref{fig4}d; see also Fig.~\ref{B_hysteresis}), indicative of orbital ferromagnetic order in the normal state~\cite{Sharpe2019,Serlin2019,Lin2021SOC}. Overall, the observed $B$-driven hysteresis closely mirrors that previously reported in R4G and R5G~\cite{Han2025chiral}.

The coexistence of transport anisotropy and the anomalous Hall effect in the normal state suggests that rotational and time-reversal symmetries are simultaneously broken, consistent with previous observations in R6G at lower displacement fields~\cite{Morissette2025rhombohedral}. In this context, a momentum-space condensation~\cite{Dong2021momentum,Jung2015momentum,Huang2023momentum,Mandal2023valleynematic,Parra2025nematicQM} provides a natural explanation for the observed transport anisotropy.

Within this picture, superconductivity emerges from a valley- and momentum-polarized Fermi sea. An out-of-plane magnetic field reverses the sign of the valley polarization, inducing hysteretic switching between two degenerate chiral configurations. This description is closely analogous to the A phase of superfluid \(^{3}\mathrm{He}\), a paradigmatic example of a chiral superconductor with \(p\)-wave pairing symmetry~\cite{Ikegami2013chiral}, suggesting that the order parameter of the \textit{SC\,i} phase is chiral in nature.

\vspace{0.1 in}
\noindent\textbf{Discussion}

Having established the principal experimental signatures of transport anisotropy, we next discuss the implications of stripe order and its coexistence with the low-temperature superconducting phase.

First, we examine the observed hysteresis in the superconducting transition. With fixed experimental parameters, such as $n$, $D$, and $B$, the superconducting transition temperature is generally expected to be independent of the measurement history. Consequently, we argue that the exceptional tunability of the superconducting transition arises from its coexistence with stripe order. In particular, changes in the underlying stripe configuration occur each time the sample is reset, whether by thermal cycling, sweeping $n$, varying the current-flow direction, or applying a d.c.\ bias (see Fig.~\ref{IV_RT}a). Such changes, for instance, a rotation of the stripe orientation, would lead to pronounced variations in the superconducting transition temperature.

It is worth noting that this impact on the superconducting transition can be very dramatic. As illustrated in Fig.~\ref{RThysteresis_SF2} and Fig.~\ref{SC_multipleRTs}, $R$--$T$ traces measured upon cooling can differ substantially between successive thermal cycles, even when the sample is cooled slowly under nominally identical conditions. Such behavior likely reflects the formation of domains in the stripe order, providing further evidence for the coexistence and intertwinement of superconductivity with stripe charge order.

Apart from the hysteretic behaviors discussed above, the presence of stripe order is also associated with pronounced nonlinearity, as illustrated in Fig.~\ref{IV_RT} and Fig.~\ref{IV_anisotropy}. However, both hysteresis and nonlinearity vanish above $T \sim 1.5~\mathrm{K}$ (see Fig.~\ref{fig3}e--g, Fig.~\ref{fig4}a,b,d, Fig.~\ref{RThysteresis_SF2}, Fig.~\ref{IV_RT}, Fig.~\ref{RT_warming}, Fig.~\ref{SC_multipleRTs}, and Fig.~\ref{HB1_RT}). At higher temperatures, although a finite transport anisotropy persists (Fig.~\ref{angle_T}), the system behaves as a conventional metal with an ohmic transport response and no detectable hysteresis or nonlinearity. These observations provide further evidence for a tunable electronic order beyond simple anisotropy, consistent with the formation of a smectic electronic phase.

It is important to consider the possible role of strain and lattice distortions, which are unavoidable in solid-state samples. Two key observations, however, indicate that electron correlations, rather than structural distortions, play the dominant role in stabilizing the stripe order. First, the anisotropy is confined to a well-defined region of the $n$--$D$ phase space, where a flat energy band is expected to strongly enhance Coulomb interactions~\cite{Bernevig2025trashcan}. Second, the observed tunability shows that the principal axis of anisotropy is not rigidly locked to a fixed crystallographic direction, but can instead rotate in response to external perturbations. Taken together, these observations indicate that the emergence of anisotropy is electronic in origin rather than arising from structural distortions.

Taken together, our observations point to an unusual and highly unconventional scenario. 
Here, the Cooper pairing instability emerges from a valley-polarized normal state, 
suggesting that the underlying superconducting order parameter may be chiral in nature~\cite{Han2025chiral}. 
At the same time, the presence of stripe (smectic) order within the superconducting phase points to a spatial modulation 
between coexisting superconducting and insulating regions. Overall, the rich interplay between superconductivity, smectic order, 
and orbital magnetism revealed here establishes a fertile platform for future investigations and is likely to motivate a broad range 
of experimental and theoretical efforts.

\section*{acknowledgments}

J.I.A.L. wishes to thank Oskar Vafek for continued support and theoretical guidance in developing the novel scheme of angle-resolved transport measurement.
J.I.A.L. also acknowledges helpful discussions with Sankar Das Sarma, Boris Shklovskii, Andrei Bernevig, Allan MacDonald, Dima Feldman, Aaron Hui, and Mathias Scheurer.
E.M. and J.I.A.L. acknowledge support from U.S. National Science Foundation under Award DMR-2143384. N.J.Z., and R.Q.N. acknowledge support from the Air Force Office of Scientific Research.  K.W. and T.T. acknowledge support from the JSPS KAKENHI (Grant Numbers 21H05233 and 23H02052) and World Premier International Research Center Initiative (WPI), MEXT, Japan. 
Part of this work was enabled by the use of pyscan (github.com/sandialabs/pyscan), scientific measurement software made available by the Center for Integrated Nanotechnologies, an Office of Science User Facility operated for the U.S. Department of Energy.

\bibliography{Li_ref}

@book{Vollhardt1990,
    author = {Vollhardt, D. and W{\"o}lfle, P.},
    title = {The Superfluid Phases of Helium 3},
    publisher = {Taylor and Francis},
    year = {1990},}

@article{Lei.13,
  title = {One-dimensional Electrical Contact to a Two-dimensional Material},
  author = {L Wang and I Meric and PY Huang and Q Gao and Y Gao and H Tran and T Taniguchi and K Watanabe and LM Campos and DA Muller and J Guo and P Kim and J Hone and K L Shepard and C R Dean},
  journal = {Science},
  volume = {342},
 issue = {6158},
  pages = {614-617},
  year = {2013},
  month = {Nov},
Publisher = {American Association for the Advancement of Science},
}

@article{Zibrov2017,
  author = "A A Zibrov and C R Kometter and H Zhou and E M Spanton and T Taniguchi and K Watanabe and  and M P Zaletel  and A F Young",
    title = "Tunable interacting composite fermion phases in a half-filled bilayer-graphene Landau level",
  journal = {Nature},
  volume = {549},
  pages = {360-364},
  year = {2017},
  month = {Sept},
  doi = {10.1038/nature23893},
  url = {http://dx.doi.org/10.1038/nature23893}
}

@article{Lilly1999stripe2,
  title = {Anisotropic States of Two-Dimensional Electron Systems in High Landau Levels: Effect of an In-Plane Magnetic Field},
  author = {Lilly, M. P. and Cooper, K. B. and Eisenstein, J. P. and Pfeiffer, L. N. and West, K. W.},
  journal = {Phys. Rev. Lett.},
  volume = {83},
  issue = {4},
  pages = {824--827},
  numpages = {0},
  year = {1999},
  month = {Jul},
  publisher = {American Physical Society},
  doi = {10.1103/PhysRevLett.83.824},
  url = {https://link.aps.org/doi/10.1103/PhysRevLett.83.824}
}

@article{Pan1999stripe,
  title = {Strongly Anisotropic Electronic Transport at Landau Level Filling Factor $\mathit{\ensuremath{\nu}}\phantom{\rule{0ex}{0ex}}=\phantom{\rule{0ex}{0ex}}9/2$ and $\mathit{\ensuremath{\nu}}\phantom{\rule{0ex}{0ex}}=\phantom{\rule{0ex}{0ex}}5/2$ under a Tilted Magnetic Field},
  author = {Pan, W. and Du, R. R. and Stormer, H. L. and Tsui, D. C. and Pfeiffer, L. N. and Baldwin, K. W. and West, K. W.},
  journal = {Phys. Rev. Lett.},
  volume = {83},
  issue = {4},
  pages = {820--823},
  numpages = {0},
  year = {1999},
  month = {Jul},
  publisher = {American Physical Society},
  doi = {10.1103/PhysRevLett.83.820},
  url = {https://link.aps.org/doi/10.1103/PhysRevLett.83.820}
}

@article{Lilly1999stripe,
  title={Evidence for an anisotropic state of two-dimensional electrons in high Landau levels},
  author={Lilly, MP and Cooper, KB and Eisenstein, JP and Pfeiffer, LN and West, KW},
  journal={Physical Review Letters},
  volume={82},
  number={2},
  pages={394},
  year={1999},
  publisher={APS}
}

@article{Du1999stripe,
  title={Strongly anisotropic transport in higher two-dimensional Landau levels},
  author={Du, RR and Tsui, DC and Stormer, HL and Pfeiffer, LN and Baldwin, KW and West, KW},
  journal={Solid State Communications},
  volume={109},
  number={6},
  pages={389--394},
  year={1999},
  publisher={Elsevier}
}

@article {Li.17b,
	author = {Li, J. I. A. and Tan, C. and Chen, S. and Zeng, Y. and Taniguchi, T. and Watanabe, K. and Hone, J. and Dean, C. R.},
	title = {Even-denominator fractional quantum Hall states in bilayer graphene},
	volume = {358},
	number = {6363},
	pages = {648--652},
	year = {2017},
	doi = {10.1126/science.aao2521},
	publisher = {American Association for the Advancement of Science},
	issn = {0036-8075},
	URL = {http://science.sciencemag.org/content/358/6363/648},
	journal = {Science}
}

@article{Sharpe2019,
  title={Emergent ferromagnetism near three-quarters filling in twisted bilayer graphene},
  author={Sharpe, Aaron L and Fox, Eli J and Barnard, Arthur W and Finney, Joe and Watanabe, Kenji and Taniguchi, Takashi and Kastner, MA and Goldhaber-Gordon, David},
  journal={arXiv preprint arXiv:1901.03520},
  year={2019}
}

@article{Serlin2019,
  title={Intrinsic quantized anomalous Hall effect in a moir$\backslash$'e heterostructure},
  author={Serlin, M and Tschirhart, CL and Polshyn, H and Zhang, Y and Zhu, J and Watanabe, K and Taniguchi, T and Balents, L and Young, AF},
  journal={arXiv preprint arXiv:1907.00261},
  year={2019}
}

@article{SI,
 journal={Please see the supplementary materials} 
}

@article{Wu2017nematic,
  title={Spontaneous breaking of rotational symmetry in copper oxide superconductors},
  author={Wu, J and Bollinger, AT and He, X and Bo{\v{z}}ovi{\'c}, I},
  journal={Nature},
  volume={547},
  number={7664},
  pages={432--435},
  year={2017},
  publisher={Nature Publishing Group}
}

@article{Wu2020nematic,
  title={Electronic nematicity in Sr2RuO4},
  author={Wu, Jie and Nair, Hari P and Bollinger, Anthony T and He, Xi and Robinson, Ian and Schreiber, Nathaniel J and Shen, Kyle M and Schlom, Darrell G and Bo{\v{z}}ovi{\'c}, Ivan},
  journal={Proceedings of the National Academy of Sciences},
  volume={117},
  number={20},
  pages={10654--10659},
  year={2020},
  publisher={National Acad Sciences}
}

@article{Kivelson2003stripe,
  title = {How to detect fluctuating stripes in the high-temperature superconductors},
  author = {Kivelson, S. A. and Bindloss, I. P. and Fradkin, E. and Oganesyan, V. and Tranquada, J. M. and Kapitulnik, A. and Howald, C.},
  journal = {Rev. Mod. Phys.},
  volume = {75},
  issue = {4},
  pages = {1201--1241},
  numpages = {0},
  year = {2003},
  month = {Oct},
  publisher = {American Physical Society},
  doi = {10.1103/RevModPhys.75.1201},
  url = {https://link.aps.org/doi/10.1103/RevModPhys.75.1201}
}

@article{Chen2020ABC,
  title={Tunable correlated chern insulator and ferromagnetism in a moir{\'e} superlattice},
  author={Chen, Guorui and Sharpe, Aaron L and Fox, Eli J and Zhang, Ya-Hui and Wang, Shaoxin and Jiang, Lili and Lyu, Bosai and Li, Hongyuan and Watanabe, Kenji and Taniguchi, Takashi and others},
  journal={Nature},
  volume={579},
  number={7797},
  pages={56--61},
  year={2020},
  publisher={Nature Publishing Group}
}

@article{Polshyn20201N2,
  title={Electrical switching of magnetic order in an orbital Chern insulator},
  author={Polshyn, H and Zhu, J and Kumar, MA and Zhang, Y and Yang, F and Tschirhart, CL and Serlin, M and Watanabe, K and Taniguchi, T and MacDonald, AH and Young, A F},
  journal={Nature},
  volume={588},
  number={7836},
  pages={66--70},
  year={2020},
  publisher={Nature Publishing Group}
}

@article{Chen20201N2,
  title={Electrically tunable correlated and topological states in twisted monolayer--bilayer graphene},
  author={Chen, Shaowen and He, Minhao and Zhang, Ya-Hui and Hsieh, Valerie and Fei, Zaiyao and Watanabe, K and Taniguchi, T and Cobden, David H and Xu, Xiaodong and Dean, Cory R and Yankowitz, Matthew},
  journal={Nature Physics},
  volume={17},
  number={3},
  pages={374--380},
  year={2021},
  publisher={Nature Publishing Group}
}

@article{Lin2021SOC,
  title={Spin-orbit--driven ferromagnetism at half moir{\'e} filling in magic-angle twisted bilayer graphene},
  author={Lin, Jiang-Xiazi and Zhang, Ya-Hui and Morissette, Erin and Wang, Zhi and Liu, Song and Rhodes, Daniel and Watanabe, K and Taniguchi, T and Hone, James and Li, JIA},
  journal={Science},
  volume={375},
  issue={6579},
  pages={437--441},
  year={2022},
  publisher={American Association for the Advancement of Science}
}

@article{Lin2022SDE,
  title={Zero-field superconducting diode effect in small-twist-angle trilayer graphene},
  author={Lin, Jiang-Xiazi and Siriviboon, Phum and Scammell, Harley D and Liu, Song and Rhodes, Daniel and Watanabe, K and Taniguchi, T and Hone, James and Scheurer, Mathias S and Li, JIA},
  journal={Nature Physics},
  volume={18},
  number={10},
  pages={1221--1227},
  year={2022},
  publisher={Nature Publishing Group UK London}
}

@article{Fradkin2010nematic,
  title={Nematic Fermi Fluids in Condensed Matter Physics},
  author={Fradkin, Eduardo and Kivelson, Steven A and Lawler, Michael J and Eisenstein, James P and Mackenzie, Andrew P},
  journal={The Annual Review of Condensed Matter Physics is},
  volume={1},
  pages={153--78},
  year={2010}
}

@article{Fernandes2014nematic,
  title={What drives nematic order in iron-based superconductors?},
  author={Fernandes, RM and Chubukov, AV and Schmalian, J},
  journal={Nature physics},
  volume={10},
  number={2},
  pages={97--104},
  year={2014},
  publisher={Nature Publishing Group}
}

@article{Vafek2023anisotropy,
  title = {Anisotropic resistivity tensor from disk geometry magnetoconductance},
  author = {Vafek, Oskar},
  journal = {Phys. Rev. Appl.},
  volume = {20},
  issue = {6},
  pages = {064008},
  numpages = {8},
  year = {2023},
  month = {Dec},
  publisher = {American Physical Society},
  doi = {10.1103/PhysRevApplied.20.064008},
  url = {https://link.aps.org/doi/10.1103/PhysRevApplied.20.064008}
}

@article{Fernandes2014nematicity,
  title={What drives nematic order in iron-based superconductors?},
  author={Fernandes, RM and Chubukov, AV and Schmalian, J},
  journal={Nature physics},
  volume={10},
  number={2},
  pages={97--104},
  year={2014},
  publisher={Nature Publishing Group UK London}
}

@article{Bohmer2022nematicity,
  title={Nematicity and nematic fluctuations in iron-based superconductors},
  author={B{\"o}hmer, Anna E and Chu, Jiun-Haw and Lederer, Samuel and Yi, Ming},
  journal={Nature Physics},
  volume={18},
  number={12},
  pages={1412--1419},
  year={2022},
  publisher={Nature Publishing Group UK London}
}

@Article{Zhang2024nonreciprocity,
author={Zhang, Naiyuan James
and Lin, Jiang-Xiazi
and Chichinadze, Dmitry V.
and Wang, Yibang
and Watanabe, Kenji
and Taniguchi, Takashi
and Fu, Liang
and Li, J. I. A.},
title={Angle-resolved transport non-reciprocity and spontaneous symmetry breaking in twisted trilayer graphene},
journal={Nature Materials},
year={2024},
month={Mar},
day={01},
volume={23},
number={3},
pages={356-362},
issn={1476-4660},
doi={10.1038/s41563-024-01809-z},
url={https://doi.org/10.1038/s41563-024-01809-z}
}

@article{Chichinadze2024nonlinearHall,
  title={Observation of giant nonlinear Hall conductivity in Bernal bilayer graphene},
  author={Chichinadze, Dmitry V and Zhang, Naiyuan James and Lin, Jiang-Xiazi and Wang, Xiaoyu and Watanabe, Kenji and Taniguchi, Takashi and Vafek, Oskar and Li, JIA},
  journal={arXiv preprint arXiv:2411.11156},
  year={2024}
}

@article{Jung2015momentum,
  title={Persistent current states in bilayer graphene},
  author={Jung, Jeil and Polini, Marco and MacDonald, Allan H},
  journal={Physical Review B},
  volume={91},
  number={15},
  pages={155423},
  year={2015},
  publisher={APS}
}

@article{Dong2021momentum,
  title = {Isospin- and momentum-polarized orders in bilayer graphene},
  author = {Dong, Zhiyu and Davydova, Margarita and Ogunnaike, Olumakinde and Levitov, Leonid},
  journal = {Phys. Rev. B},
  volume = {107},
  issue = {7},
  pages = {075108},
  numpages = {10},
  year = {2023},
  month = {Feb},
  publisher = {American Physical Society},
  doi = {10.1103/PhysRevB.107.075108},
  url = {https://link.aps.org/doi/10.1103/PhysRevB.107.075108}
}

@article{Huang2023momentum,
  title={Spin and orbital metallic magnetism in rhombohedral trilayer graphene},
  author={Huang, Chunli and Wolf, Tobias MR and Qin, Wei and Wei, Nemin and Blinov, Igor V and MacDonald, Allan H},
  journal={Physical Review B},
  volume={107},
  number={12},
  pages={L121405},
  year={2023},
  publisher={APS}
}

@article{Chu2012divergent,
  title={Divergent nematic susceptibility in an iron arsenide superconductor},
  author={Chu, Jiun-Haw and Kuo, Hsueh-Hui and Analytis, James G and Fisher, Ian R},
  journal={Science},
  volume={337},
  number={6095},
  pages={710--712},
  year={2012},
  publisher={American Association for the Advancement of Science}
}

@article{Mandal2023valleynematic,
  title={Valley-polarized nematic order in twisted moir{\'e} systems: In-plane orbital magnetism and crossover from non-Fermi liquid to Fermi liquid},
  author={Mandal, Ipsita and Fernandes, Rafael M},
  journal={Physical Review B},
  volume={107},
  number={12},
  pages={125142},
  year={2023},
  publisher={APS}
}

@article{Zhou2021RTG,
  title={Half-and quarter-metals in rhombohedral trilayer graphene},
  author={Zhou, Haoxin and Xie, Tian and Ghazaryan, Areg and Holder, Tobias and Ehrets, James R and Spanton, Eric M and Taniguchi, Takashi and Watanabe, Kenji and Berg, Erez and Serbyn, Maksym and others},
  journal={Nature},
  volume={598},
  number={7881},
  pages={429--433},
  year={2021},
  publisher={Nature Publishing Group UK London}
}

@article{Zhou2022BLG,
  title={Isospin magnetism and spin-polarized superconductivity in Bernal bilayer graphene},
  author={Zhou, Haoxin and Holleis, Ludwig and Saito, Yu and Cohen, Liam and Huynh, William and Patterson, Caitlin L and Yang, Fangyuan and Taniguchi, Takashi and Watanabe, Kenji and Young, Andrea F},
  journal={Science},
  volume={375},
  number={6582},
  pages={774--778},
  year={2022},
  publisher={American Association for the Advancement of Science}
}

@article{Zhou2021RTG_SC,
  title={Superconductivity in rhombohedral trilayer graphene},
  author={Zhou, Haoxin and Xie, Tian and Taniguchi, Takashi and Watanabe, Kenji and Young, Andrea F},
  journal={Nature},
  volume={598},
  number={7881},
  pages={434--438},
  year={2021},
  publisher={Nature Publishing Group UK London}
}

@article{Parra2025nematicQM,
  title={Band Renormalization, Quarter Metals, and Chiral Superconductivity in Rhombohedral Tetralayer Graphene},
  author={Parra-Martinez, Guillermo and Jimeno-Pozo, Alejandro and Phong, Vo Tien and Sainz-Cruz, Hector and Kaplan, Daniel and Emanuel, Peleg and Oreg, Yuval and Pantaleon, Pierre A and Silva-Guillen, Jose Angel and Guinea, Francisco},
  journal={arXiv preprint arXiv:2502.19474},
  year={2025}
}

@article{Berg2009stripe,
  title={Theory of the striped superconductor},
  author={Berg, Erez and Fradkin, Eduardo and Kivelson, Steven A},
  journal={Physical Review B—Condensed Matter and Materials Physics},
  volume={79},
  number={6},
  pages={064515},
  year={2009},
  publisher={APS}
}

@article{Han2025chiral,
  title={Signatures of chiral superconductivity in rhombohedral graphene},
  author={Han, Tonghang and Lu, Zhengguang and Hadjri, Zach and Shi, Lihan and Wu, Zhenghan and Xu, Wei and Yao, Yuxuan and Cotten, Armel A and Sharifi Sedeh, Omid and Weldeyesus, Henok and others},
  journal={Nature},
  volume={643},
  number={8072},
  pages={654--661},
  year={2025},
  publisher={Nature Publishing Group UK London}
}

@article{Morissette2025rhombohedral,
  title={Intertwined Nematicity, Multiferroicity, and Nonlinear Hall Effect in Rhombohedral Pentalayer Graphene},
  author={Morissette, Erin and Qin, Peiyu and Watanabe, K and Taniguchi, T and Li, JIA},
  journal={arXiv preprint arXiv:2503.09954},
  year={2025}
}

@article{Ikegami2013chiral,
  title={Chiral symmetry breaking in superfluid 3He-A},
  author={Ikegami, H and Tsutsumi, Y and Kono, K},
  journal={Science},
  volume={341},
  number={6141},
  pages={59--62},
  year={2013},
  publisher={American Association for the Advancement of Science}
}

@article{Zhang2025angular,
  title={Angular Interplay of Nematicity, Superconductivity, and Strange Metallicity in a Moir{\'e} Flat Band},
  author={Zhang, Naiyuan J and Nosov, Pavel A and Sommer, Ophelia Evelyn and Wang, Yibang and Watanabe, Kenji and Taniguchi, Takashi and Khalaf, Eslam and Li, JIA},
  journal={arXiv preprint arXiv:2503.15767},
  year={2025}
}

@article{Choi2025SC,
  title={Superconductivity and quantized anomalous Hall effect in rhombohedral graphene},
  author={Choi, Youngjoon and Choi, Ysun and Valentini, Marco and Patterson, Caitlin L and Holleis, Ludwig FW and Sheekey, Owen I and Stoyanov, Hari and Cheng, Xiang and Taniguchi, Takashi and Watanabe, Kenji and others},
  journal={Nature},
  pages={1--6},
  year={2025},
  publisher={Nature Publishing Group UK London}
}

@article{Yang2024chiralSC,
  title={Topological incommensurate Fulde-Ferrell-Larkin-Ovchinnikov superconductor and Bogoliubov Fermi surface in rhombohedral tetra-layer graphene},
  author={Yang, Hui and Zhang, Ya-Hui},
  journal={arXiv preprint arXiv:2411.02503},
  year={2024}
}

@article{Chou2024chiralSC,
  title={Intravalley spin-polarized superconductivity in rhombohedral tetralayer graphene},
  author={Chou, Yang-Zhi and Zhu, Jihang and Sarma, Sankar Das},
  journal={arXiv preprint arXiv:2409.06701},
  year={2024}
}

@article{Christos2025chiralSC,
  title={Finite-momentum pairing and superlattice superconductivity in valley-imbalanced rhombohedral graphene},
  author={Christos, Maine and Bonetti, Pietro M and Scheurer, Mathias S},
  journal={arXiv preprint arXiv:2503.15471},
  year={2025}
}

@article{Geier2025chiralSC,
  title={Chiral and topological superconductivity in isospin polarized multilayer graphene},
  author={Geier, Max and Davydova, Margarita and Fu, Liang},
  journal={Nature Communications},
  year={2025},
  publisher={Nature Publishing Group UK London}
}

@article{Chen2025chiralSC,
  title={Intrinsic superconducting diode effect and nonreciprocal superconductivity in rhombohedral graphene multilayers},
  author={Chen, Yinqi and Schrade, Constantin},
  journal={arXiv preprint arXiv:2503.16391},
  year={2025}
}

@article{Patri2025chiralSC,
  title={Family of multilayer graphene superconductors with tunable chirality: Momentum-space vortices nucleated by a ring of Berry curvature},
  author={Patri, Adarsh S and Franz, Marcel},
  journal={Physical Review B},
  volume={112},
  number={21},
  pages={214505},
  year={2025},
  publisher={APS}
}

@article{Han2024rhombohedral,
  title={Large quantum anomalous Hall effect in spin-orbit proximitized rhombohedral graphene},
  author={Han, Tonghang and Lu, Zhengguang and Yao, Yuxuan and Yang, Jixiang and Seo, Junseok and Yoon, Chiho and Watanabe, Kenji and Taniguchi, Takashi and Fu, Liang and Zhang, Fan and others},
  journal={Science},
  volume={384},
  number={6696},
  pages={647--651},
  year={2024},
  publisher={American Association for the Advancement of Science}
}

@article{Sha2024rhombohedral,
  title={Observation of a Chern insulator in crystalline ABCA-tetralayer graphene with spin-orbit coupling},
  author={Sha, Yating and Zheng, Jian and Liu, Kai and Du, Hong and Watanabe, Kenji and Taniguchi, Takashi and Jia, Jinfeng and Shi, Zhiwen and Zhong, Ruidan and Chen, Guorui},
  journal={Science},
  volume={384},
  number={6694},
  pages={414--419},
  year={2024},
  publisher={American Association for the Advancement of Science}
}

@article{Yoon2025QMSC,
  title={Quarter Metal Superconductivity},
  author={Yoon, Chiho and Xu, Tianyi and Barlas, Yafis and Zhang, Fan},
  journal={arXiv preprint arXiv:2502.17555},
  year={2025}
}

@article{Berg2009PDW,
  doi = {10.1038/nphys1422},
  url = {https://doi.org/10.1038/nphys1422},
  year = {2009},
  month = {nov},
  publisher = {Springer Science and Business Media {LLC}},
  volume = {5},
  number = {11},
  pages = {830--833},
  author = {Erez Berg and Eduardo Fradkin and Steven A. Kivelson},
  title = {Charge-4e superconductivity from pair-density-wave order in certain high-temperature superconductors},
  journal = {Nature Physics}
}

@article{Fogler1996stripe,
  title = {Ground state of a two-dimensional electron liquid in a weak magnetic field},
  author = {Fogler, M. M. and Koulakov, A. A. and Shklovskii, B. I.},
  journal = {Phys. Rev. B},
  volume = {54},
  issue = {3},
  pages = {1853--1871},
  numpages = {0},
  year = {1996},
  month = {Jul},
  publisher = {American Physical Society},
  doi = {10.1103/PhysRevB.54.1853},
  url = {https://link.aps.org/doi/10.1103/PhysRevB.54.1853}
}

@article{Koulakov1996stripe,
  title = {Charge Density Wave in Two-Dimensional Electron Liquid in Weak Magnetic Field},
  author = {Koulakov, A. A. and Fogler, M. M. and Shklovskii, B. I.},
  journal = {Phys. Rev. Lett.},
  volume = {76},
  issue = {3},
  pages = {499--502},
  numpages = {0},
  year = {1996},
  month = {Jan},
  publisher = {American Physical Society},
  doi = {10.1103/PhysRevLett.76.499},
  url = {https://link.aps.org/doi/10.1103/PhysRevLett.76.499}
}

@article{Li2025trashcan,
  title={Berry Trashcan With Short Range Attraction: Exact $ p\_x+ i p\_y $ Superconductivity in Rhombohedral Graphene},
  author={Li, Ming-Rui and Kwan, Yves H and Yao, Hong and Bernevig, B Andrei},
  journal={arXiv preprint arXiv:2509.16312},
  year={2025}
}

@article{Bernevig2025trashcan,
  title={" Berry Trashcan" Model of Interacting Electrons in Rhombohedral Graphene},
  author={Bernevig, B Andrei and Kwan, Yves H},
  journal={arXiv preprint arXiv:2503.09692},
  year={2025}
}

@article{Vojta2009stripe,
  title={Lattice symmetry breaking in cuprate superconductors: stripes, nematics, and superconductivity},
  author={Vojta, Matthias},
  journal={Advances in Physics},
  volume={58},
  number={6},
  pages={699--820},
  year={2009},
  publisher={Taylor \& Francis}
}

@article{Gruner1988CDW,
  title = {The dynamics of charge-density waves},
  author = {Gr\"uner, G.},
  journal = {Rev. Mod. Phys.},
  volume = {60},
  issue = {4},
  pages = {1129--1181},
  numpages = {0},
  year = {1988},
  month = {Oct},
  publisher = {American Physical Society},
  doi = {10.1103/RevModPhys.60.1129},
  url = {https://link.aps.org/doi/10.1103/RevModPhys.60.1129}
}

@article{Simon1999comment,
  title={Comment on``Evidence for Anisotropic State of Two-Dimensional Electrons in High Landau Levels''},
  author={Simon, Steven H},
  journal={arXiv preprint cond-mat/9903086},
  year={1999}
}

@article{Zeng2023solid,
  title={Evidence for a Superfluid-to-solid Transition of Bilayer Excitons},
  author={Zeng, Yihang and Shi, Q and Okounkova, A and Sun, Dihao and Dean, CR and Li, JIA},
  journal={arXiv preprint arXiv:2306.16995},
  year={2023}
}

@article{Liu2021anisotropy,
author = {Changjiang Liu  and Xi Yan  and Dafei Jin  and Yang Ma  and Haw-Wen Hsiao  and Yulin Lin  and Terence M. Bretz-Sullivan  and Xianjing Zhou  and John Pearson  and Brandon Fisher  and J. Samuel Jiang  and Wei Han  and Jian-Min Zuo  and Jianguo Wen  and Dillon D. Fong  and Jirong Sun  and Hua Zhou  and Anand Bhattacharya },
title = {Two-dimensional superconductivity and anisotropic transport at KTaO<sub>3</sub> (111) interfaces},
journal = {Science},
volume = {371},
number = {6530},
pages = {716-721},
year = {2021},
doi = {10.1126/science.aba5511},
URL = {https://www.science.org/doi/abs/10.1126/science.aba5511},
eprint = {https://www.science.org/doi/pdf/10.1126/science.aba5511}}

\newpage
\clearpage

\section*{Methods}

\renewcommand{\thefigure}{M\arabic{figure}}
\def\theequation{M\arabic{equation}} 
\def\thetable{M\Roman{table}}
\setcounter{figure}{0}
\setcounter{equation}{0}

In this section, we provide a detailed discussion to further substantiate the results presented in the main text. We elaborate on the extraction of intrinsic transport anisotropy using angle-resolved transport measurements and examine how this anisotropy evolves across the phase space. We also present additional evidence of hysteretic behavior associated with the transport anisotropy, further supporting the conclusions in the main text regarding the tunability of the underlying order parameter and its impact on the 
superconducting transition. We discuss three R6G samples used in this work, and present transport measurements from all of them.

\begin{table}[h!]
\centering
\renewcommand{\arraystretch}{1.4}
\begin{tabular}{|| c | c | c ||} 
 \hline 
\hspace{2pt} Sample \hspace{2pt} & \hspace{2pt} geometry \hspace{2pt} & \hspace{2pt} Layer \hspace{2pt} \\  
 \hline\hline
SF1 & sunflower & R6G \\ 
 \hline SF2 & sunflower & R6G  \\ 
 \hline HB1 & hall-bar & R6G  \\ 
 \hline
\end{tabular}
\caption{{\bf{List of devices.}} This table describes the rhombohedral graphene devices used in this work, including the geometry and layer number. Furthermore, we summarize the sample used for each data set in the following. Fig.~\ref{fig1}, Fig.~\ref{fig2}, Fig.~\ref{fig3}, Fig.~\ref{fig4}b, c, d, Fig.~\ref{resistivity}, Fig.~\ref{Activated}a, Fig.~\ref{angle_T}, Fig.~\ref{CW}, Fig.~\ref{Isotropic}, Fig.~\ref{fermiology}, Fig.~\ref{B_hysteresis}, Fig.~\ref{RT_warming}, Fig.~\ref{SC_IV}, Fig.~\ref{BKT}, Fig.~\ref{B_IV}, Fig.~\ref{Activate_SI}, Fig.~\ref{Layer}, Fig.~\ref{IV_anisotropy}a are taken from sample SF1. Fig.~\ref{RTdiverge_SF2}, Fig.~\ref{RThysteresis_SF2}, Fig.~\ref{IV_RT}, Fig.~\ref{Activated}b, Fig.~\ref{SC_multipleRTs}, and Fig.~\ref{diode} are taken from sample SF2. Fig.~\ref{fig4}a, Fig.~\ref{Activated}c,  Fig.~\ref{IV_anisotropy}b, Fig.~\ref{RT_control}, Fig.~\ref{HB1_RT}, Fig.~\ref{hardaxis}, and Fig.~\ref{HB1_RTs} are taken from sample HB1. }
\label{table:1}
\end{table}

\subsection{I. Stripe order in the superconducting phase}

Stripe order in the superconducting phase is defined here by a clear and stringent criterion: the observation of superconducting transport along one in-plane direction coexisting with insulating behavior along the orthogonal direction. Such a transport response implies a spatial modulation between superconducting and insulating regions, forming stripe-like structures. To the best of our knowledge, this transport signature has not been previously reported in any material system.

In R6G, the superconducting phase emerges from a metallic state that also exhibits stripe-like transport anisotropy. The pronounced anisotropy of this metallic state, as shown in Fig.~1 and Fig.~2, rivals the transport signature of the quantum Hall stripe phase~\cite{Lilly1999stripe,Lilly1999stripe2,Du1999stripe}. While transport anisotropy has been reported in the normal state of other superconductors, the ratio between resistivities measured along orthogonal directions is typically orders of magnitude smaller~\cite{Zhang2025angular,Liu2021anisotropy,Wu2017nematic}. 

The hysteretic superconducting transition offers another distinct observation. Comparable transition hysteresis has only been reported in the superfluid transition between the $^3$He A and B phases, where it is understood as a manifestation of a first-order transition between two competing $p$-wave order parameters~\cite{Vollhardt1990}. In the present case, however, there is no evidence for more than one superconducting phase. We therefore argue that the observed hysteresis must be directly linked to the underlying stripe order, providing a direct window into the relationship between superconductivity and stripe order.

It is also important to emphasize that the observation of extreme transport anisotropy alone does not provide definitive evidence for stripe formation. This is particularly relevant given that the stripe phase is expected to melt into a nematic phase, which can also exhibit strong anisotropy. Within this constraint, evidence for stripe formation can instead be inferred from the distinctive thermal hysteresis behavior. 

Most of the hysteretic transitions—induced by sweeping $n$, varying the current-flow direction, or sweeping the applied current—arise simultaneously with the thermal hysteresis. As shown in Fig.~\ref{fig3}e--g and Fig.~\ref{fig4}a--b, the onset of hysteresis closely coincides with the superconducting transition. Taken together, these observations suggest that the stripe-like behavior is not merely inherited from the normal state, nor does it represent a passive background on which superconductivity develops. Instead, the extreme anisotropy appears to be an intrinsic property of the superconducting phase itself, providing an important constraint for theoretical descriptions of the system.

\subsection{II. Activated behavior perpendicular to the easy axis}

\begin{figure*}
\includegraphics[width=0.99\linewidth]{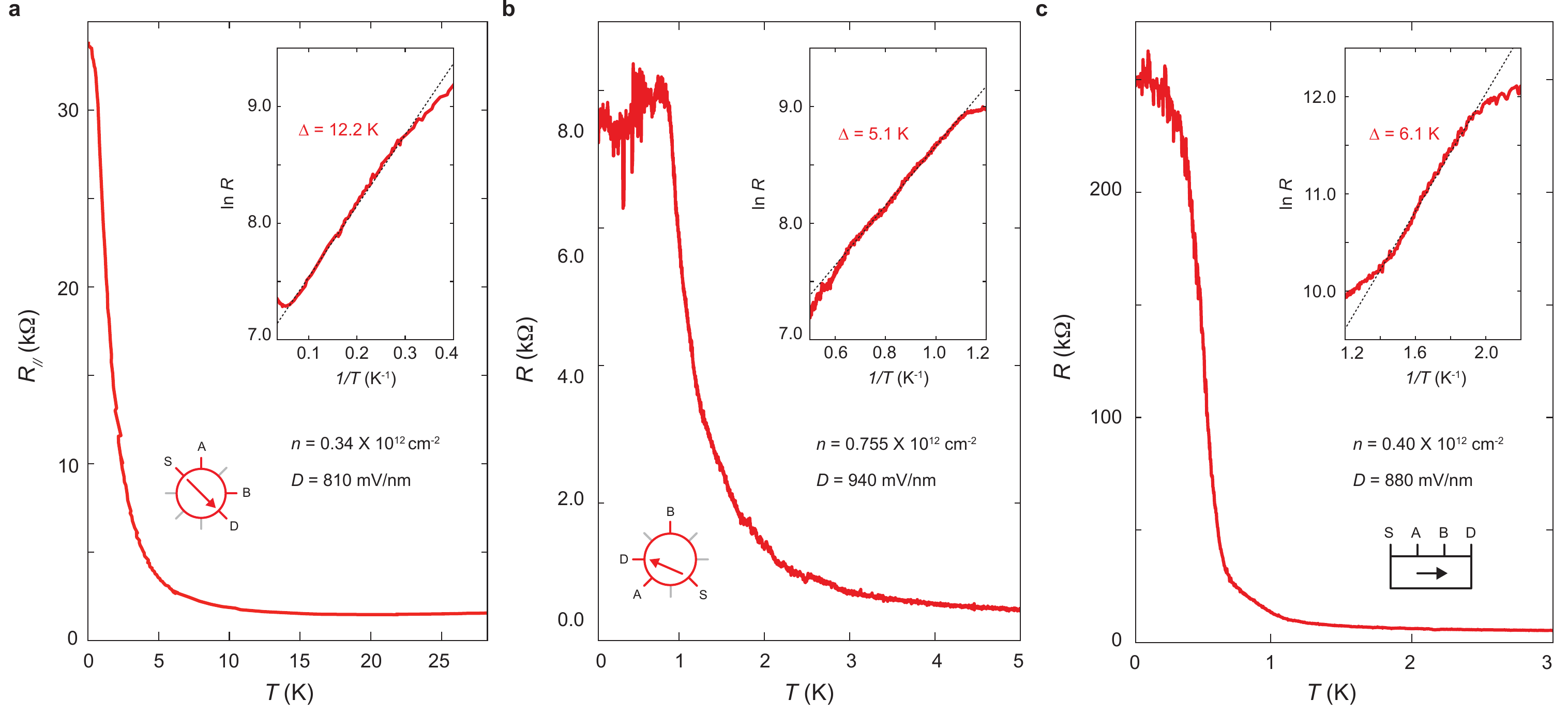}
\caption{\label{Activated} 
\textbf{Activated behavior perpendicular to the easy axis.} 
$R$--$T$ traces measured from (a)~SF1, (b)~SF2, and (c)~HB1, with current applied perpendicular to the easy axis. Insets show the corresponding Arrhenius plots, where the slope provides an estimate of the activation gap associated with insulating behavior along the hard axis. See also Fig.~\ref{Activate_SI}. 
}
\end{figure*}

An important signature of the observed transport anisotropy is the insulating behavior perpendicular to the easy axis. 
Figure~\ref{Activated} plots this insulating behavior for all three R6G samples, measured within the anisotropic regime highlighted in Fig.~\ref{fig1}c.

By fitting the slope of the Arrhenius plots, we estimate the energy gap associated with the activated transport, which can be as large as $12~\mathrm{K}$ (inset of Fig.~\ref{Activated}a). 
The coexistence of activated transport along one direction and metallic behavior along the orthogonal direction bears a striking resemblance to the transport signatures of the quantum Hall stripe phase~\cite{Fradkin2010nematic}, providing strong evidence that a smectic electronic order underlies the observed anisotropy.

\begin{figure}[!b]
\includegraphics[width=0.95\linewidth]{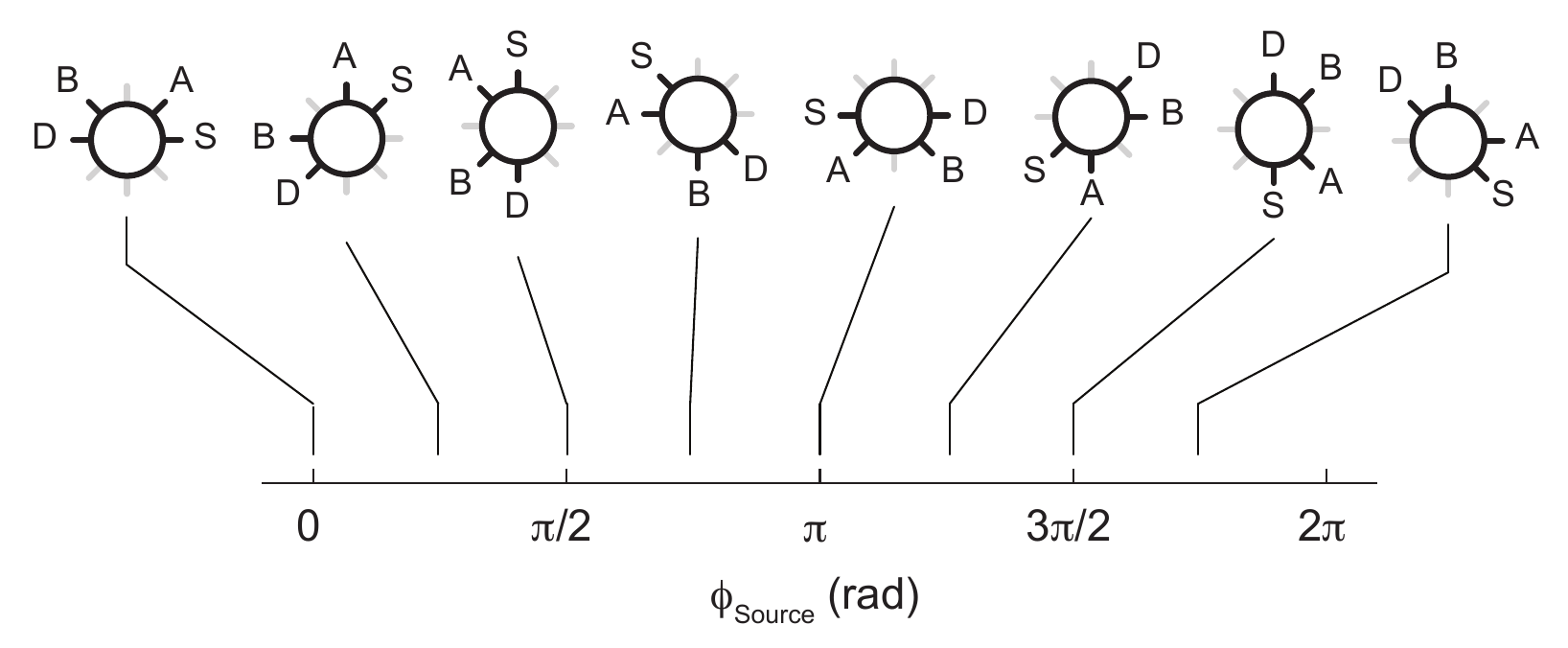}
\caption{\label{ARTM}{\bf{Definition of $\phi$.}} 
Schematic of the angle-resolved transport measurement setup used to extract $R_{\parallel}$  in a sample shaped into the so-called ``sunflower'' geometry.
}
\end{figure}

\begin{figure*}
\includegraphics[width=0.8\linewidth]{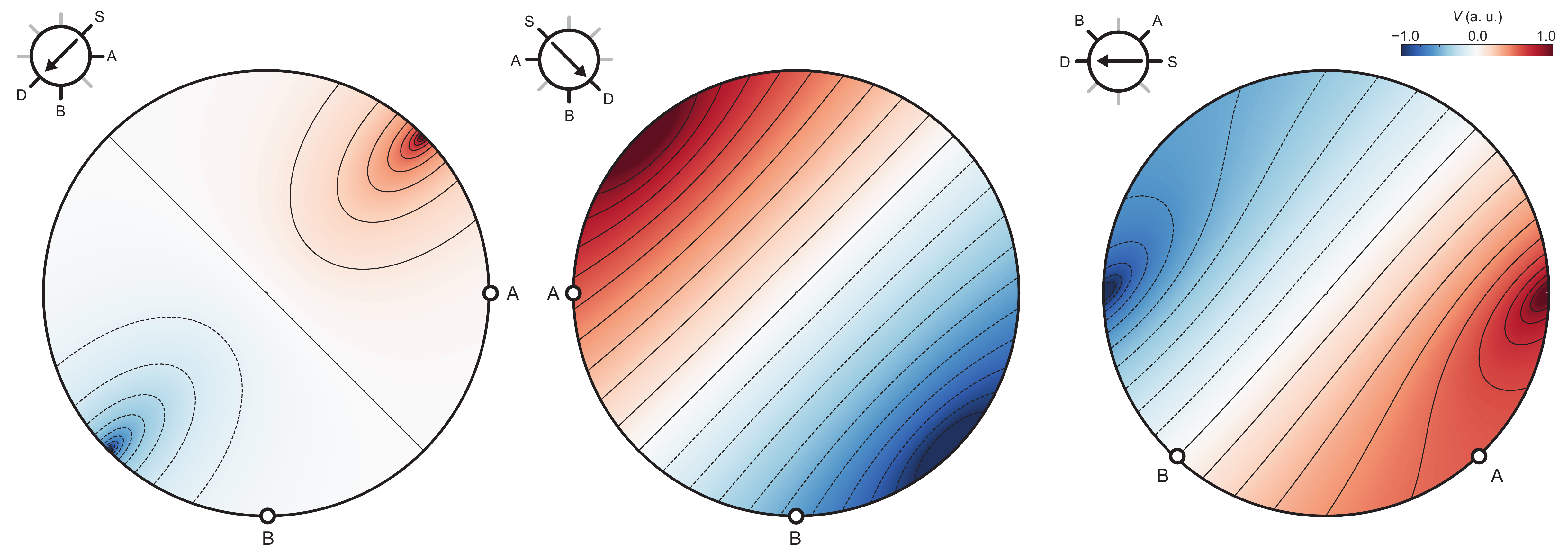}
\caption{\label{Potential}{\bf{Potential distribution across a disk-shaped sample in the presence of extreme anisotropy.}} 
Simulated potential distribution across a disk-shaped sample with different source and drain configurations, assuming an intrinsic resistivity anisotropy of \( \rho_{\max} / \rho_{\min} = 5 \), where \( \rho_{\max} \) and \( \rho_{\min} \) represent the resistivity along the easy and hard axes. The voltage difference between contacts A and B provide a measurement for \Rpara\ for each respective configurations. 
}
\end{figure*}

\begin{figure*}
\includegraphics[width=0.98\linewidth]{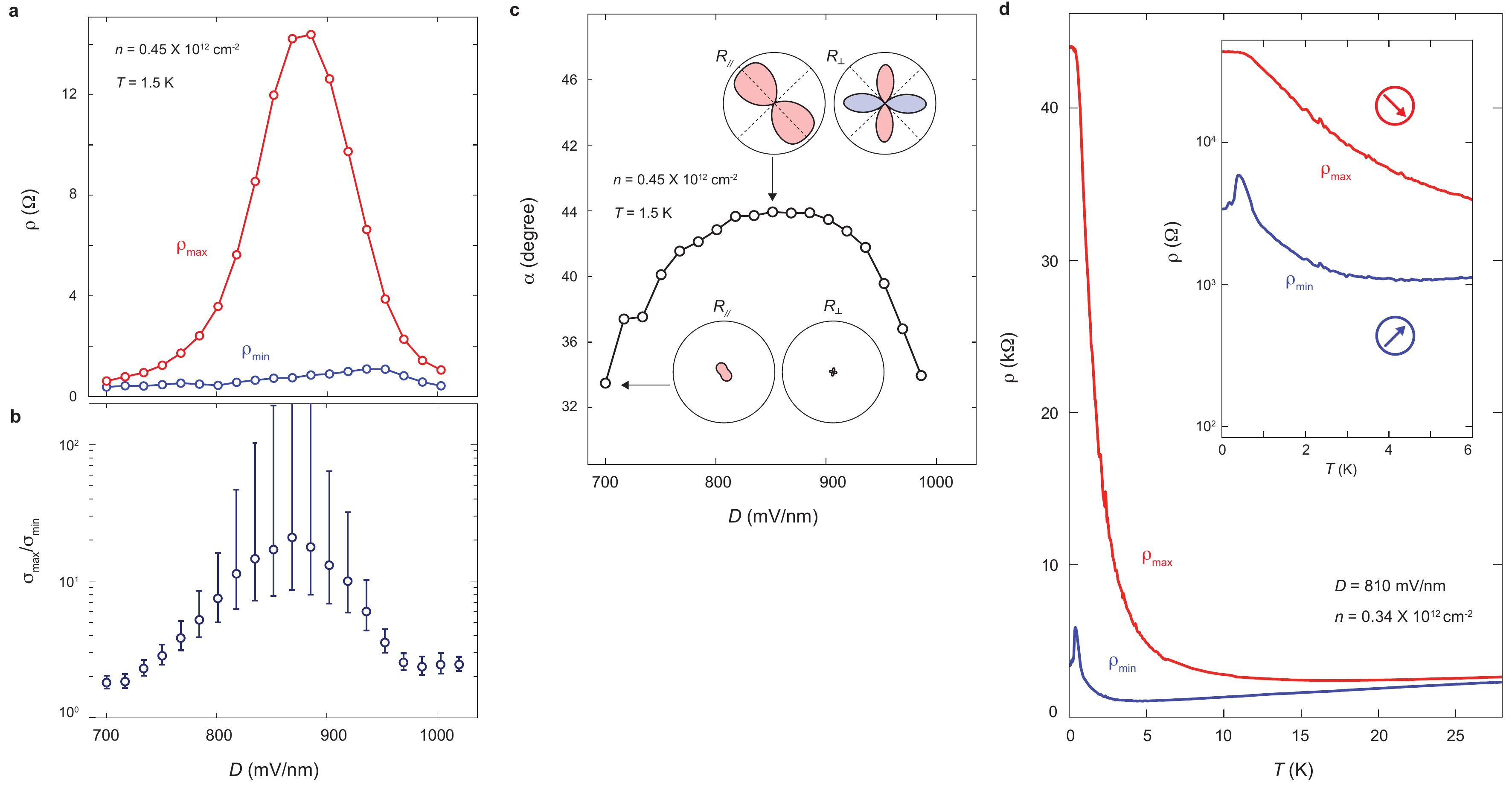}
\caption{\label{resistivity} 
\textbf{Intrinsic anisotropy in resistivity.} 
(a) Resistivity along orthogonal principal axes $\rho_{\mathrm{max}}$ and $\rho_{\mathrm{min}}$,  (b) strength of anisotropy, defined by the ratio of resistivity  $\rho_{max}/\rho_{min}$, and (c) the orientation of the principal axis with minimum resistivity, $\alpha$, as functions of $D$ at $n = 0.45 \times 10^{12}\,\mathrm{cm}^{-2}$.  $\rho$ and $\alpha$ define the resistivity matrix, extracted from angle-resolved measurements as illustrated in Fig.~\ref{fig2}. 
(d) Resistivity versus temperature corresponding to the same data in Fig.~\ref{fig1}f. Inset: logarithmic scale. 
}
\end{figure*}

\subsection{I. Angle-resolved transport measurement}

In angle-resolved measurements, the angle $\phi$ is defined by the orientation of the source (S) and drain (D) contacts. In the sunflower geometry, measurement configurations differ from one another by rotations of $\pi/4$, enabling eight independent measurements corresponding to successive rotations of $\phi$ over $2\pi$. An example of this rotation scheme is shown in Fig.~\ref{ARTM}, which illustrates the measurement configurations used to extract the angular dependence of $R_{\parallel}$.

In the following, we discuss the fitting procedure applied to the extensive set of angle-resolved measurements presented in Fig.~\ref{fig2}.
This procedure relies on the ability to simulate potential distribution across the sample based on the underlying resistivity matrix. 

Fig.~\ref{Potential} provides an illustration. Assuming an intrinsic resistivity anisotropy of $\rho_{\max}/\rho_{\min} = 5$, with the easy axis aligned along $\phi = 45^{\circ}$, the color-scale maps plots the resulting potential distribution across a disk-shaped sample. Since the locations of the source (S) and drain (D) contacts offer boundary condition for the potential distribution,

The three panels in Fig.~\ref{Potential} display the  distributions of electric potential for configurations with $S$ and $D$ aligned along different directions. The underlying anisotropy is evident in the pattern of equipotential lines, which consistently tend to align along the easy axis regardless of the chosen $S$ and $D$ contacts. This preferential alignment of equal-potential contours is the microscopic origin of the observed transport anisotropy. Based on the simulated potential distribution, $R_{\parallel}$ corresponds simply to the potential difference between contacts A and B, as illustrated in Fig.~\ref{Potential}. More generally, a simulated potential map for a given pair of $S$ and $D$ contacts provides a prediction for all measurement configurations that share the same $S$ and $D$ locations.

For example, at each value of $\phi$, panels (a) and (b) of Fig.~\ref{fig2} share the same $S$ and $D$ contacts and can therefore be modeled using the same potential-distribution map.

What follows is the procedure used to vary the components of the resistivity matrix in order to obtain the best fit to all measured data points. The algorithm for determining the optimal fit has been described previously in the Methods section of previous reports ~\cite{Zhang2024nonreciprocity}. As such, we do not reproduce those details here.

It is worth noting that the angular response in Fig.~\ref{fig2} reflects a unique consequence of extreme transport anisotropy, clearly distinguishable from previous experimental reports 
~\cite{Wu2017nematic,Wu2020nematic,Zhang2024nonreciprocity,Zhang2025angular,Morissette2025rhombohedral}. In the presence of moderate anisotropy (e.g., $\rho_{\max}/\rho_{\min} < 2$), the angular oscillation of any measurement configuration can be accurately described by cosine functions with a periodicity of $\pi$ (see Eqs.~M1–S2 for example), and contributions from higher harmonic terms with smaller angular periods are negligible. 
However, this description breaks down once the anisotropy exceeds a certain threshold 
(e.g., $\rho_{\max}/\rho_{\min} > 5$). In this strong-anisotropy regime, higher harmonic components must be included to correctly capture the potential distribution across the device, resulting in the distinctly non-cosine angular oscillations represented by the solid black lines in Fig.~\ref{fig2}.

Due to distinct shape of the angular dependence, \( R_{\parallel} \) is expected to form a  plateau around the principal axes. Consequently, \Rpara\ remains vanishingly small even when the current flow is misaligned from the easy axis. As shown in Fig.~\ref{fig2}, this plateau could span an angular range of $45^{\circ}$.

Fig.~\ref{resistivity}a--c shows the evolution of the extracted resistivity matrix as a function $D$ across the anisotropy regime. A resistivity matrix is characterized by four parameters, $\rho_{max}$, $\rho_{min}$, $\alpha$, and $\rho_H$. Here, $\alpha$ describes the orientation of the easy axis, $\rho_{min}$ ($\rho_{max}$) is the resistivity parallel (perpendicular) to the easy axis. $\rho_H$ is the off-diagonal antisymmetric term, which is also known as the Hall coefficient. In the presence of strong anisotropy, both $\rho_{min}$ and $\rho_H$ are much smaller than $\rho_{max}$, the resulting matrix is essentially determined by $\rho_{max}$ and $\alpha$. 

As shown in Fig.~\ref{resistivity}, anisotropy is the strongest in the range of $800 < D < 900$ mV/nm. As $D$ is detuned from this regime, $\rho_{max}$ and $\rho_{min}$ start to converge, suggesting a more isotropic transport response. At the same time, the principal axis begins to rotate. both  the variation of anisotropy strength and principal axis orientation are consistent with an electronic origin in observed anisotropy. 

Notably, the extracted anisotropy ratio exhibits pronounced uncertainty in the regime of extreme anisotropy, as illustrated in Fig.~\ref{resistivity}b. This uncertainty arises from two main effects.

First, when the transport anisotropy is large, the measured resistance along the easy axis substantially underestimates the true resistivity. This effect is illustrated by the simulated potential distribution in the left panel of Fig.~\ref{Potential}: 
when the S and D contacts are aligned along the easy axis, the equipotential contours preferentially align in the same direction, causing the voltage drop between contacts A and B to be strongly suppressed. The discrepancy between the measured resistance and the intrinsic resistivity along the easy axis is further demonstrated in Fig.~\ref{resistivity}d, which plots the extracted resistivity $\rho$ versus temperature for the data shown in Fig.~\ref{fig1}f. 
While the measured resistance along the easy axis is approximately $100~\Omega$, the corresponding resistivity obtained from the angular fit exceeds $1000~\Omega$. 
In contrast, for currents flowing perpendicular to the easy axis, the measured resistance is in reasonable agreement with the extracted resistivity.

\begin{figure}[!b]
\includegraphics[width=0.85\linewidth]{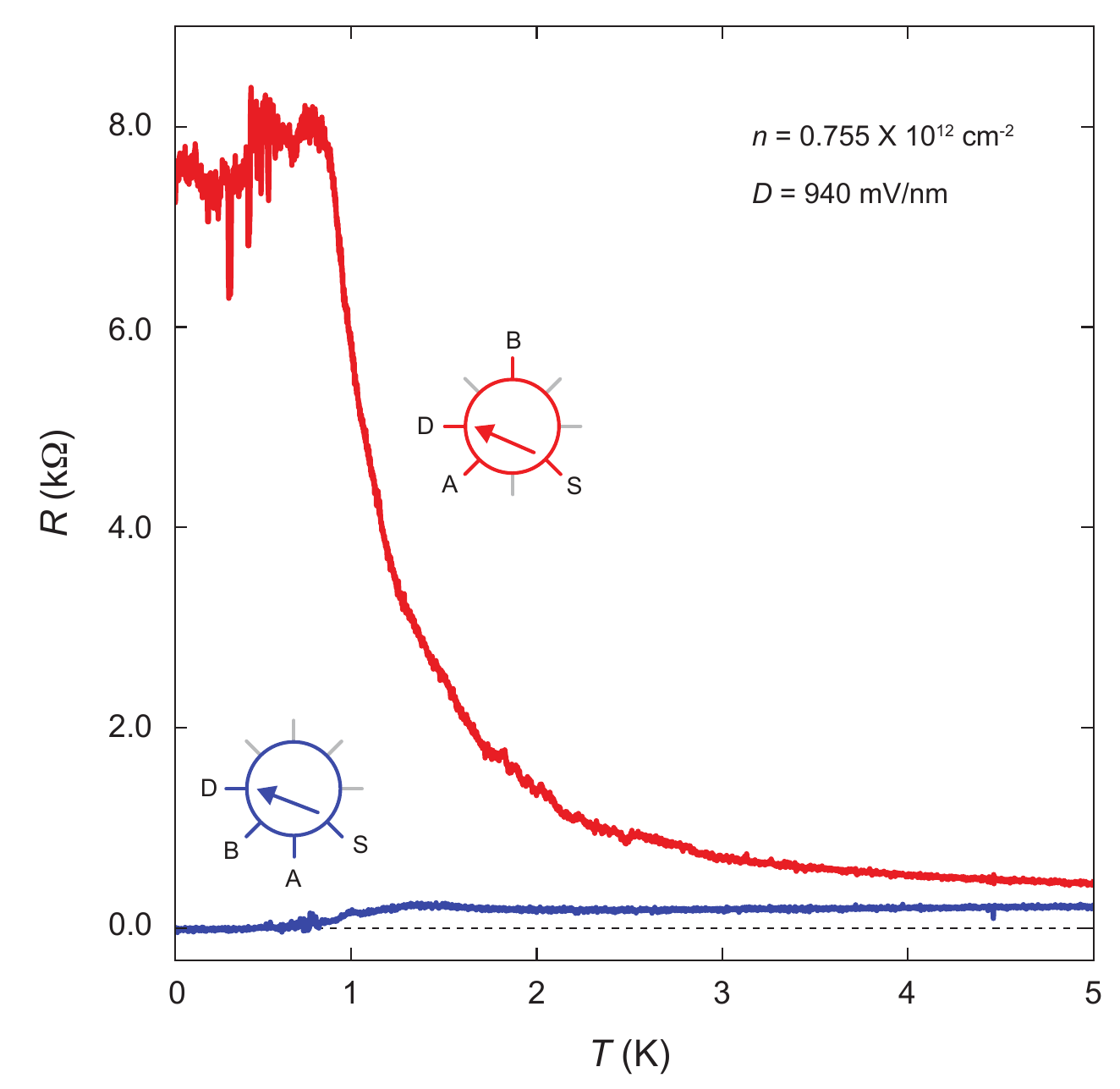}
\caption{\label{RTdiverge_SF2} \textbf{Extreme anisotropy in SF2.} 
Resistance versus temperature measured at $n = 0.755 \times 10^{12}~\mathrm{cm}^{-2}$ and $D = 940~\mathrm{mV/nm}$, inside the SC\,i phase, from a second sunflower sample (SF2). 
Inset shows measurement configurations, the blue and red traces correspond to measurements along two orthogonal directions. 
While the traces converge at $T > 5~\mathrm{K}$, they diverge strongly at low temperature: the blue trace exhibits vanishing resistance consistent with a low-temperature superconducting phase, whereas the red trace shows insulating behavior that 
remains highly resistive upon cooling. This behavior is consistent with the anisotropy observed in SF1, shown in Fig.~\ref{fig3}. 
}
\end{figure}

Second, varying the anisotropy ratio between $100$ and $1000$ primarily affects the sharpness of the corners in the expected angular response (black solid lines in Fig.~\ref{fig2}), without substantially altering the overall functional form. 
As a result, the angular fits become relatively insensitive to the precise value of $\rho_{\max}/\rho_{\min}$ in this extreme-anisotropy regime, contributing an additional source of uncertainty to the extracted anisotropy ratio.

Interestingly, the effect of extreme anisotropy on resistance measurements along the easy axis was first recognized in the context of quantum Hall stripe phases ~\cite{Simon1999comment,Lilly1999stripe}. In our case, the ability to perform extensive angle-resolved measurements provides new insight into the intrinsic difficulty of determining the anisotropy ratio, a limitation that arises from this same geometric effect ~\cite{Vafek2023anisotropy}.

\subsection{II. Transport responses from SF2}

\begin{figure*}
\includegraphics[width=0.7\linewidth]{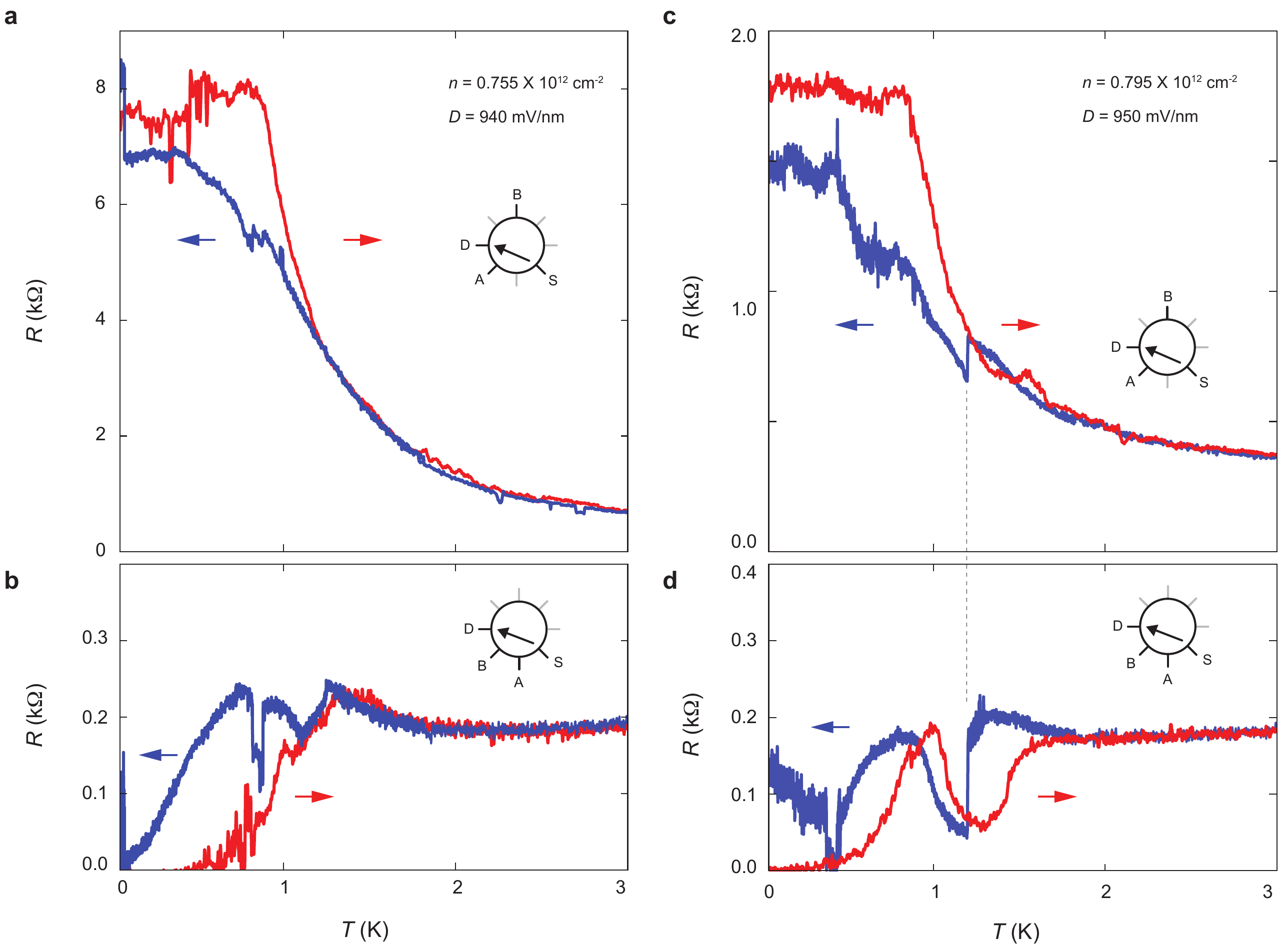}
\caption{\label{RThysteresis_SF2} \textbf{Temperature-driven hysteresis in SF2.} 
$R$--$T$ traces measured on warming (red) and cooling (blue) from SF2 at 
(a--b) $n = 0.755 \times 10^{12}~\mathrm{cm}^{-2}$ and $D = 940~\mathrm{mV/nm}$, and 
(c--d) $n = 0.795 \times 10^{12}~\mathrm{cm}^{-2}$ and $D = 950~\mathrm{mV/nm}$. Inset shows measurement configurations.
Both sets of $(n, D)$ values lie inside the low-temperature superconducting regime.
}
\end{figure*}

In the following, we present transport measurements from two additional R6G samples, SF2 and HB1. 

Fig.~\ref{RTdiverge_SF2} shows the extreme transport anisotropy measured from a second R6G sample patterned into the sunflower geometry, referred to as SF2. With current flowing along $\phi = 157.5^{\circ}$, the voltage response is measured along two orthogonal directions, yielding the red and blue $R$--$T$ traces. 
These measurements reveal a pronounced anisotropy similar to that reported in Fig.~\ref{fig3}: as the temperature is lowered, the two traces diverge strongly. 
The blue trace exhibits vanishing resistance below $T < 1~\mathrm{K}$, consistent with a low-temperature superconducting phase, whereas the red trace shows a sharp upturn around the same temperature, indicating insulating behavior that remains highly resistive down to the base temperature of the dilution refrigerator.

Due to two of the eight contacts being non-transparent, a full angle-resolved measurement cannot be performed on SF2. Nevertheless, the $R$--$T$ traces in Fig.~\ref{RTdiverge_SF2} indicate that $\phi = 157.5^{\circ}$ lies within the low-resistance plateau expected in the angular dependence of $R_{\parallel}$ (top panel of Fig.~\ref{fig2}a). 
Consequently, the orientation of the easy axis in SF2 must fall within the range 
$135^{\circ} < \alpha < 180^{\circ}$.

Beyond the anisotropy, SF2 also exhibits pronounced hysteresis between warming and cooling, as illustrated in Fig.~\ref{RThysteresis_SF2}. While the overall hysteretic behavior is consistent with that observed in SF1, it displays several notable distinctions. 
First, the onset of hysteresis occurs at a higher temperature, approaching $2~\mathrm{K}$ in Figs.~\ref{RThysteresis_SF2}c--d. Second, the temperature hysteresis is accompanied by sharp, discontinuous jumps in the measured resistance. For example, the vertical dashed line in Figs.~\ref{RThysteresis_SF2}c--d highlights simultaneous jumps in both blue 
curves, taken from the same cooling sweep. The jumps point to changes in the underlying order parameter associated with transport anisotropy.

What also stands out is the behavior of the blue trace in Fig.~\ref{RThysteresis_SF2}d, which repeatedly switches between low- and high-resistance states. 
At $T = 30~\mathrm{mK}$, this trace saturates at a non-zero resistance. 
A vanishing resistance can be recovered, however, by tuning the carrier density $n$, applying a large d.c.\ current bias $I_{\mathrm{dc}}$, or changing the current-flow direction. This tunable behavior is further demonstrated in Fig.~\ref{SC_multipleRTs}. Each time the sample is thermally reset at 3~K, a distinct $R$--$T$ trace is obtained on cooling, highlighting the sensitivity of the superconducting transition. 


\begin{figure*}
\includegraphics[width=0.7\linewidth]{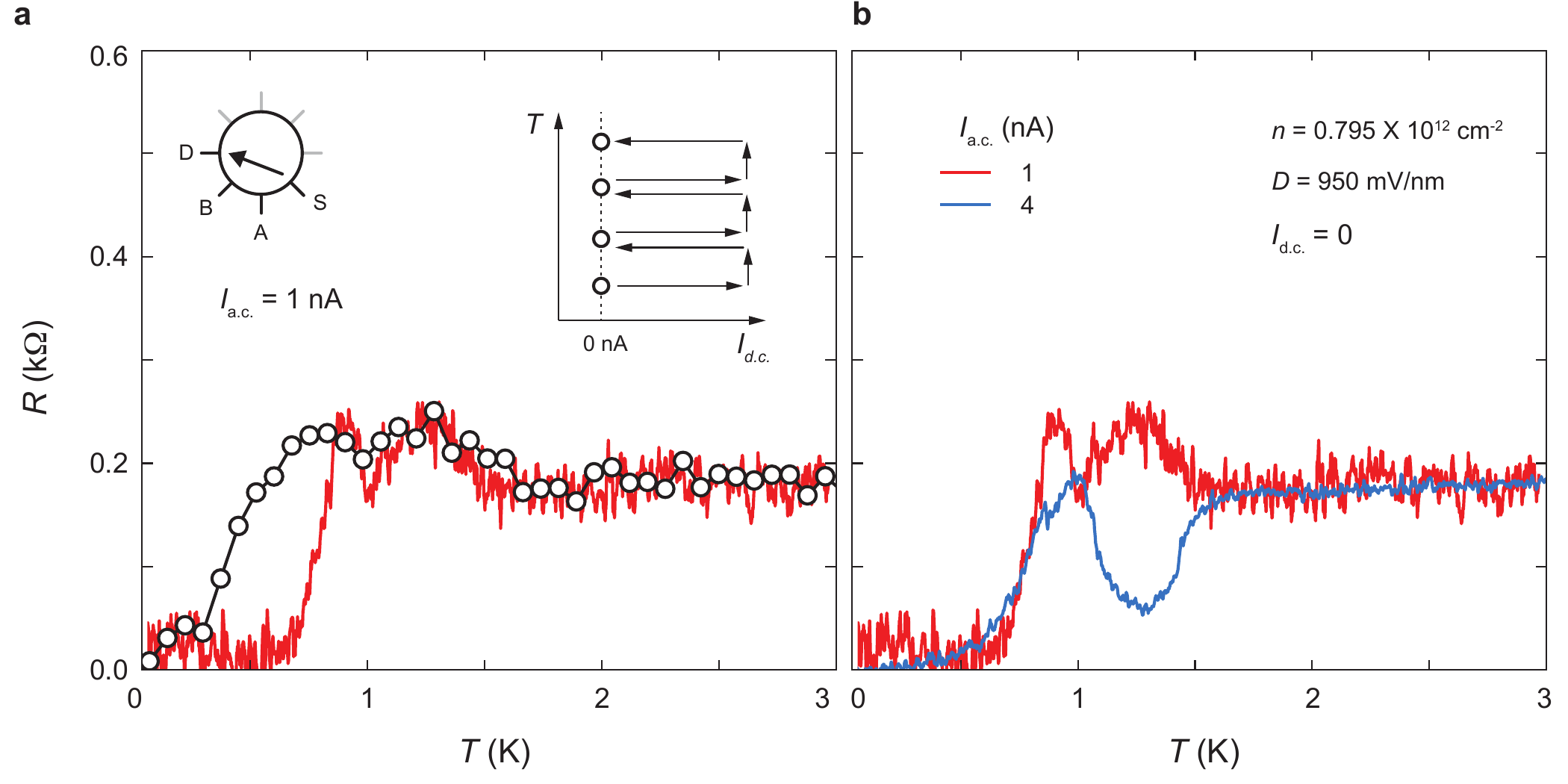}
\caption{\label{IV_RT} \textbf{The effect of applied current on superconducting transition. } 
(a)  $R$--$T$ curves measured in the superconducting phase of SF2 with $I_{a.c.} = 1$ nA and zero d.c. bias ($I_{d.c.} = 0$).
The red trace corresponds to continuous warming and cooling sweeps, respectively, whereas the black open circles are measured at the same $(n, D)$ point but with $I_{d.c.}$ swept to 40 nA between consecutive temperature points.
(b) $R$--$T$ curves measured in the superconducting phase of SF2 with $I_{a.c.} = 1$ nA (red trace )and $I_{a.c.} = 4$ nA (blue trace). All measurements are performed at $n = 0.795 \times 10^{12}$ cm$^{-2}$ and $D = 950$ mV/nm.
}
\end{figure*}

\begin{figure*}
\includegraphics[width=0.8\linewidth]{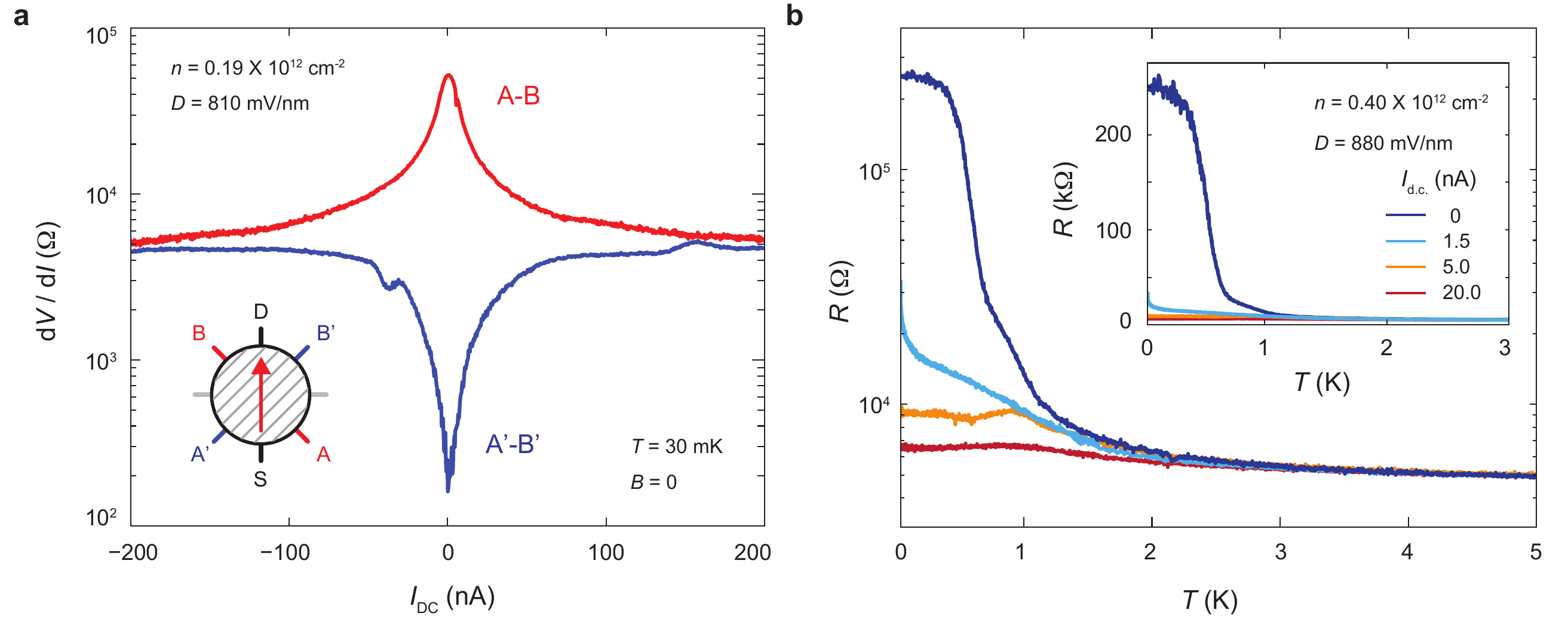}
\caption{\label{IV_anisotropy} \textbf{The effect of applied current on metallic anisotropy. } 
(a) Differential resistance $dV/dI$ measured from SF1 as a function of d.c.\ current bias $I_{\mathrm{DC}}$ applied parallel (blue) and perpendicular (red) to the easy axis. (b) $R$--$T$ curves measured from HB1 under different d.c.\ current biases. Inset: logarithmic scale. Both panels are measured in the anisotropic regime but outside the superconducting pocket. 
}
\end{figure*}

\begin{figure*}
\includegraphics[width=0.9\linewidth]{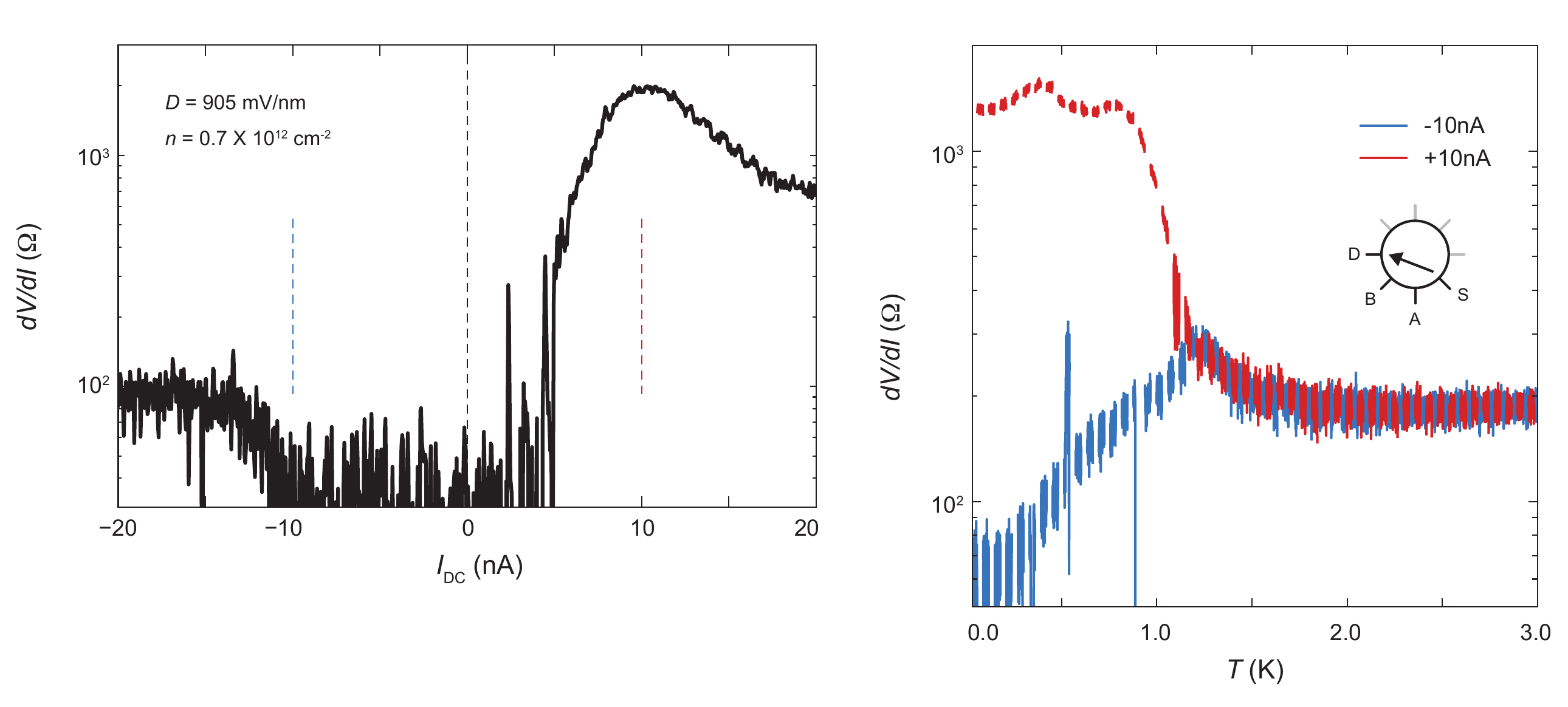}
\caption{\label{diode}{\bf Nonreciprocity in superconducting transport.}
Left panel: Differential resistance $dV/dI$ as a function of d.c. current bias $I_{\mathrm{d.c.}}$ measured in the \textit{SC\,i} regime of SF2. The critical supercurrent is identified by the onset of differential resistance. The magnitude of the critical current is highly asymmetric between forward and reverse current bias, demonstrating the zero-field superconducting diode effect~\cite{Lin2022SDE}. 
Right panel: Temperature dependence of the differential resistance measured at $+10$ and $-10$~nA, shown as red and blue traces, respectively. The data reveal a characteristic temperature of approximately $1.2$~K at which the nonreciprocity onsets. Above this temperature, the nonreciprocity vanishes and the sample behaves as a conventional, reciprocal metal. The resistance also exhibits a slight upturn immediately above this onset temperature, resembling similar behavior observed in Fig.~\ref{fig3}g and Fig.~4a. The onset of nonreciprocity occurs close to the stripe formation temperature and the emergence of orbital ferromagnetism, suggesting a link between these orders.}
\end{figure*}

\subsection{III. Tunability of applied current}

Figures~\ref{fig4}c--d illustrate how a d.c.~current bias modifies the anisotropic metallic state. In this subsection, we extend this analysis to investigate how an applied current influences the superconducting transition. 

Fig.~\ref{IV_RT}a shows $R$--$T$ curves of the superconducting phase measured from SF2 with a small a.c. current bias of 1nA and zero d.c. current. While both traces are measured on warming, the red trace denotes a continuous warming measurement. In contrast, the black open circles are measured at the same $(n, D)$ point but with $I_{d.c.}$ swept to 40 nA between consecutive temperature points.

Figure~\ref{IV_RT}a shows $R$--$T$ curves of the superconducting phase measured from SF2 using a small a.c.~current bias of 1~nA and zero d.c.~bias. Both traces are acquired during warming, but under different measurement protocols. The red trace corresponds to a continuous warming sweep at fixed $(n, D)$ inside the superconducting regime, whereas the black open circles are measured at the same $(n, D)$ point but with 
the d.c.~current $I_{\mathrm{d.c.}}$ swept up to 40~nA between consecutive temperature points. Because 40~nA exceeds the critical supercurrent, this protocol effectively resets the superconducting state between measurements.

The effect of this measurement protocol closely resembles the behaviors shown in Fig.~\ref{fig4}a--b. Taken together, these observations demonstrate that resetting the superconducting phase—whether by sweeping $n$, applying a large $I_{\mathrm{d.c.}}$, or varying the current-flow direction—consistently leads to a pronounced suppression of the superconducting transition temperature. To account for this phenomenon, two crucial factors are needed: first, the details of transport anisotropy, such as easy axis orientation, must be tunable with varying experimental parameters; second, the superconducting transition must depend on the orientation of current flow relative to the principal axis. 

The effect of this measurement protocol closely resembles the behaviors shown in Fig.~\ref{fig4}a--b. Taken together, these observations demonstrate that resetting the superconducting phase—whether by sweeping $n$, applying a large $I_{\mathrm{d.c.}}$, or varying the current-flow direction—consistently leads to a pronounced suppression of the superconducting transition temperature. To account for this phenomenon, two key ingredients are required. First, the details of the transport anisotropy—such as the orientation of the easy axis—must be tunable under changes in experimental parameters. 
Second, the superconducting transition must depend sensitively on the orientation of the applied current relative to the principal axis of anisotropy. 

Figure~\ref{IV_RT}b shows two $R$--$T$ traces, both measured continuously on warming at the same $(n, D)$ values. The red and blue traces correspond to measurements performed with slightly different a.c.\ current biases, $1~\mathrm{nA}$ and $4~\mathrm{nA}$, respectively. In both cases, the superconducting transition appears at the same temperature; however, the larger a.c.\ current induces a pronounced dip in the resistance just above $1~\mathrm{K}$.  This behavior is consistent with the current-induced modification of the metallic phase shown in Fig.~\ref{fig4}c--d, except that here the effect of tuning the current is most prominent in the intermediate 
temperature window $1~\mathrm{K} < T < 1.5~\mathrm{K}$.

Further evidence that the nature of the anisotropy can be modified by the applied current is obtained by examining the metallic state within the anisotropic regime but outside the superconducting phase. Figure~\ref{fig4}c shows the $I$–$V$ characteristics measured using the same configuration as in Fig.~\ref{fig3}c, which enables 
voltage responses to be recorded simultaneously along directions parallel and perpendicular to the easy axis. At small d.c.\ bias, the measurement reveals a strongly anisotropic response, with high (low) resistance observed perpendicular (parallel) to the easy axis. However, this large contrast rapidly diminishes with increasing d.c.\ bias, indicating that the principal axis of anisotropy effectively rotates toward the direction of current flow, thereby suppressing the difference between the two voltage signals.

Extreme sensitivity to d.c.\ current bias is also observed in the Hall-bar sample, despite the fact that the direction of current flow cannot be controlled in this geometry. With an a.c.\ excitation of $0.5~\mathrm{nA}$ and zero d.c.\ bias, 
Fig.~\ref{fig4}d shows an $R$--$T$ trace indicating a strongly insulating state emerging below $T < 1~\mathrm{K}$. Remarkably, this insulating behavior is entirely suppressed by the application of a modest d.c.\ current of only $1.5~\mathrm{nA}$. Because such a small d.c.\ bias is unlikely to destroy an intrinsically insulating phase, we conclude that the observed nonlinear response arises from a rotation of the principal axis of anisotropy in response to the applied d.c.\ bias.

\subsection{V. The onset temperature of the potential smectic order}

\begin{figure*}
\includegraphics[width=0.74\linewidth]{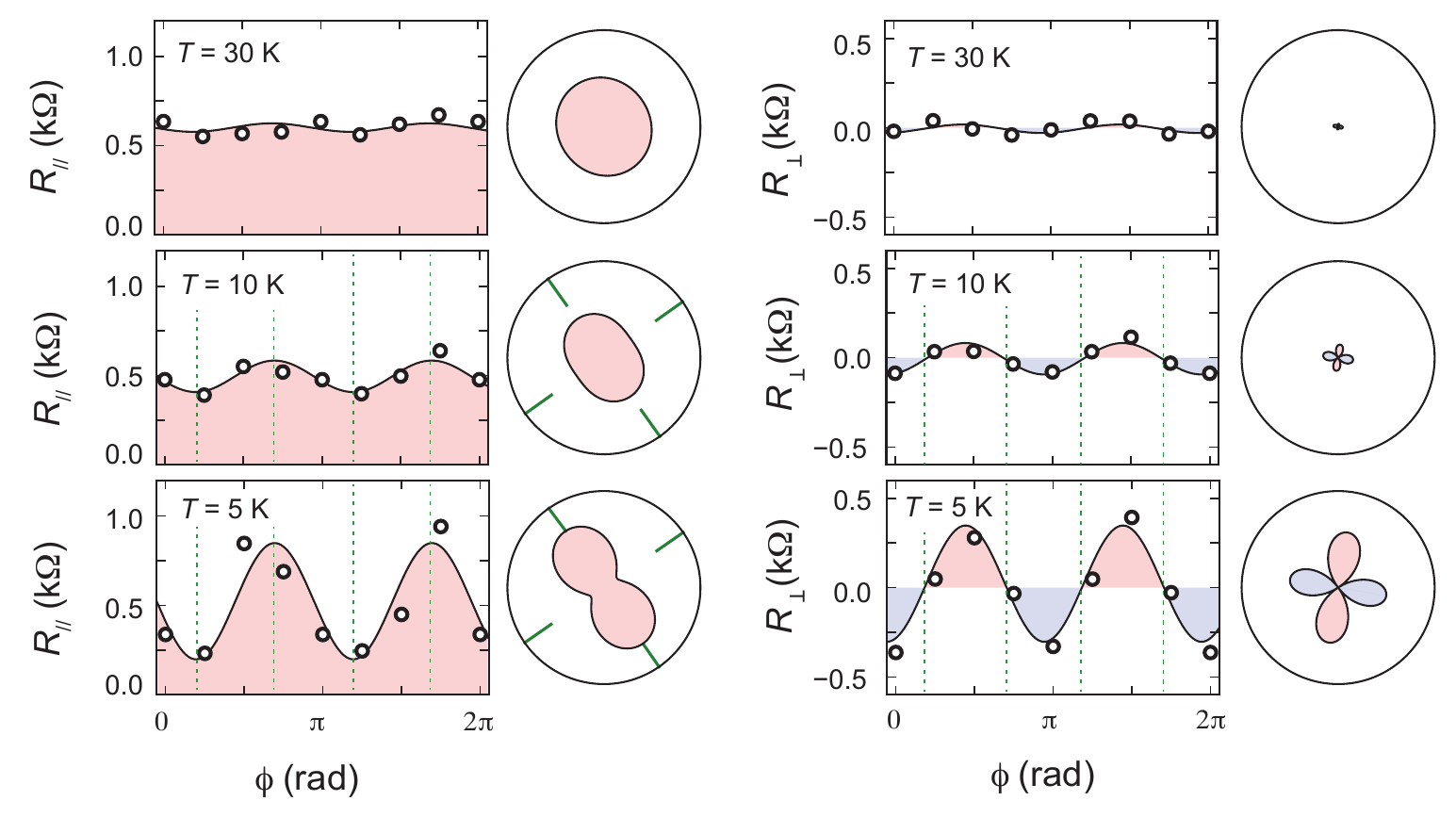}
\caption{\label{angle_T} 
\textbf{Angular dependence of the transport response at elevated temperatures.} 
Angle-resolved measurements of $R_{\parallel}$ (left) and $R_{\perp}$ (right) at 
$T = 30~\mathrm{K}$ (top), $10~\mathrm{K}$ (middle), and $5~\mathrm{K}$ (bottom). 
Solid curves denote best fits using Eqs.~(M1) and (M2). The angular oscillations is consistent with the orientation of easy and hard axes in the anisotropy regime of SF1 (see Fig.~\ref{fig1}, Fig.~\ref{fig2}, and Fig.~\ref{fig3}). 
Measurements are performed at $n = 0.74 \times 10^{12}~\mathrm{cm}^{-2}$ and $D = 954~\mathrm{mV/nm}$, within the \textit{SC\,i} region, from sample SF1.}
\end{figure*}

Across all three samples, $T \sim 1.5~\mathrm{K}$ emerges as a characteristic temperature marking the onset of hysteresis 
and versatile tunability within the anisotropic regime 
(see Fig.~\ref{fig3}e--g, Fig.~\ref{fig4}a,b,d, Fig.~\ref{RThysteresis_SF2}, Fig.~\ref{IV_RT}, 
Fig.~\ref{RT_warming}, Fig.~\ref{SC_multipleRTs}, and Fig.~\ref{HB1_RT}). 
Above this temperature, all three samples behave as conventional metals with an ohmic transport response and no 
detectable hysteresis or tunability. Notably, a finite transport anisotropy persists to much higher temperatures 
(Fig.~\ref{angle_T}). 

This separation of temperature scales highlights the emergence of an additional low-temperature order beyond simple 
anisotropy. We therefore associate this characteristic temperature with the onset of smectic order.

We note that the precise onset temperature varies modestly depending on the specific values of $n$ and $D$ within the 
anisotropic regime. In addition, thermal hysteresis does not always appear at exactly the same temperature between consecutive warming and cooling measurements
(see Fig.~\ref{IV_RT} and Fig.~\ref{RT_warming}), consistent with the random nature of domain formation and easy-axis 
selection upon cooling. Nevertheless, $T \sim 1.5~\mathrm{K}$ provides a robust upper bound that delineates the onset of 
hysteresis and tunability across all three samples for the corresponding $(n,D)$ values examined in this work.

The temperature hysteresis observed here stands in stark contrast to the behavior measured in an unpolarized metallic regime, where the warming and cooling $R$--$T$ traces collapse onto the same curve (Fig.~\ref{RT_control}). This comparison highlights that the hysteresis is not a generic feature of the measurement protocol, but instead emerges only in the anisotropic regime.

It is also worth noting that temperature hysteresis associated with solid--melting transitions has been reported in other correlated electron systems. For example, in quantum Hall graphene bilayers, thermal hysteresis has been shown to be directly linked to a first-order melting transition of an exciton solid, whereas the superfluid transition of the exciton fluid phase exhibits no measurable hysteresis ~\cite{Zeng2023solid}. Although this represents a distinct physical platform, the association between thermal hysteresis and crystalline order in that system provides supporting context for our identification of a smectic electronic order in the present work.

\subsection{VI. Microscopic coexistence and macroscopic phase separation}

Smectic order is a distinctive electronic phase that intrinsically involves spatial modulation between coexisting conducting and insulating regions. This naturally raises an important question for transport experiments: how can one distinguish an intrinsic stripe (smectic) order from macroscopic phase separation arising from extrinsic spatial inhomogeneity? While transport measurements alone do not provide direct real-space information regarding the characteristic length scale, several observations strongly support the interpretation of an intrinsic smectic electronic order, rather than sample inhomogeneity.

First, the extensive angle-resolved transport measurements presented in Fig.~\ref{fig2} provide a stringent constraint on the underlying transport tensor. All twelve measurement configurations, spanning the full angular range, can be 
quantitatively captured by a single resistivity tensor. Such a high degree of internal consistency is incompatible with macroscopic phase separation, which would generically produce configuration-dependent responses that cannot be reconciled within a unified tensor description.

Second, the observed hysteresis is highly tunable and can be systematically controlled by temperature cycling, sweeping carrier density $n$, varying current-flow direction, or applying a d.c.\ bias. Such controlled and reversible manipulation of the transport response, which onsets sharply around $1.5$ K, is difficult to reconcile with static disorder or frozen phase separation, but is naturally explained by reconfiguration of an intrinsic order parameter with multiple degenerate orientations, such as the director of a smectic phase.

Third, the extreme anisotropy and associated hysteretic behavior are reproducible across multiple R6G samples. This reproducibility argues strongly against accidental inhomogeneity, which would be expected to vary significantly between devices.

Finally, it is instructive to compare our results with those from magic-angle graphene moir\'e systems, which are known to 
exhibit substantial spatial inhomogeneity. For example, the observation of Fraunhofer interference patterns in these systems is widely understood to arise from accidental coexistence of superconducting and metallic regions on mesoscopic length scales. Importantly, however, such phase separation does not give rise to the extreme transport anisotropy or hysteresis observed here.

A representative example is provided in Ref.~\cite{Zhang2025angular}, where transport anisotropy is reported in several magic-angle twisted trilayer graphene (tTLG) samples. Despite the presence of anisotropy in both the superconducting and metallic phases,   measured resistance vanishes at the same temperature regardless of the direction of current flow. This behavior is fundamentally different from that observed in R6G, where the superconducting transition depends sensitively on the orientation of current flow relative to the easy axis.

More generally, spatial inhomogeneity in graphene moir\'e systems is typically reflected in a broad superconducting transition. In contrast, the superconducting transitions observed in R6G are remarkably sharp (Fig.~\ref{fig4}a--b, Fig.~\ref{RThysteresis_SF2}, and Fig.~\ref{HB1_RT}), further distinguishing our observations from 
scenarios dominated by macroscopic phase separation. 

Together, these observations point to an intrinsic anisotropy in the superconducting and metallic phases rather than extrinsic inhomogeneity.

\begin{figure}
\includegraphics[width=0.85\linewidth]{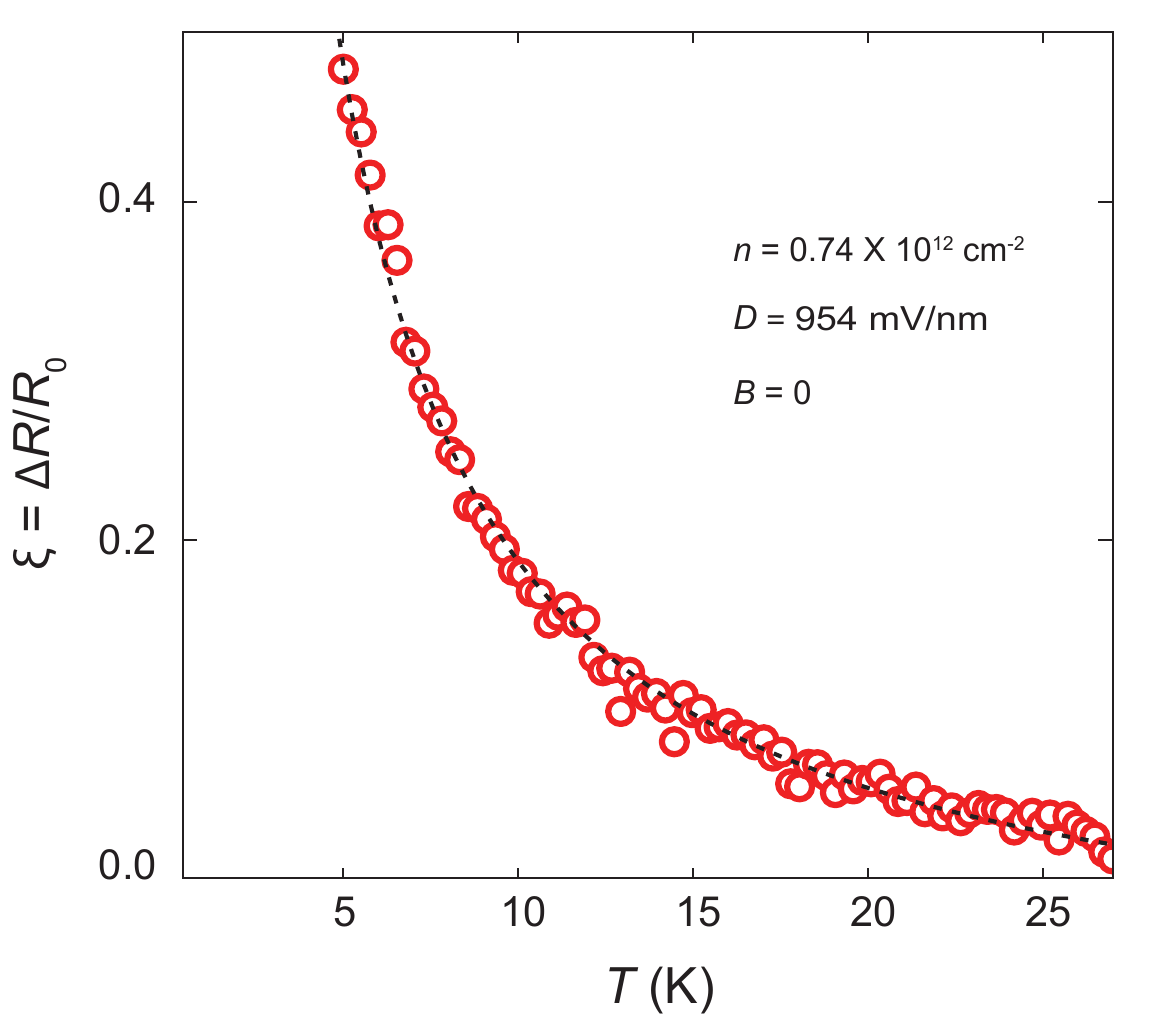}
\caption{\label{CW} 
\textbf{Curie-Weiss behavior.} 
At elevated temperature, the onset of anisotropy is well captured by a Curie-Weiss behavior (dashed black line) in $\xi = \Delta R/R_0$.
}
\end{figure}

\subsection{VI. Angular oscillation at $T > 5$ K}

Fig.~\ref{angle_T} plots angular dependence of \Rpara\ and \Rperp\ measured at higher temperature. As anisotropy decreases with increasing temperature, the observed angular oscillation recovers cosine behaviors, described by:

\begin{eqnarray}
R_{\parallel}(\phi) &=& R_0 - \Delta R \cos 2(\phi - \alpha), \\
R_{\perp}(\phi) &=& R_H + \Delta R \sin 2(\phi - \alpha).
\end{eqnarray}

Here, \( \Delta R \) is the oscillation amplitude, \( R_0 \) is the mean value of \( R_{\parallel} \), \( R_H \) is the average of \( R_{\perp} \), and \( \alpha \) shares the same definition as the main text. In the regime of moderate anisotropy, Eqs.~(M1) and (M2) fully determine the resistivity matrix. The ratio \( \Delta R / R_0 \) provides a quantitative measure of the anisotropy strength, which is referred to as $\xi$, while a non-zero \( R_H \) reflects antisymmetric off-diagonal components of the conductivity tensor, signaling an anomalous Hall effect arising from orbital ferromagnetism~\cite{Morissette2025rhombohedral}. 

In the temperature regime above 5 K, the angular oscillations recover a cosine form because higher harmonic contributions become negligible when the anisotropy is not extremely large.

Figure~\ref{CW} characterizes the temperature evolution of the transport anisotropy in the regime of moderate anisotropy by plotting $\xi$ as a function of temperature. We find that the temperature dependence of $\xi$ is well described by a Curie--Weiss form (black dashed line), consistent with the widely observed Curie--Weiss–like enhancement of nematic susceptibility in correlated electron systems~\cite{Chu2012divergent,Fernandes2014nematic,Bohmer2022nematicity}. The Curie--Weiss fit extrapolates to a divergence at approximately $T_{\mathrm{Curie}} \sim 2$~K, which lies above the temperature scale at which smectic (stripe) order is identified from transport measurements. We note that the uncertainty in the extracted value of $T_{\mathrm{Curie}}$ arises from the fact that the Curie--Weiss fit is performed only at higher temperatures, in order to avoid the regime of extreme anisotropy where the anisotropy ratio cannot be reliably determined.

At temperatures below \(5~\mathrm{K}\), the emergence of extreme anisotropy renders Eqs.~(M1) and (M2) inadequate for describing the angular dependence. 
In this regime, higher harmonic contributions become essential, and the full angular structure must be captured by the tensor-based fitting procedure illustrated in Fig.~\ref{fig2}.

\clearpage

\newpage
\begin{widetext}
\section{Supplementary Materials}

\begin{center}
\textbf{\large Stripe Order in the Metallic and Superconducting Phases of Rhombohedral Hexalayer Graphene}\\
\vspace{10pt}

Peiyu Qin$^{\ast}$,
Hai-Tian Wu$^{\ast}$,
Ron Q. Nguyen$^{\ast}$,
Erin Morissette$^{\ast}$,
Naiyuan J. Zhang,
Kenji Watanabe, Takashi Taniguchi, and J.I.A. Li$^{\dag}$

\vspace{10pt}
$^{\dag}$ Corresponding author. Email: jia.li@austin.utexas.edu
\end{center}


\renewcommand{\vec}[1]{\boldsymbol{#1}}

\renewcommand{\thefigure}{S\arabic{figure}}
\def\theequation{S\arabic{equation}}
\def\thetable{S\Roman{table}}
\setcounter{figure}{0}
\setcounter{equation}{0}

\vspace{0.1 in}

\begin{figure*}[!b]
\includegraphics[width=0.85\linewidth]{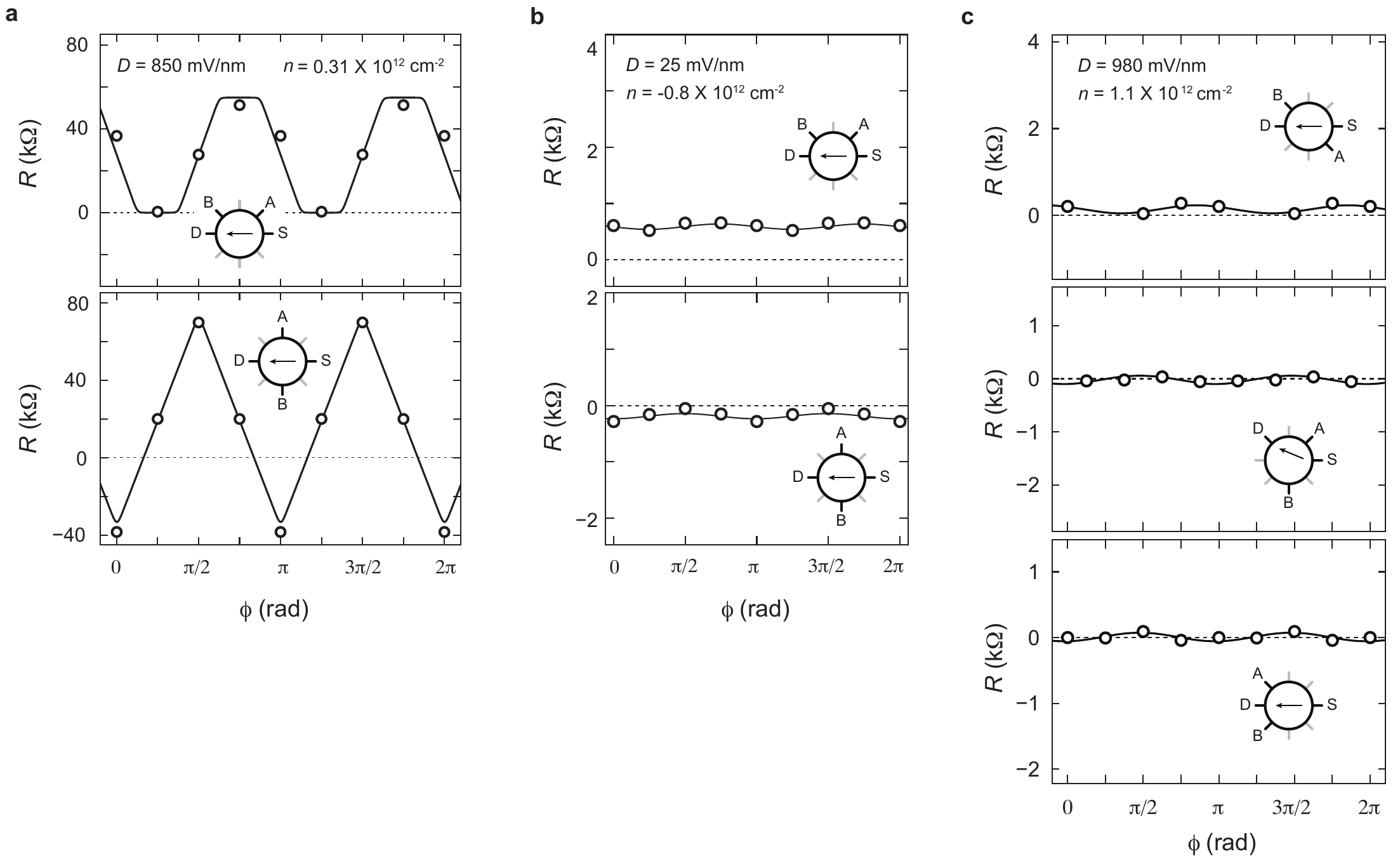}
\caption{\label{Isotropic}{\bf{Anisotropic and isotropic transport responses.}} 
Angle-resolved transport measurements performed in different regions of the low-temperature $n$--$D$ phase space: 
(a) inside the anisotropic regime, and (b--c) outside the anisotropic regime. Insets show the corresponding measurement 
configurations. Solid black lines indicate the best fits to the angular dependence. Owing to the strong anisotropy, the 
angular fit in panel (a) requires higher-harmonic components. In contrast, the angular oscillations in panels (b) and (c) 
are fully captured by Eqs.~(M1) and (M2), yielding an anisotropy ratio close to unity and indicating a nearly isotropic 
transport response. All measurements were performed on sample SF1. (a) is measured $T = 1.5$ K, whereas (b--c) are measured at $T = 30$ mK.
}
\end{figure*}

\begin{figure*}
\includegraphics[width=0.95\linewidth]{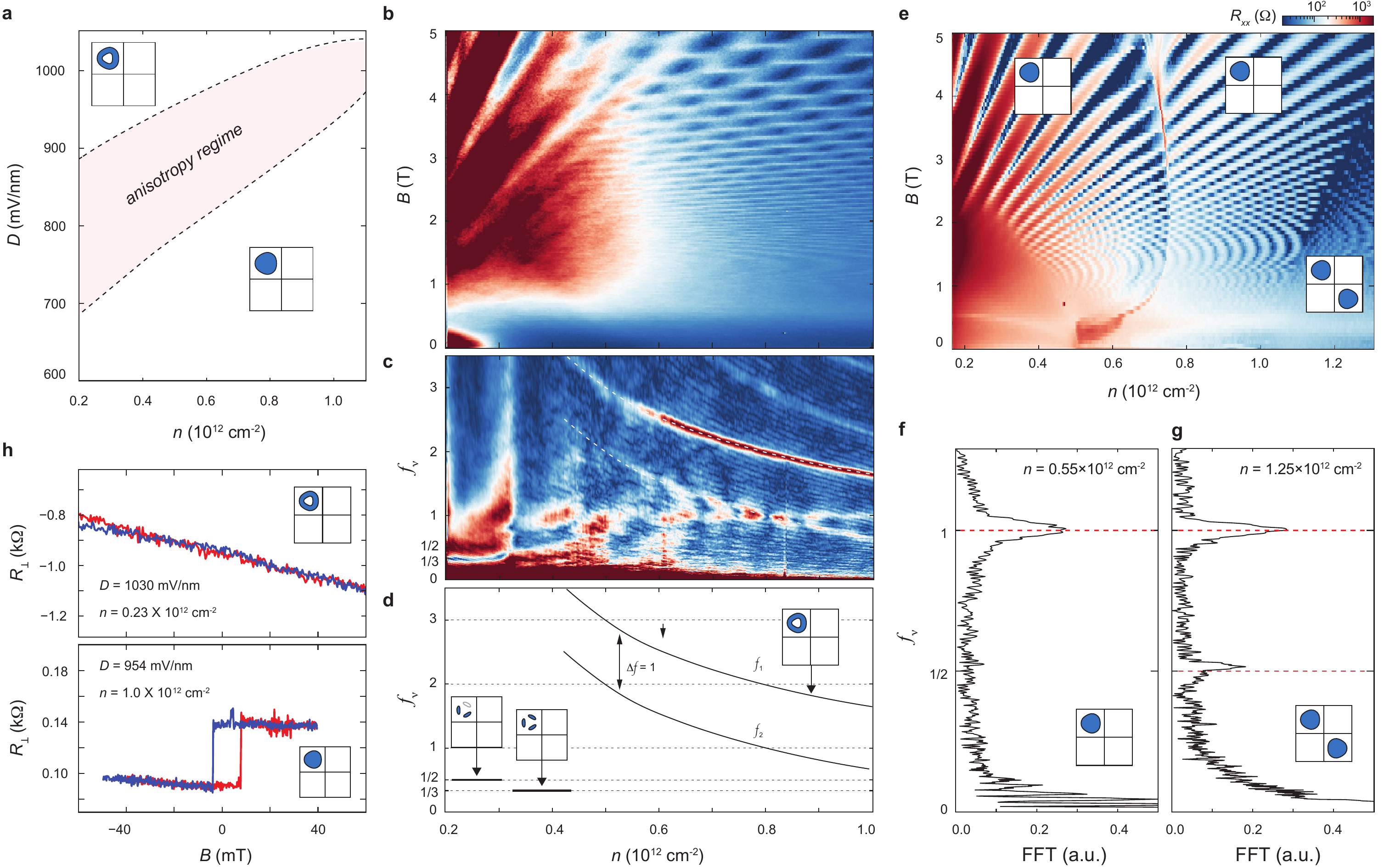}
\caption{\label{fermiology}{\bf{Fermiology around the anisotropy regime.}} 
(a) Schematic \( n \)--\( D \) map showing the stripe regime as the shaded red area.  
(b) \Rpara\ as a function of \( n \) and \( B \), measured in the upper-left corner of the \( n \)--\( D \) map at \( D = 1065\,\mathrm{mV/nm} \). 
(c) Fourier transform of quantum oscillations in (b). 
(d) Solid lines indicate the center frequencies of the most prominent peaks in \( f_\nu \) from (c). Inset: illustration of possible Fermi surface contours across different density regimes, including an annular Fermi sea on the low-density side of the stripe regime. 
(e) \Rpara\ as a function of \( n \) and \( B \), measured in the lower-right corner of the \( n \)--\( D \) map at \( D = 650\,\mathrm{mV/nm} \). 
(f--g) Fourier spectrum amplitudes of data from (e), highlighting the quarter-metal state at \( n = 0.55 \times 10^{12}\,\mathrm{cm}^{-2} \) in (f), and the half-metal state at \( n = 1.25 \times 10^{12}\,\mathrm{cm}^{-2} \) in (g).
(h) Resistance \Rperp\ as a function of magnetic field swept back and forth, exhibiting hysteresis on the high-density side of the stripe regime, but no hysteresis on the low-density side.
}
\end{figure*}

\subsection{Fermiology}

Fermiology provides a powerful means of characterizing the Fermi-surface geometry and degeneracy through Shubnikov--de Haas (SdH) quantum oscillations. However, such oscillations are conspicuously absent within the regime exhibiting pronounced transport anisotropy. This suppression of SdH oscillations can naturally arise from the presence of an extremely flat energy band, which both enhances Coulomb interactions and facilitates the stabilization of the anisotropic electronic state.

Here, we instead analyze the SdH oscillations measured in regions surrounding the anisotropic phase. Figure~\ref{fermiology}a presents a schematic of the $n$--$D$ phase space near the stripe regime, highlighted by the red shaded area. We focus on two adjacent regimes where well-defined oscillations are recovered: the region with lower $n$ and higher $D$ (to the top left)  and the region with higher $n$ and lower $D$ (to the bottom right of the anisotropy regime). 
These two regimes provide complementary insight into the evolution of the Fermi surface across the regime of large anisotropy.

Figure~\ref{fermiology}b shows the Shubnikov–de Haas (SdH) oscillations measured in the upper-left region of the $n$–$D$ phase map (lower $n$ and higher $D$). 
The resulting fish-net pattern reflects the coexistence of electron- and hole-type Fermi surfaces. A Fourier analysis reveals two dominant frequencies, $f_1$ and $f_2$, which satisfy the relation $f_1 - f_2 = 1$ over a broad density range (Fig.~\ref{fermiology}c–d). 
We attribute these oscillations to an annular Fermi sea, in which a larger electron pocket coexists with a smaller hole pocket. In this regime, the anomalous Hall effect is absent  (top panel of Fig.~\ref{fermiology}h), indicating that the Berry curvature of the underlying band is concentrated within the region enclosed by the hole surface.

We note that a nonzero resistance measured in the transverse channel does not imply a nonzero Hall coefficient, which corresponds to the antisymmetric off-diagonal component of the resistivity tensor. In the presence of transport anisotropy, a finite transverse resistance can arise even in the absence of a Hall effect when the current flow is misaligned with the principal axes of the anisotropic resistivity tensor. This behavior is a natural and well-established consequence of anisotropic transport~\cite{Wu2017nematic,Zhang2025angular,Morissette2025rhombohedral}.

\begin{figure*}[!b]
\includegraphics[width=0.95\linewidth]{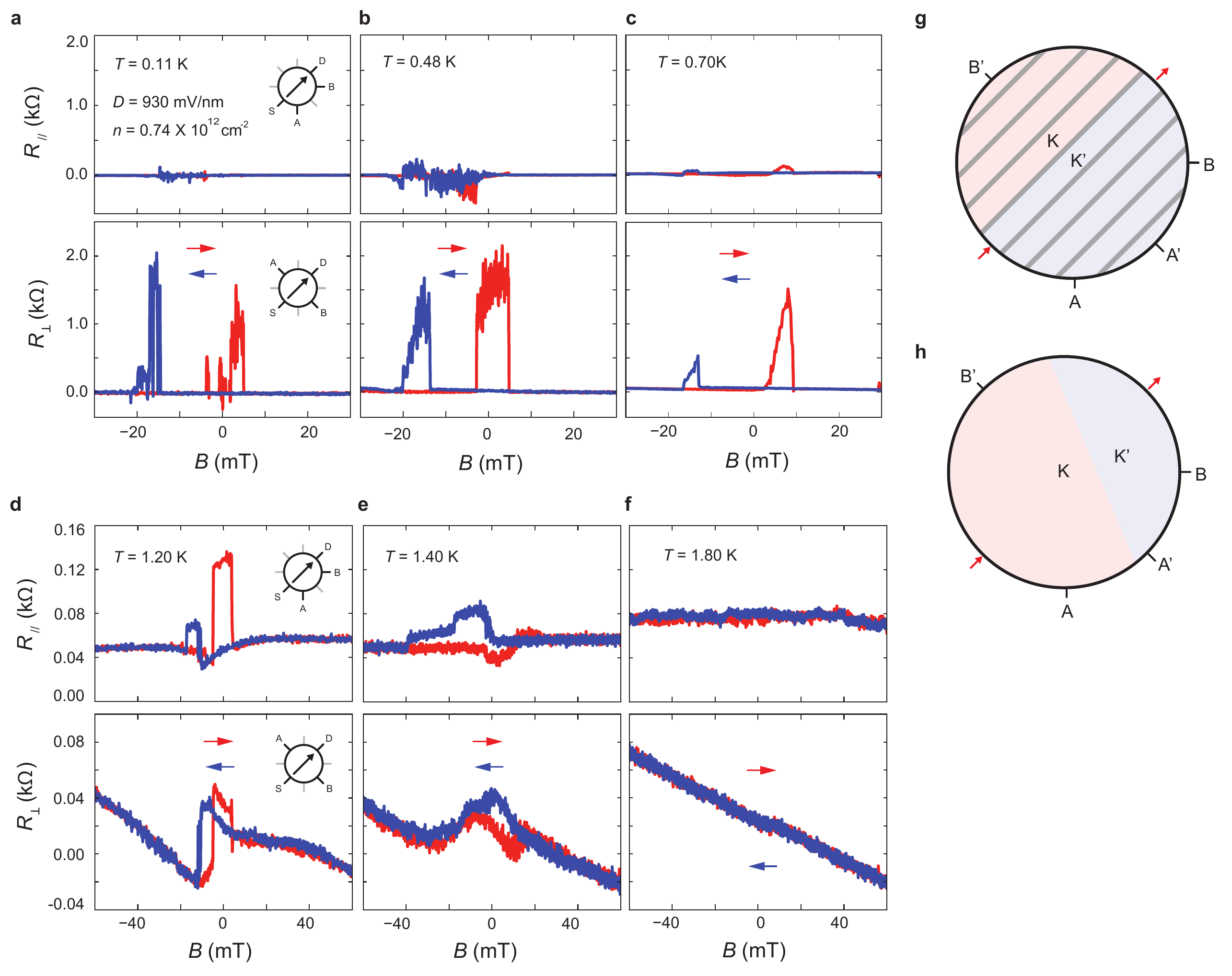}
\caption{\label{B_hysteresis}{\bf{Field-driven hysteresis in the superconducting and metallic phases.}} 
(a--f) \Rpara\ (top) and \Rperp\ (bottom), measured as the out-of-plane magnetic field $B$ is swept back and forth, with current flowing parallel to the easy axis along $\phi = 45^{\circ}$. Data in (a--c) and (d--f) are taken below and above the superconducting transition temperature, respectively. All measurements are performed at $D = 930$~mV/nm and $n = 0.74 \times 10^{12}$~cm$^{-2}$. The $B$-driven hysteresis points to an orbital ferromagnet emerging below $1.4$ K ~\cite{Sharpe2019,Serlin2019,Lin2021SOC,Chen2020ABC,Chen20201N2,Polshyn20201N2}. 
(g--h) Schematic illustration: in the superconducting phase, domain boundaries between valley $K$ and $K'$ align with the stripe orientation, generating resistive responses in \Rperp\ but not \Rpara. In the normal state, these domain boundaries are no longer locked to a principal axis, leading to hysteresis in both \Rperp\ and \Rpara. 
}
\end{figure*}

On the bottom right of the anisotropy regime, the system enters a quarter-metal phase with only electron-type carrier. Here, SdH oscillations display a dominant peak at \( f_{\nu} = 1 \) (Fig.~\ref{fermiology}f), consistent with a fully spin- and valley-polarized Fermi sea. This interpretation is supported by the magnetic-field–induced hysteresis observed in transport measurements (bottom panel, Fig.~\ref{fermiology}h). At even higher carrier densities, the quarter metal evolves into a half-metallic state. This is indicated by a peak at \( f_{\nu} = 1/2 \) in the Fourier spectrum (Fig.~\ref{fermiology}e and g).

Fermiology reveals that the regime of strong anisotropy lies between two distinct metallic phases: the metal with an annular Fermi sea and the quarter-metal phase. 
This placement suggests two possible microscopic scenarios for the emergence of rotational symmetry breaking. In the first scenario, anisotropy originates from three trigonally distorted Fermi pockets, where a Coulomb-driven flocking instability causes charge carriers to spontaneously condense into a single pocket, thereby breaking both rotational and time-reversal symmetries~\cite{Dong2021momentum}. Alternatively, anisotropy may arise from an annular Fermi sea that itself undergoes spontaneous momentum polarization, driven by strong Coulomb interactions~\cite{Jung2015momentum,Huang2023momentum}. Both scenarios are compatible with a momentum-condensed state.

\subsection{B-driven hysteresis at different temperature}

Fig.~\ref{B_hysteresis} examines the hysteretic transition driven by sweeping an out-of-plane $B$-field, as temperature warms slowly through the superconducting transition.

The magnetic-field–driven transition exhibits distinct transport responses below and above the superconducting transition. As shown in Fig.~\ref{B_hysteresis}a--c, the transition is marked by pronounced peaks in \( R_{\perp} \), whereas \( R_{\parallel} \) remains largely featureless across the same field range. Previous work has proposed that such resistive features arise when a domain wall of valley polarization intersects 
the voltage probes~\cite{Han2025chiral}. 

Within this picture, the contrasting behaviors of \( R_{\parallel} \) and \( R_{\perp} \) imply that valley domain walls preferentially align along the easy axis, as illustrated schematically in Fig.~\ref{B_hysteresis}g. In this geometry, a potential drop across the domain wall is always detected between \( A' \) and \( B' \), while it is 
largely absent between \( A \) and \( B \).  

At temperatures above \( T \approx 1.2~\mathrm{K} \), the field-driven hysteresis becomes visible in both \( R_{\parallel} \) and \( R_{\perp} \) (Fig.~\ref{B_hysteresis}e) before disappearing entirely above \( 1.4~\mathrm{K} \). 
This behavior suggests that the domain walls gradually decouple from the principal axis of transport anisotropy as the anisotropy weakens with increasing temperature. 
Taken together, these observations indicate that valley domain walls have a strong tendency to align with the principal axis in the presence of extreme anisotropy.

\begin{figure*}[!b]
\includegraphics[width=0.6\linewidth]{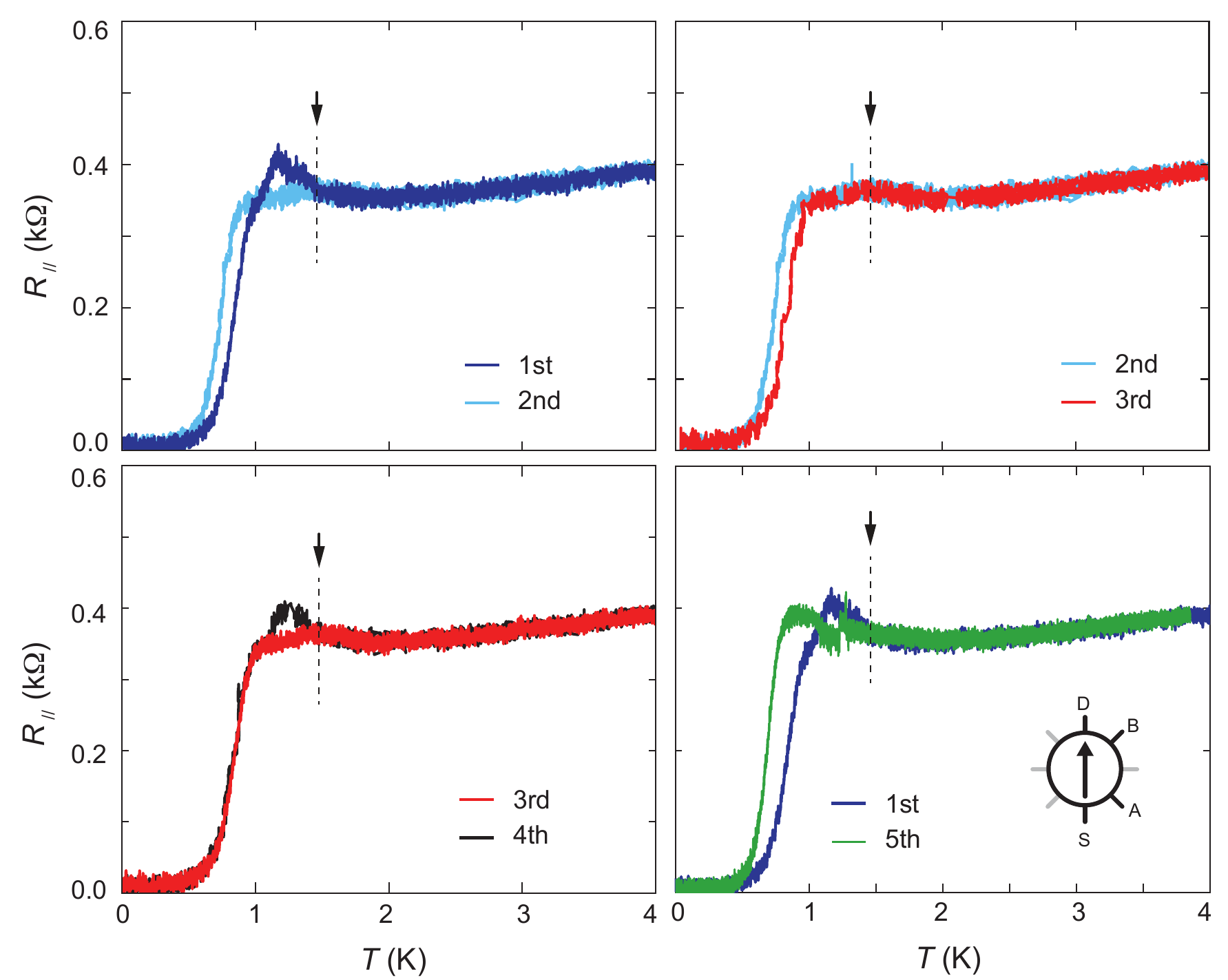}
\caption{\label{RT_warming}{\bf{$R-T$ traces from consecutive warming measurements at a fixed $n,D$.}} 
$R$--$T$ curves from five consecutive measurements, all performed at identical carrier density ($n$), displacement field ($D$), and using the same warming rate. The traces exhibit several distinct forms of hysteresis: shifts in the superconducting transition temperature ($T_c$), the presence or absence of a resistance peak just above $T_c$, and discontinuous jumps in the $R$--$T$ curve (green trace in the bottom right panel). 
}
\end{figure*}

\subsection{Superconducting transitions on warming}

Figure~\ref{RT_warming} shows $R$–$T$ curves from five consecutive measurements taken at the same $(n, D)$ values, all performed during continuous warming with an identical warming rate. While the curves exhibit highly reproducible behavior at higher temperatures, they bifurcate below the temperature indicated by the vertical dashed line.

These $R$–$T$ traces reveal two distinct forms of hysteresis. One appears as variations in the apparent superconducting transition temperature, while the other manifests as a small 
but reproducible increase in resistance preceding the transition. The latter closely resembles the behavior observed in Fig.~\ref{fig4}a.

The variations among consecutive warming traces provide a direct manifestation of changes in the underlying anisotropy between thermal cycles, offering additional support for the observations and interpretations presented in the main text.

\subsection{Supplementary Data}

In the following, we present additional supplementary data accompanied by self-contained captions for clarity.

\begin{figure*}
\includegraphics[width=0.55\linewidth]{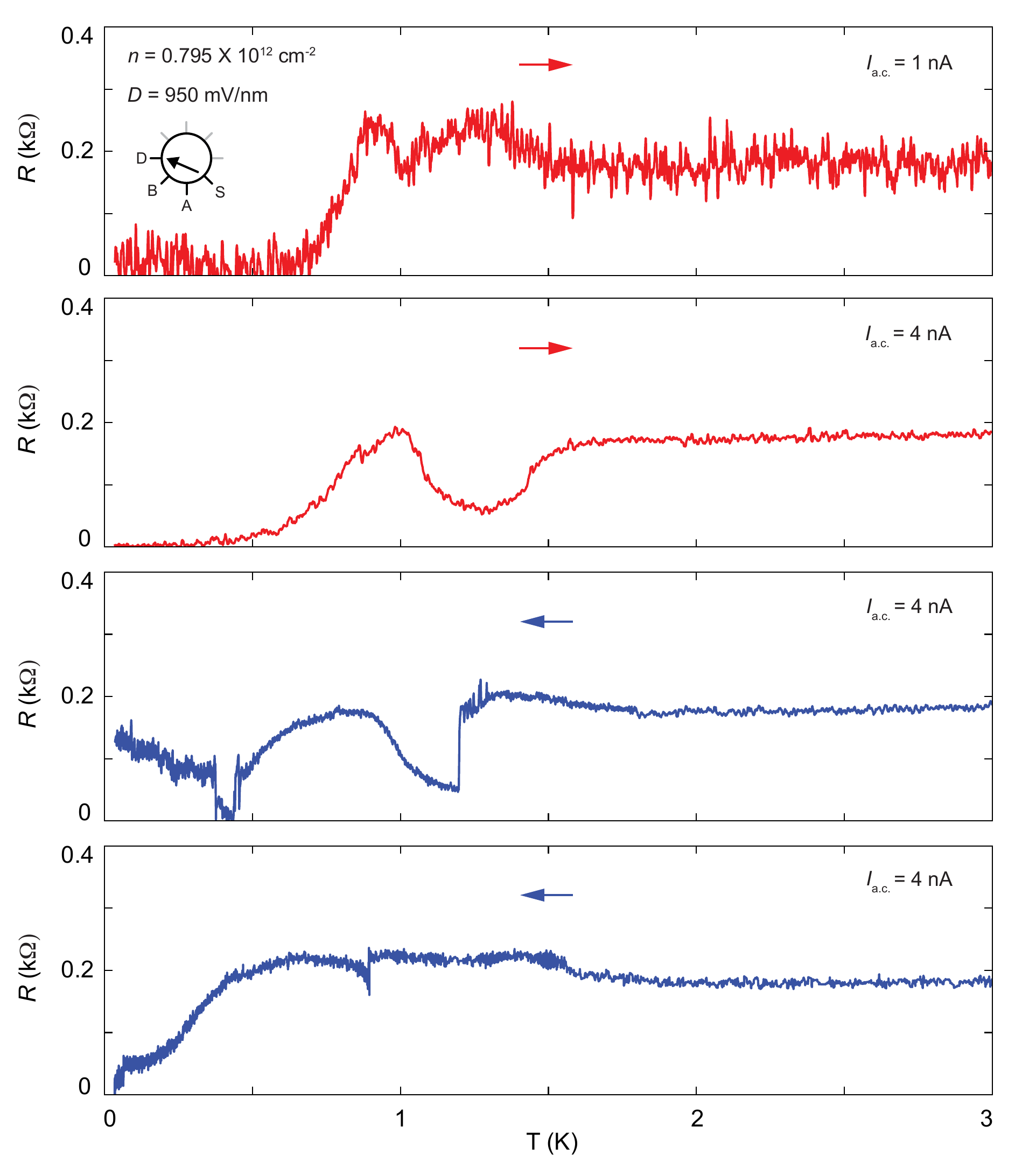}
\caption{\label{SC_multipleRTs}
\textbf{$R-T$ traces from consecutive warming and cooling measurements at a fixed $n,D$.} 
$R$--$T$ traces measured within the superconducting regime during continuous warming (red) and cooling (blue). Each time the sample is thermally reset at 3~K, a distinct $R$--$T$ trace is obtained.
}
\end{figure*}

\begin{figure*}
\includegraphics[width=0.3\linewidth]{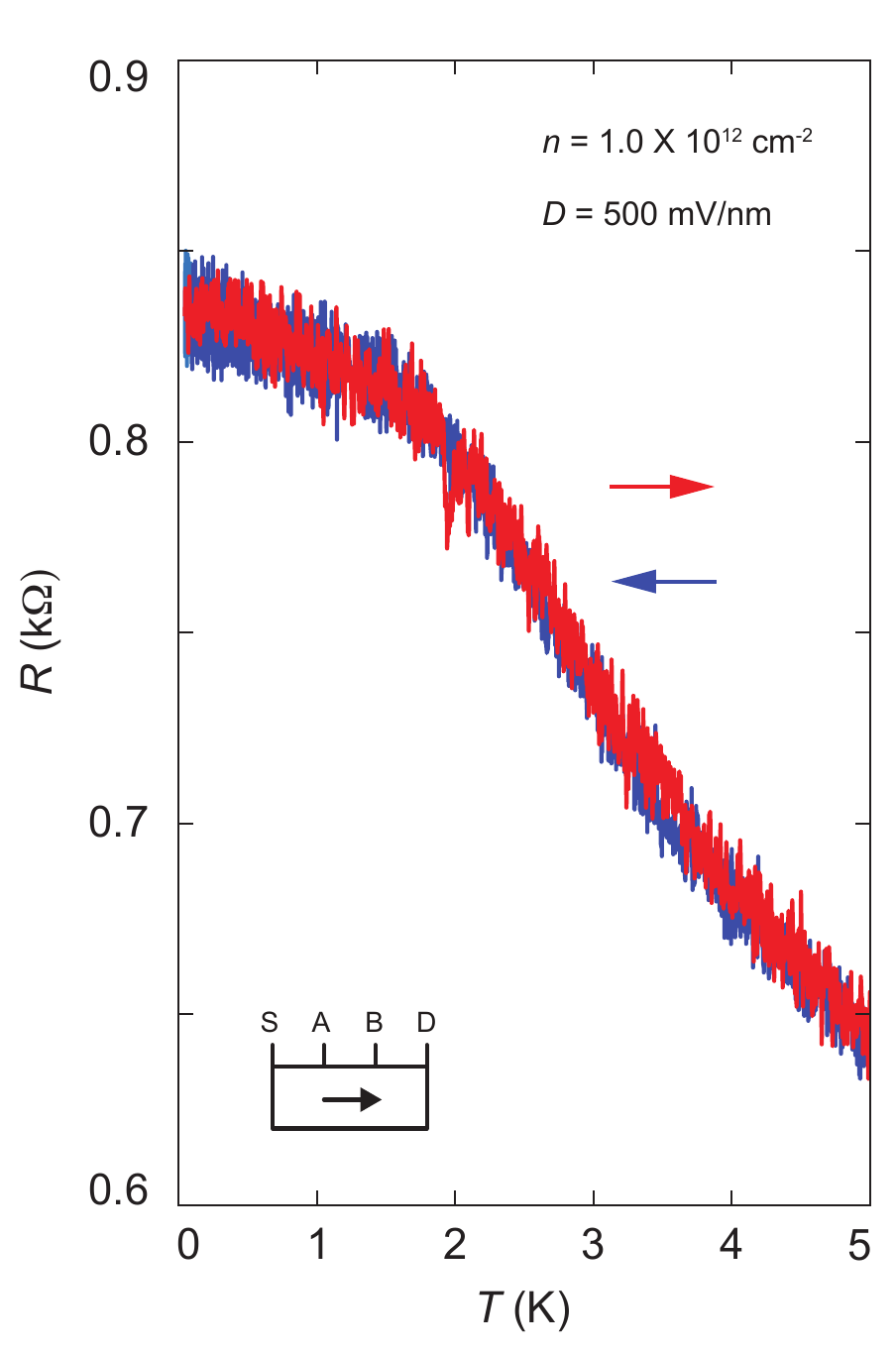}
\caption{\label{RT_control}
\textbf{Temperature sweep control.}
$R$--$T$ traces of an unpolarized metallic phase measured continuously on warming (red) and cooling (blue). 
}
\end{figure*}

\begin{figure*}
\includegraphics[width=0.5\linewidth]{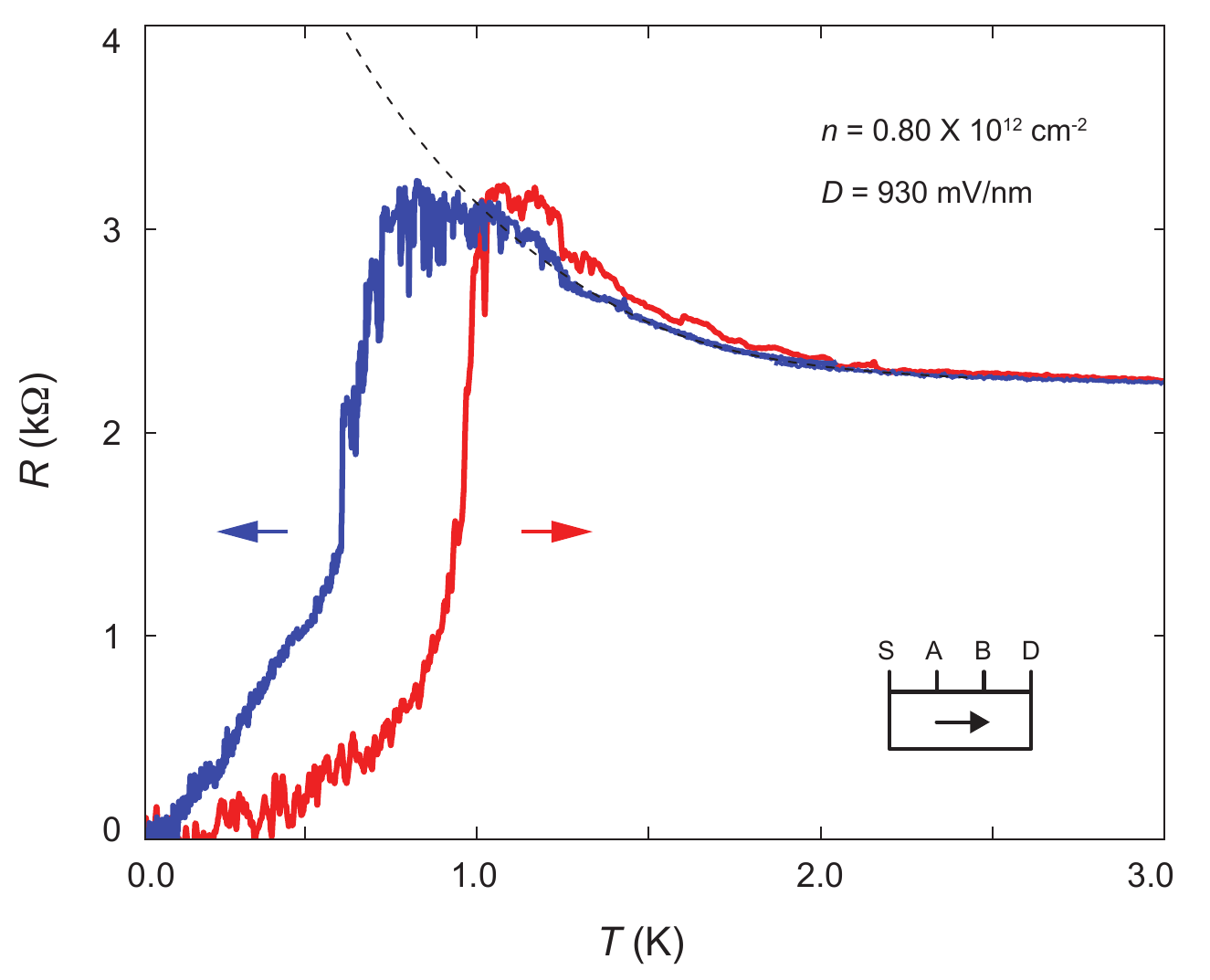}
\caption{\label{HB1_RT}{\bf{Superconducting transition in HB1.}} 
$R$--$T$ traces measured from the Hall-bar sample HB1 within the \textit{SC\,i} regime at 
$n = 0.80 \times 10^{12}~\mathrm{cm}^{-2}$ and $D = 930~\mathrm{mV/nm}$.
}
\end{figure*}

\begin{figure*}
\includegraphics[width=0.8\linewidth]{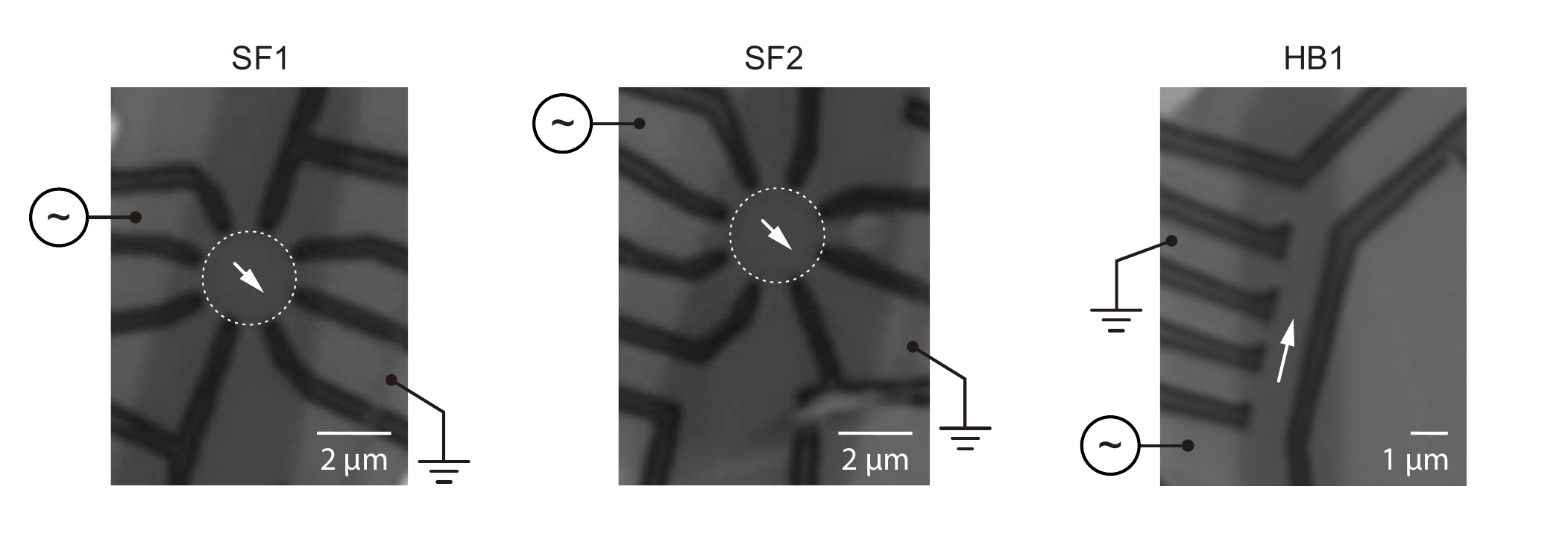}
\caption{\label{Optical}
\textbf{Sample geometries.} Optical images of the RHG samples used in this work. Two are patterned into the "sunflower" geometry with a diameter of $2.5~\mu$m, and one is patterned into a Hall-bar with a width of $1~\mu$m.
}
\end{figure*}

\begin{figure*}[h]
\includegraphics[width=0.5\linewidth]{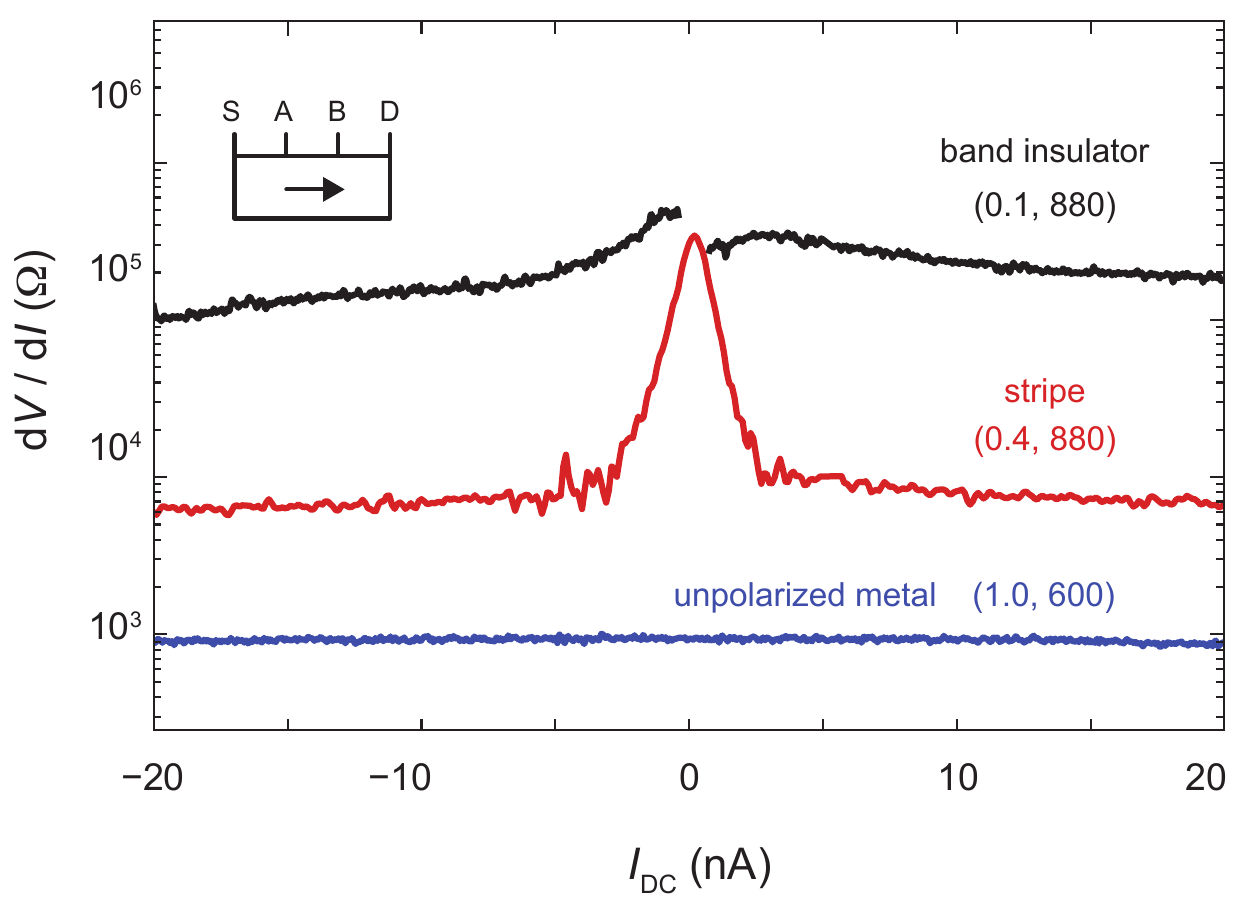}
\caption{\label{hardaxis}{\bf{IV curves of the stripe phase, band insulator, and unpolarized metal.}} 
\( dV/dI \) as a function of d.c.~current measured from HB1. The red trace is measured from inside the anisotropic regime, where as the black (blue) trace is measured from a band insulator (unpolarized metal). 
}
\end{figure*}

\begin{figure*}
\includegraphics[width=0.95\linewidth]{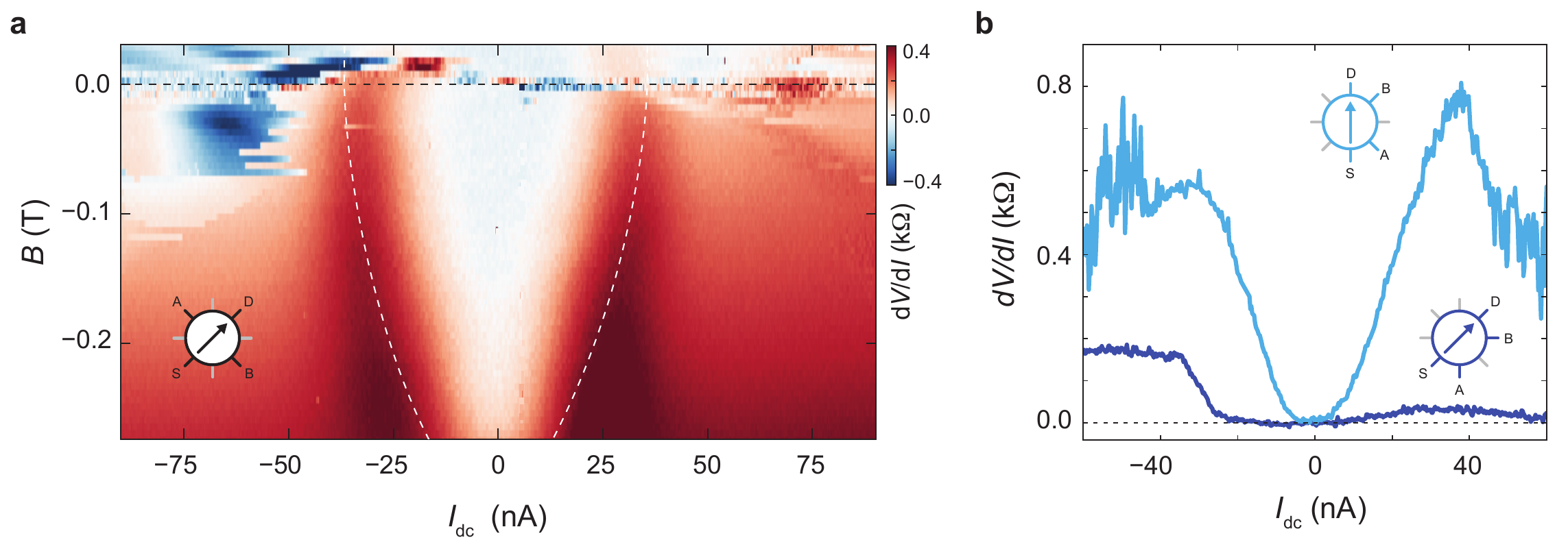}
\caption{\label{SC_IV}{\bf{Current--voltage characteristics of the \textit{SC\,i} phase.}} 
(a) Color-scale map of the differential resistance $dV/dI$ measured as a function of d.c.\ current bias and out-of-plane magnetic field $B$. 
The measurement is performed at $n = 0.74 \times 10^{12}~\mathrm{cm}^{-2}$ and $D = 954~\mathrm{mV/nm}$, with current flowing along the stripe (easy-axis) direction. 
(b) $dV/dI$ as a function of d.c.\ current measured at $B = 0$ and the same n and D as panel (a) for different current orientations $\phi$. In panel (a), enhanced noise is observed near $B = 0$, which is likely associated with current-driven switching of the underlying transport anisotropy 
(see Fig.~\ref{fig4}c--d and Fig.~\ref{IV_RT}). 
Such current-driven reconfiguration of the anisotropic state presents a unique challenge for extracting intrinsic $I$--$V$ characteristics of the superconducting phase.
}
\end{figure*}

\begin{figure*}
\includegraphics[width=0.98\linewidth]{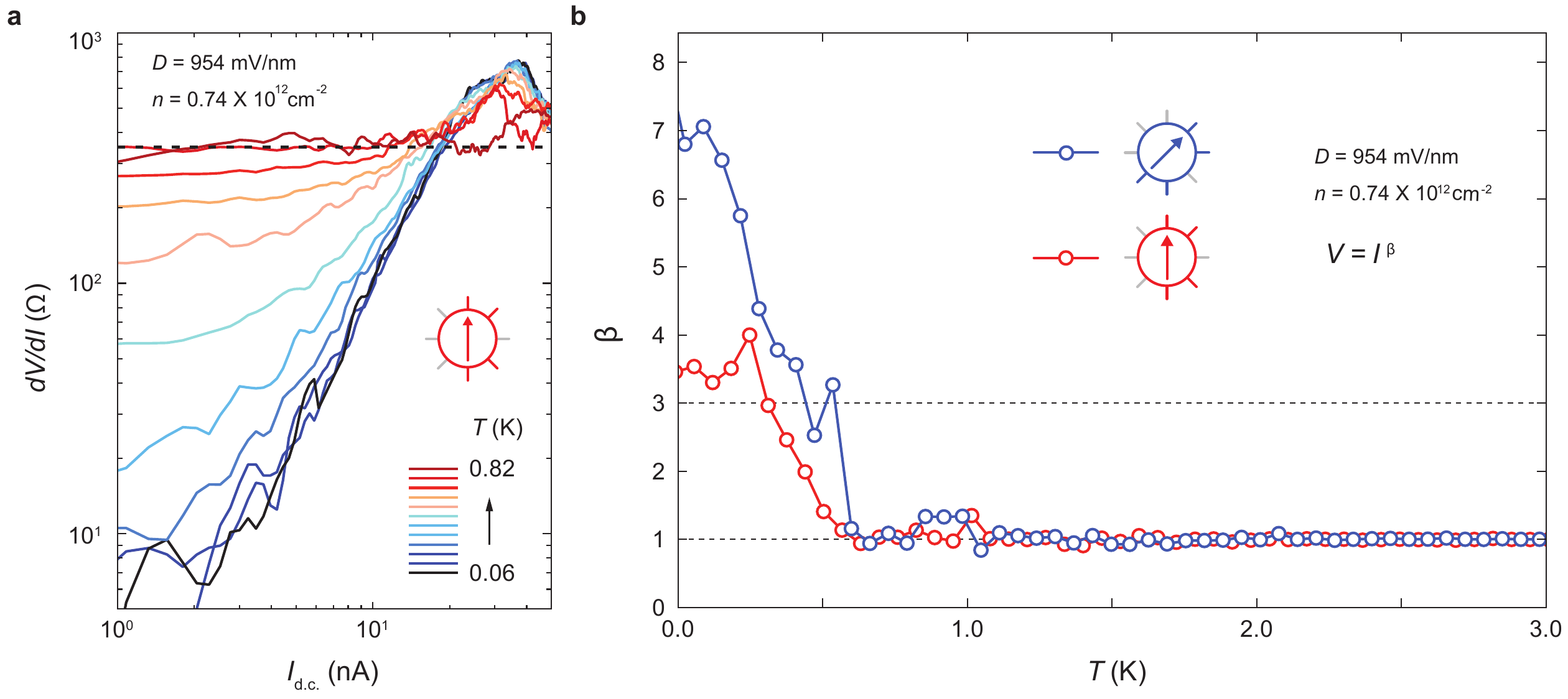}
\caption{\label{BKT}{\bf{Power-law behavior of the current--voltage characteristics.}} 
(a) Current--voltage ($I$--$V$) curves measured in the \textit{SC\,i} phase of sample SF1 at different temperatures. 
(b) Temperature dependence of the power-law exponent $\beta$, defined by $V \propto I^{\beta}$, extracted from the $I$--$V$ characteristics. Blue traces correspond to measurements with the current aligned along the easy axis, whereas the red trace is measured with the current misaligned from the easy axis. When the current is applied along the easy axis, $\beta$ exhibits a sharp jump to $\beta = 3$ around 
$T \approx 0.5~\mathrm{K}$, indicating a BKT transition at this temperature. 
This value is consistent with the superconducting transition temperature measured alongside d.c.\ current sweeps (Fig.~\ref{IV_RT}a).
}
\end{figure*}

\begin{figure*}
\includegraphics[width=0.4\linewidth]{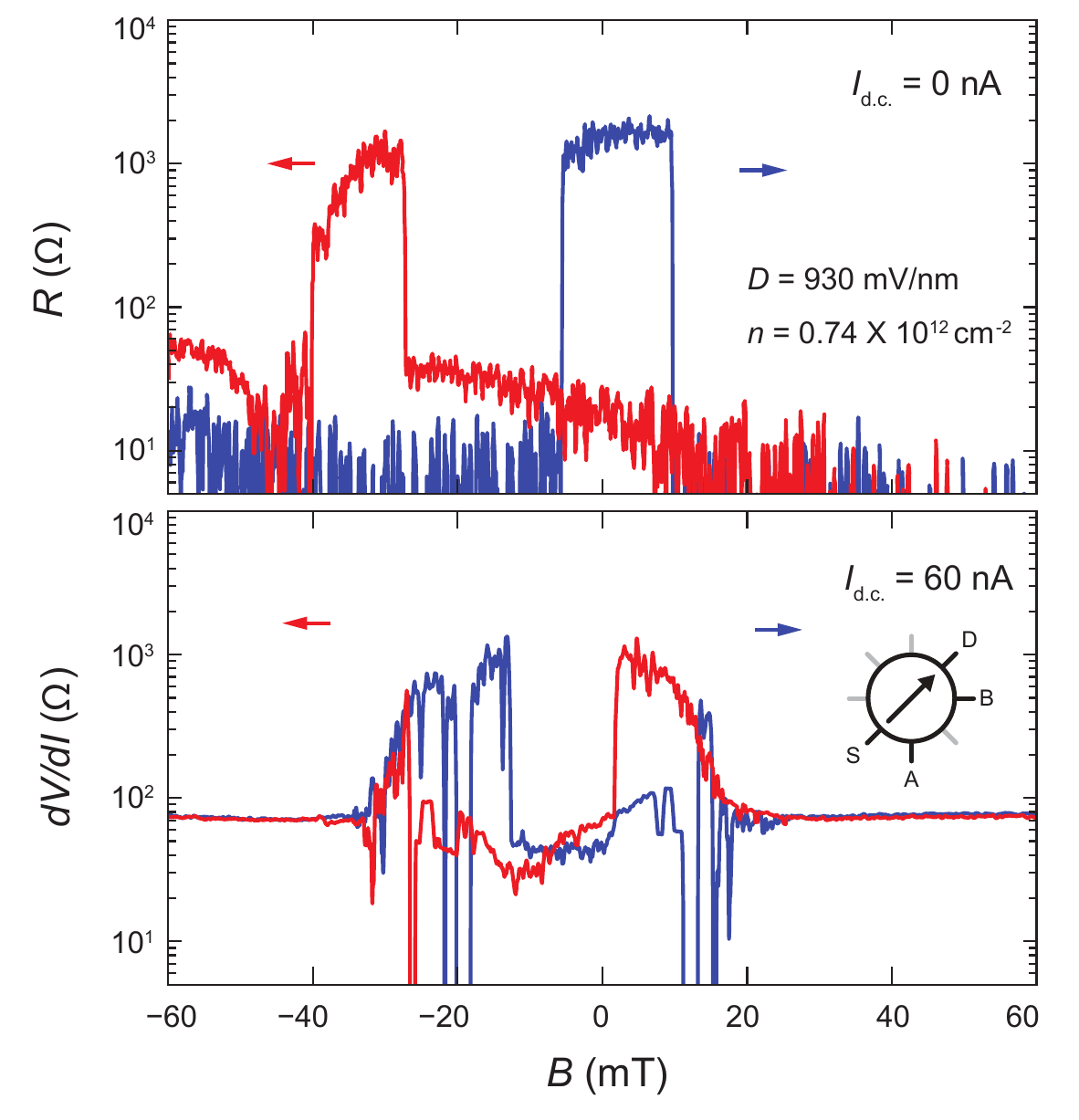}
\caption{\label{B_IV}{\bf{$B$-driven hysteresis in the presence of a d.c.\ current bias.}} 
Transport responses measured while an out-of-plane magnetic field is swept back and forth within the \textit{SC\,i} 
regime at $n = 0.74\times10^{12}~\mathrm{cm}^{-2}$ and $D = 930~\mathrm{mV/nm}$. 
The top panel is measured using a small current bias, revealing vanishing resistance away from the resistive transition. The increase in resistance in the red trace is due to temperature response from sweeping the magnetic field.
The bottom panel is measured with a d.c.\ current bias of $I_{\mathrm{d.c.}} = 60$~nA, which is sufficient to suppress 
superconductivity. In both cases, the current is applied along the easy axis to minimize its influence on the transport 
anisotropy. Nevertheless, the nature of the magnetic-field-driven hysteresis in the absence of superconductivity remains an open question, as both the applied 
current and the out-of-plane magnetic field can influence valley polarization and the underlying anisotropic order. A full 
understanding of the system’s tunability under large d.c.\ current bias is beyond the scope of the present work.
}
\end{figure*}

\begin{figure*}
\includegraphics[width=0.7\linewidth]{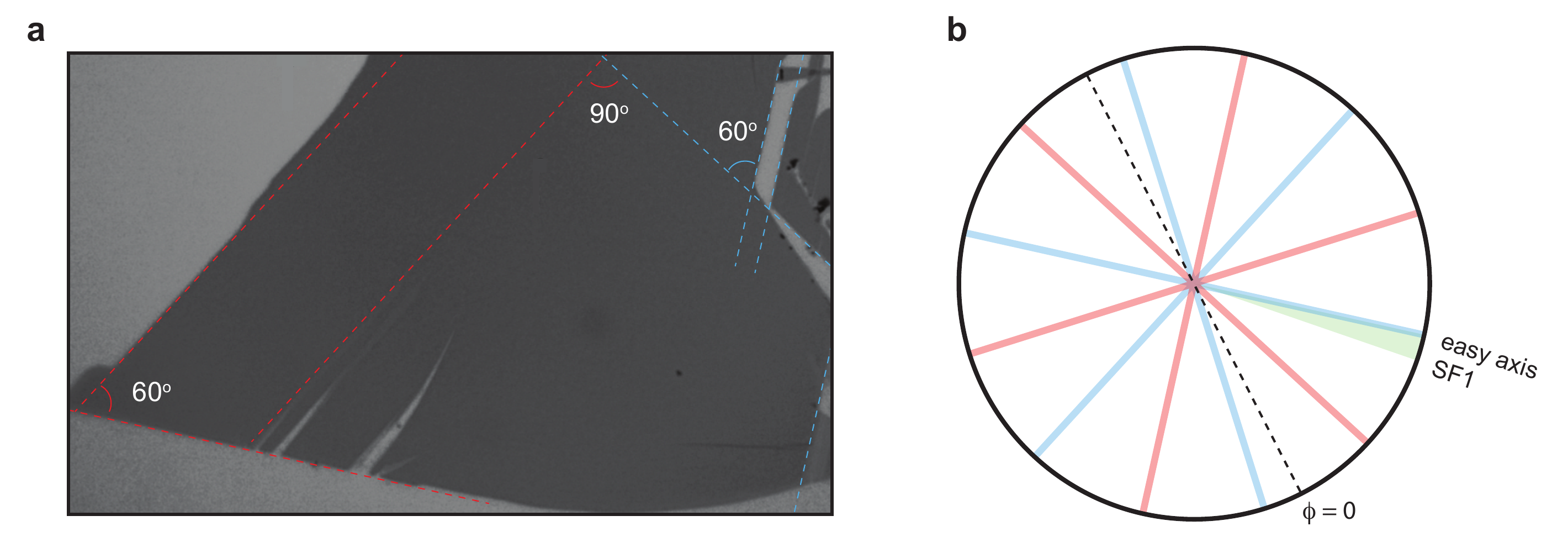}
\caption{\label{Crystal}{\bf{Relative orientation between the easy axis and crystallographic axes.}} 
(a) Optical image of the hexalayer graphene flake. Dashed red and blue lines mark straight edges, which correspond 
to crystallographic axes. The angle between the red and blue lines is $30^\circ$ modulo $60^\circ$, indicating that they 
correspond to different crystallographic directions, although their absolute crystallographic indices cannot be uniquely 
identified from the optical image alone. 
(b) Schematic comparison between the crystallographic axes and the angle-resolved transport results. The black dashed 
line defines the reference direction $\phi = 0$, while the green shaded cone highlights the orientation of the easy axis 
extracted from angle-resolved transport measurements in SF1 (see Fig.~2).
}
\end{figure*}

\begin{figure*}
\includegraphics[width=0.35\linewidth]{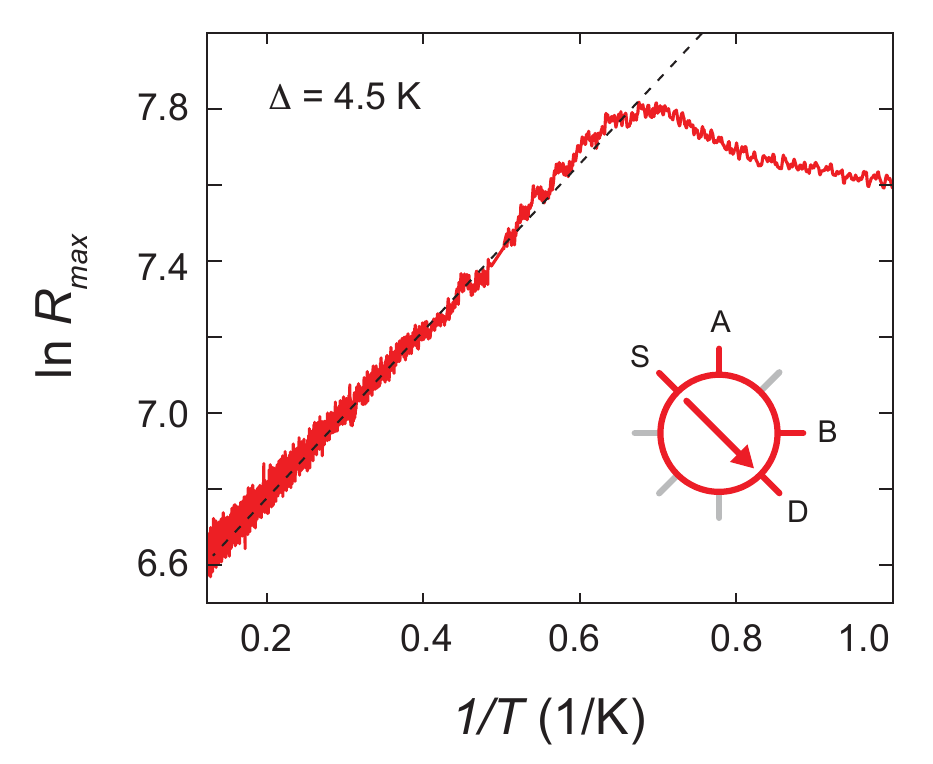}
\caption{\label{Activate_SI}
\textbf{Activated behavior}. Arrhenius plot corresponding to the insulating trace in Fig.~\ref{fig3}b. The measurement is performed at $n = 0.74\times10^{12}~\mathrm{cm}^{-2}$ and $D = 954~\mathrm{mV/nm}$.
}
\end{figure*}

\begin{figure*}
\includegraphics[width=0.78\linewidth]{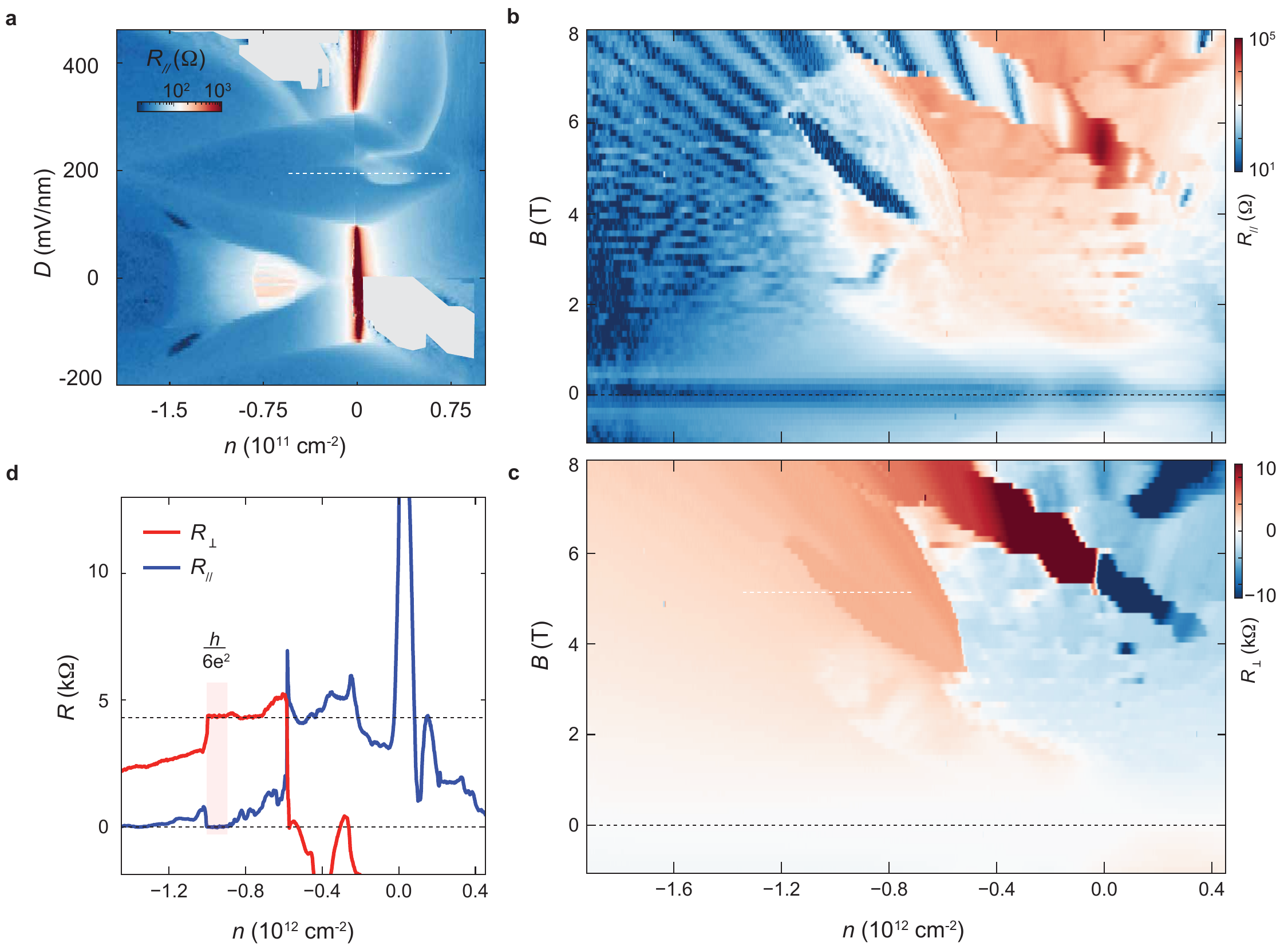}
\caption{\label{Layer} 
\textbf{Identification of rhombohedral hexalayer graphene.} 
Previous studies have established that the number of layers in a rhombohedral multilayer is encoded in the Chern number of a certain orbital ferromagnetic state~\cite{Han2024rhombohedral,Sha2024rhombohedral}. 
(a)~Color-scale map of the longitudinal resistance, with the white dashed line cutting through  this state. 
(b,c)~$n$--$B$ Landau fan diagrams of (b)~$R_{\parallel}$ and (c)~$R_{\perp}$ measured along the dashed line in panel~(a). The orbital ferromagnetic state is continuously connected to a Chern state that emerges between $4$ and $6$~T. 
(d)~Line cut through this Chern state, highlighted by the shaded red stripe, showing a quantized plateau  of $R_{\perp}=h/(6e^2)$, accompanied by vanishing longitudinal resistance. The appearance of a Chern-number 6 plateau identifies the sample as rhombohedral hexalayer graphene.
}
\end{figure*}

\begin{figure*}
\includegraphics[width=0.5\linewidth]{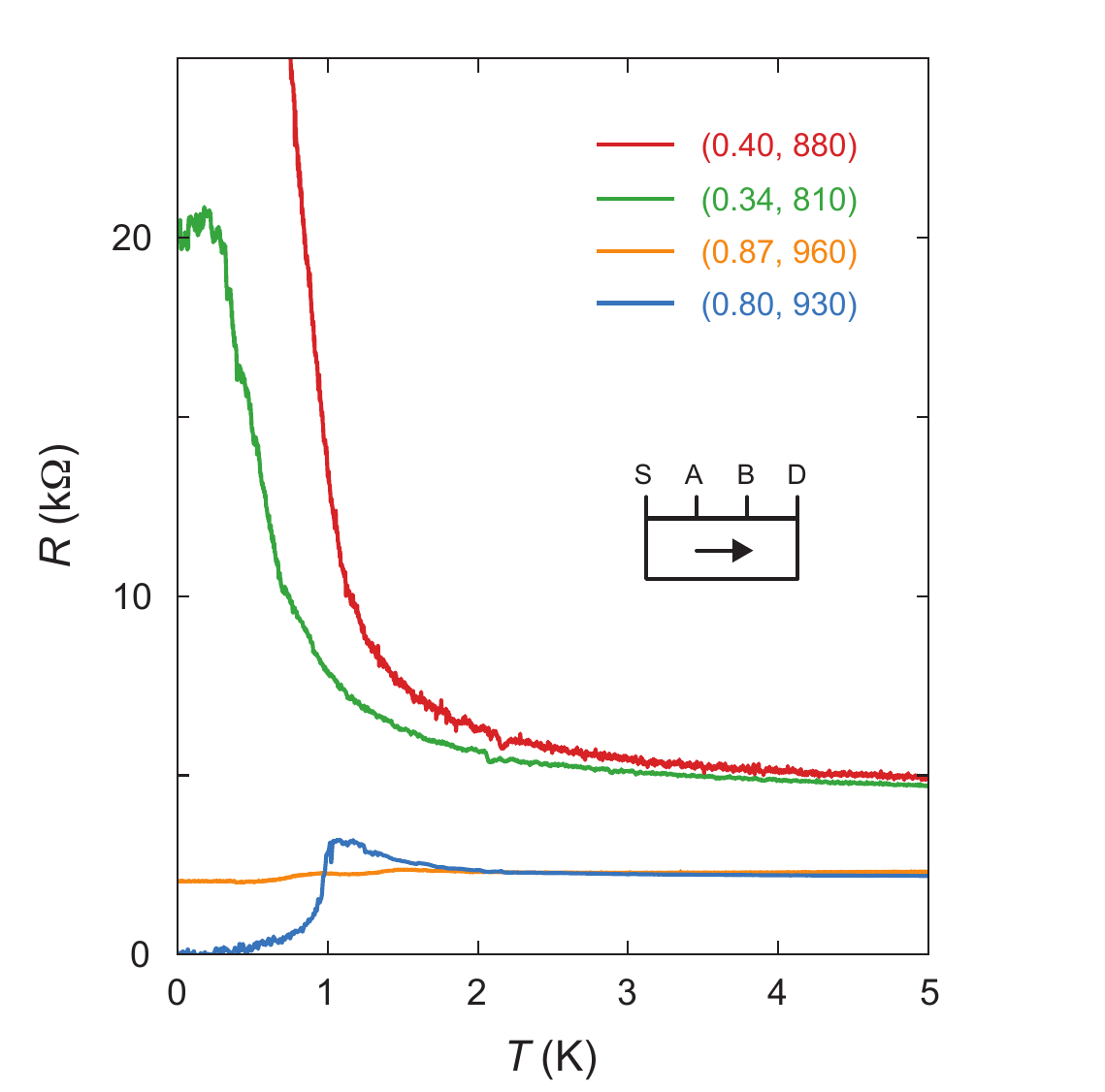}
\caption{\label{HB1_RTs} 
\textbf{Distinct behaviors between sunflower and Hall-bar samples.}
$R$--$T$ traces measured at different $(n, D)$ values across the stripe regime from device HB1. Notably, depending on the $(n, D)$ values, the transport response measured along the same current-flow direction exhibits distinct trends at low temperature. For example, the blue curve shows superconducting behavior at low temperature, whereas the red and green traces diverge at low temperature. This behavior is distinct from that observed in the sunflower samples. As illustrated in Fig.~\ref{fig1}c, when current flows along $45^{\circ}$ the sample remains highly conductive across the stripe regime, whereas when current flows along $135^{\circ}$ the sample consistently exhibits insulating behavior regardless of the $(n, D)$ values. These observations suggest that the stripe orientation varies with $(n, D)$ in the Hall-bar sample, while remaining largely independent of $n$ and $D$ in the sunflower samples.
}
\end{figure*}

\end{widetext}

\end{document}